\documentclass[aps,prd,twocolumn,superscriptaddress,nofootinbib]{revtex4-2}

\usepackage{amsmath}
\usepackage{amssymb}
\usepackage{graphicx}
\usepackage{float}
\usepackage{multirow}
\usepackage[pdftex,breaklinks,colorlinks,
linkcolor=Blue,
citecolor=teal,
anchorcolor=red,
urlcolor=cyan,
pdfencoding=auto]{hyperref}
\usepackage[dvipsnames]{xcolor}
\usepackage{array}
\usepackage{dcolumn}
\usepackage{mathtools}
\usepackage{orcidlink}
\usepackage[normalem]{ulem}
\usepackage{mathrsfs}
\usepackage[scr=boondoxo]{mathalpha}
\usepackage{bm}

\usepackage{accents}

\newcommand{\e}{\varepsilon}
\newcommand{\ee}{\epsilon}
\newcommand{\E}{\sigma}
\newcommand{\trunctwo}{+\mathcal{O}(\varepsilon^2)}
\newcommand{\truncthree}{+\mathcal{O}(\varepsilon^3)}

\begin{document}

\title{Post-adiabatic self-force waveforms: slowly spinning primary and precessing secondary}

\author{Josh Mathews\,\orcidlink{0000-0002-5477-8470}}
\affiliation{Department of Physics, National University of Singapore, 21 Lower Kent Ridge Rd, Singapore 119077}

\author{Barry Wardell\,\orcidlink{0000-0001-6176-9006}}
\affiliation{School of Mathematics \& Statistics, University College Dublin, Belfield, Dublin 4, Ireland, D04 V1W8}
\author{Adam Pound\,\orcidlink{0000-0001-9446-0638}}
\affiliation{School of Mathematical Sciences and STAG Research Centre, University of Southampton, Southampton, United Kingdom, SO17 1BJ}
\author{Niels Warburton\,\orcidlink{0000-0003-0914-8645}}
\affiliation{School of Mathematics \& Statistics, University College Dublin, Belfield, Dublin 4, Ireland, D04 V1W8}

\begin{abstract}
Recent progress in gravitational self-force theory has led to the development of a first post-adiabatic (1PA) waveform model for nonspinning, quasicircular compact binaries [Phys. Rev. Lett. 130, 241402 (2023)]. In this paper, we extend that model to allow for a slowly spinning primary black hole and a generic, precessing spin on the secondary object, restricting to the case of small misalignment between the primary spin and the orbital angular momentum. We demonstrate excellent agreement between our waveforms and fully nonlinear numerical relativity simulations for mass ratios $q\gtrsim 5$ and primary spins $|\chi_1|\lesssim 0.1$ and arbitrary secondary spin $\chi_2 \lesssim 1$. In particular we present the re-summed 1PAT1R waveform model, which significantly improves the accuracy of the original 1PAT1 waveforms for comparable masses and increasing primary spin. Our models are publicly available in the WaSABI package.
\end{abstract}

\maketitle

\tableofcontents

\section{Introduction}
\label{sec:introduction}
The detection of gravitational wave signals from the merger of compact-object binaries is now routine, with nearly 100 events observed so far by the LIGO-Virgo-KAGRA collaboration \cite{KAGRA:2021vkt}, and many more expected in the current ``O4'' observing run. From a waveform modelling perspective most of these events were qualitatively very similar, involving comparable-mass systems undergoing a quasi-circular inspiral. As detectors have become more sensitive and more events have been observed, however, observations of qualitatively different systems involving precession \cite{Hannam:2021pit} and mass-asymmetry \cite{LIGOScientific:2020stg} have begun to emerge. These observations bring with them the dual consequences of potentially interesting new physical processes to explore, while simultaneously introducing significant new challenges for waveform modelling efforts. With the development of next-generation ground-based gravitational wave detectors underway \cite{Abac:2025saz,Evans:2021gyd}, and with the recent adoption of the LISA mission~\cite{LISA:2024hlh}, it is expected that observations of such systems will become commonplace, and significant developments on the waveform modelling side will be required to keep pace.

With a view toward producing fast and accurate models of precessing, asymmetric binaries, post-adiabatic self-force models have emerged as a promising solution to the waveform modelling problem posed by next-generation gravitational wave detectors~\cite{LISAConsortiumWaveformWorkingGroup:2023arg}. Although self-force models were initially devised for systems with highly asymmetric masses, such as extreme-mass-ratio inspirals (EMRIs)~\cite{LISA:2022yao}, there is a growing body of evidence for good convergence of the small-mass-ratio expansion used in self-force theory, even for binaries with only moderately asymmetric masses. This has been illustrated via comparisons between numerical relativity (NR) and self-force predictions of the binding energy and redshift~\cite{LeTiec:2011dp, Albalat:2022lfz}, periastron advance~\cite{LeTiec:2011bk, LeTiec:2013uey, Ramos-Buades:2022lgf}, gravitational energy flux~\cite{Warburton:2021kwk, Ramos-Buades:2022lgf, Warburton:2024xnr} and the inspiral and waveform~\cite{ vandeMeent:2020xgc, Wardell:2021fyy, Albertini:2022rfe, NavarroAlbalat:2022tvh, Ramos-Buades:2022lgf,Islam:2025tjj}. While these studies are limited to a small part of the parameter space, there is certainly cause to be optimistic about the performance of self-force theory in modelling generic binaries. 

The most accurate self-force waveform model, constructed in Ref.~\cite{Wardell:2021fyy}, is based on second-order self-force calculations~\cite{Miller:2020bft,Warburton:2021kwk,Miller:2023ers,Spiers:2023mor,Cunningham:2024dog}, expanding the spacetime metric to second order in the binary's small mass ratio. Within a multiscale formulation of the field equations~\cite{Miller:2020bft,Miller:2023ers}, this leads to a ``first post-adiabatic'' (1PA) waveform model, which is the target order of accuracy for modelling EMRIs for LISA~\cite{LISAConsortiumWaveformWorkingGroup:2023arg}. However, the waveform model in Ref.~\cite{Wardell:2021fyy}, dubbed 1PAT1, was restricted to nonspinning, quasicircular binaries. In this paper, we extend the 1PAT1 model to binaries with a generic (precessing) spinning secondary and a slowly spinning primary, restricting to quasicircular orbits and primary spins that have a small misalignment with the orbital angular momentum. 

This represents a specialized application of the multiscale framework for spinning binaries with generic spin orientations presented in Ref.~\cite{Mathews:2025nyb} (hereafter `Paper~I'), which built on Refs.~\cite{Pound:2021qin,Drummond:2023wqc,Grant:2024ivt,Piovano:2024yks,Witzany:2024ttz,Skoupy:2025nie} (among others). As explained in  Paper~I and elsewhere, the precessing component of the secondary spin enters the waveform as a second post-adiabatic (2PA) effect, making it subdominant relative to the other spin effects we compute; the inclusion of the \emph{non}precessing components is therefore the main advance in this paper. Nevertheless, we include the precession because its modulation of the waveform is distinct from from all the 2PA terms we omit. We refer to Paper I for a detailed discussion and a summary of other recent progress toward fully generic binary configurations.

Through comparisons with NR waveforms, we find that our waveform model is equally accurate for spinning as for nonspinning binaries (assuming sufficiently small primary spin). We also show that a re-summed variant, labelled 1PAT1R, significantly increases the phase accuracy of 1PA models at higher values of the primary spin and for binaries with comparable masses. This re-summation, which is based on an exact law of general relativity, provides a simple way of extending the coverage of self-force waveform models toward equal-mass systems.

The multiscale form of our model makes it readily incorporable into rapid-waveform generators~\cite{Katz:2021yft,Nasipak:2023kuf,Chapman-Bird:2025xtd}, enabling fast and accurate waveform generation for spinning binaries with mass ratios as low as $ q \gtrsim 5$ (and moderate accuracy for $q$ even closer to unity). While our model is restricted to the inspiral stage of the binary evolution, it should be suitable for data analysis studies of intermediate- and extreme-mass-ratio systems~\cite{Burke:2023lno} in preparation for next-generation GW detectors~\cite{LISA:2017pwj,Colpi:2024xhw,TianQin:2015yph,Li:2024rnk,Gong:2021gvw, Luo:2019zal,Kawamura:2011zz, Punturo:2010zz, Reitze:2019iox}. It should also be extendable through the merger and ringdown using the methods of Refs.~\cite{Kuchler:2024esj,Kuchler:2025hwx,Roy:2025kra}. We expect it can additionally be used to calibrate other inspiral-merger-ringdown models (see, e.g., Refs.~\cite{vandeMeent:2023ols,Albertini:2024rrs, Leather:2025nhu}) with more extensive coverage of the parameter space.

\subsection{Outline and conventions}
Our presentation begins with the multiscale expansion of the equations of motion and Einstein field equations and ends with a summary of our waveform models. In Sec.~\ref{sec:sf} we summarise self-force theory with a spinning compact secondary including a convenient parameterization of the secondary's spin introduced in Paper I. In Sec.~\ref{sec:tt}, we specialise the multiscale expansion of Paper I to a secondary with a generic precessing spin in a quasi-circular inspiral about a slowly spinning primary. The primary's spin axis is restricted to have a small, ${\cal O}(1/q)$ opening angle relative to the orbital angular momentum. This specialisation is chosen to build a spinning-binary waveform model compatible with the currently available second-order self-force flux data~\cite{Warburton:2021kwk}. In Sec.~\ref{sec:fluxbal}, we derive an approximate balance law through 1PA order for the binary configuration described in the previous section, relying upon a particular approximation to the binary's binding energy.
In Sec.~\ref{sec:implementation}, we give an overview of the numerical calculations of the required self-force data at the level of the field equations. In Sec.~\ref{sec:waveform}, we employ the balance law in a series of five different inspiral and waveform models and compare them with NR simulations. In Sec.~\ref{sec:conclusions} we summarise our findings and avenues for future extensions. We also comment on the next steps for the 1PA waveform program of research. 

We keep the conventions of Paper I, working in geometric units with $G=c=1$. We denote the individual masses as $m_i$ with $m_1\geq m_2$. We use $\e\equiv 1$ as a counting parameter of the small mass ratio $\ee\equiv m_1 / m_2$ and define $q\equiv\ee^{-1}$. The total mass is $M\equiv m_1+m_2$. $\chi_i$ are the dimensionless spin magnitudes $S_i/m_i^2$, where $S_i$ are the spin angular momenta of the two bodies. For astrophysical \emph{compact} binaries, $S_i\sim m_i^2$ and so in the small-mass-ratio expansion $S_1\sim\e^0$ and  $S_2\sim \e^2$. For each dimensionless spin this implies $\chi_i\sim \e^0$. If the body is a Kerr black hole, then we have the more precise restriction $0\leq|\chi_i|\leq1$. When considering a slowly spinning primary black hole, we impose the extra restriction that $\chi_1\sim\e$.
 
\section{Overview of Self-force theory with a spinning secondary}
\label{sec:sf}

In this section we provide a brief overview of self-force theory with a spinning secondary through second order in perturbation theory. For a more detailed review, see Paper I.  For compactness, we adopt the self-consistent formulation~\cite{Pound:2009sm, Miller:2020bft} and leave the background spacetime and binary configuration generic. In Sec.~\ref{sec:tt}, we reformulate the problem in a two-timescale expansion. There we also specialise the binary configuration, restricting to a slowly spinning primary with a small opening angle between its spin and the orbital angular momentum, and neglecting eccentricity.

\subsection{Field equations and effective metric}

The Einstein field equations governing the compact binary's metric ($\textbf{g}_{\mu\nu}$) are
\begin{equation}
\label{eq:FE}
G_{\mu \nu}(\textbf{g})=8\pi T_{\mu \nu},
\end{equation}
where $T_{\mu\nu}$ is an effective stress-energy tensor for the secondary object, described below. Taking the usual black hole perturbation theory approach, we solve the field equations by expanding the binary's metric in powers of $\e$: 
\begin{equation}
\label{eq:metric}
\textbf{g}_{\mu \nu}=g_{\mu \nu}+\e h_{\mu \nu}^{1}+\e^2 h_{\mu \nu}^{2} \truncthree,
\end{equation}
where $g_{\mu \nu}$ is the spacetime of the primary black hole, and $h_{\mu \nu}^{1}$ and $h_{\mu \nu}^{2}$ are the first and second order metric perturbations, respectively.

Since it absorbs gravitational radiation during the binary's evolution, the primary black hole itself slowly evolves. We account for this by splitting the mass and spin into constant terms ($m_1^{(0)}$ and $\chi_1^{(0)}$) and small, dynamical corrections ($\delta m_1$ and $\delta\chi_1$). We specialize to a slowly spinning primary by choosing $\chi_1^{(0)}=0$, such that $g_{\mu\nu}$ is a Schwarzschild metric with mass parameter $m_1^{(0)}$. The perturbations $h^{n}_{\mu\nu}$ then include terms that are $n$th order in $\delta m_1$ and $\delta\chi_1$. 

The secondary object, which can be a black hole or a material body, is assumed to be sufficiently compact compared to the external scale of spacetime curvature to allow us to adopt the pole-dipole approximation through second order in $\e$~\cite{Mathews:2021rod}. Using the method of matched asymptotic expansions, Refs.~\cite{Gralla:2008fg,Pound:2012dk,Upton:2021oxf} showed that for such an object, the stress-energy tensor can be written as
\begin{equation}
\label{eq:pole-dipole}
 T^{\alpha \beta}=\e T_{(m)}^{\alpha \beta} + \e^2 T_{(d)}^{\alpha \beta} + \mathcal{O}(\e^3),
\end{equation}
where $T_{(m)}^{\alpha \beta}$ is a mass-monopole term and $T_{(d)}^{\alpha \beta} $ is a spin-dipole term; quadrupole and higher moments would appear at higher orders in $\e$.
The two multipole terms have the form of a ``gravitational skeleton''~\cite{Mathisson:1937zz,Mathisson:2010opl},
\begin{subequations}
\begin{align}
\label{eq:SEmono}
T_{(m)}^{\alpha \beta}&=m_2\int d \hat\tau' \, \frac{\delta^{4}\left[x^{\mu}- z^{\mu}(\tau' )\right]}{\sqrt{-\hat g'}} \hat u^{\alpha}(\tau') \hat u^{\beta}(\tau' ),\\
\label{eq:SEdipole}
T_{(d)}^{\alpha \beta} &= m_2^2 \hat \nabla_{\rho} \left[ \int d\hat \tau'  \, \frac{\delta^{4}\left[x^{\mu}- z^{\mu}(\tau' )\right]}{\sqrt{-\hat g'}} \hat u^{(\alpha}(\tau' ) {\hat S}^{\beta)\rho}(\tau') \right]\!,
\end{align}
\end{subequations}
where $\delta^{4} $ is the four-dimensional Dirac delta function, $z^\mu$ is the object's effective center-of-mass worldline, $\hat S^{\alpha\beta}$ is the object's dimensionless (mass-normalized) spin tensor, and primes are used to indicate evaluation at $z^\mu(\tau')$. 

Importantly, the stress-energy terms~\eqref{eq:SEmono} and \eqref{eq:SEdipole} take the form of a spinning particle in a certain \emph{effective} vacuum metric $\hat g_{\alpha\beta}$ rather than in the external background $g_{\alpha\beta}$. The proper time $\hat\tau$, four-velocity $\hat u^\alpha\equiv \frac{dz^\alpha}{d\hat\tau}$, metric determinant $\hat g$, and covariant derivative $\hat\nabla_\alpha$ are all defined from $\hat g_{\alpha\beta}$. The effective metric itself is defined by subtraction of suitably defined singular self-fields $h^{{\rm S}n}_{\alpha\beta}$ from the physical metric~\cite{Harte:2011ku, Pound:2012dk, Mathews:2021rod}.

We then express the effective metric as
\begin{equation}
\label{eq:effectivemet}
\hat g_{\mu \nu}=g_{\mu \nu}+ h_{\mu \nu}^{\mathrm{R}},
\end{equation}
where $h_{\mu \nu}^{\mathrm{R}}$ is the regularized metric perturbation.

In Sec.~\ref{sec:tt}, we divide the Einstein field equation~\eqref{eq:FE} into a sequence of equations for each $h^{n}_{\mu\nu}$ 
using a multiscale expansion. 

\subsection{MPD-Harte equations of motion}

For an appropriate effective metric $\hat g_{\mu\nu}$, the dynamics of the secondary are equivalent to those of a test-body in $\hat g_{\mu\nu}$~\cite{Pound:2012nt,Harte:2011ku,Pound:2017psq} and are described by the Mathisson-Papapetrou-Dixon (MPD) test-body equations~\cite{Mathisson:1937zz,Papapetrou:1951pa,Dixon:1974xoz}. We may neglect terms that are quadratic order (or higher) in the secondary's spin and the effects of the secondary's quadrupole moment~\cite{Mathews:2025nyb}. Both effects enter the equations of motion conservatively (at least assuming the secondary to be a Kerr black hole~\cite{Vines:2016unv, Ramond:2024ozy,Rahman:2021eay}) at $\mathcal{O}(\e^2, s^2)$ and first impact the waveform at 2PA order, where we use ``${\cal O}(s^n)$'' to denote powers of the secondary's spin. We thus adopt the pole-dipole approximation, and the relevant MPD equations in $\hat g_{\mu\nu}$ are
\begin{subequations}
\label{eq:MPDhat}
\begin{align}
\label{eqn:MPD1hat}
\frac{\hat D \hat u^{\alpha}}{d \hat\tau} &=-\frac{m_2}{2 } \hat R^{\alpha}{}_{\beta \gamma \delta} \hat u^{\beta} \hat S^{\gamma \delta}+ {\cal O}(s^2), \\
\label{eqn:MPD2hat}
\frac{\hat D \hat S^{\gamma \delta}}{d \hat\tau} &= {\cal O}(s^2),
\end{align}
\end{subequations}
where $\hat D/d\hat \tau \equiv \hat u^\alpha\hat\nabla_\alpha$ and we have  imposed the Tulczyjew-Dixon (TD) spin supplementary condition (SSC),
\begin{equation}
\label{eq:TDssc}
\hat P^{\alpha}\hat S_{\alpha \beta} = 0.
\end{equation}
Here the quantity $\hat P^{\mu}$ is the secondary's linear momentum in the effective spacetime. 

We call Eqs.~\eqref{eq:MPDhat} the MPD-Harte equations, following the terminology of Ref.~\cite{Mathews:2021rod}. We refer the reader to that earlier work for an assessment of the validity of these equations (and to  Ref.~\cite{Pound:2015tma} for additional historical context).

Following the same notation as Paper I, we use the effective metric to raise and lower indices on hatted quantities. We continue to define $\hat S_{\alpha \beta}$ as the mass-normalised effective spin tensor of the secondary, such that
\begin{equation}\label{eq:chi2=hatS.hatS}
(\chi_2)^2=\frac{1}{2}\hat S_{\alpha \beta}\hat S^{\alpha \beta}=\left(\frac{S_2}{m_2^2}\right)^2.
\end{equation}
We define an effective (dimensionless) spin vector,
\begin{equation}
\hat{S}^{\mu} = -\frac{1}{2} \hat \epsilon^{\mu}{}_{\alpha \beta \gamma} \hat u^{\alpha} \hat{S}^{\beta \gamma},
\end{equation}
with inverse relation
\begin{equation}
\label{eq:SpinTensorFromVector}
\hat{S}^{\mu \nu}=- \hat \epsilon^{\mu \nu \alpha \beta}\hat{S}_{\alpha}\hat u_{\beta}.
\end{equation}
The TD SSC implies that $\hat P^\alpha=m_2 \hat u^\alpha+{\cal O}(s^2)$ and $\hat u^{\alpha} \hat S_{\alpha}=0$. 
Imposing Eq.~\eqref{eq:MPDhat} implies the spin vector must satisfy the parallel transport equation
\begin{equation}
\label{eq:PTeqn}
\frac{\hat D\hat{S}^{\alpha}}{d \hat \tau}={\cal O}(s^2).
\end{equation}

In practice, we re-express the equations of motion in Eq.~\eqref{eq:MPDhat} in terms of the \emph{background} metric and the regular perturbations via Eq.~\eqref{eq:effectivemet} and expand in powers of $\e$. In this form, the metric perturbations appear in the equations of motion as a self-acceleration and self-torque~\cite{Mathews:2021rod}:
\begin{subequations}
\label{eq:selfforceEOMs}
\begin{align}
\frac{D u^{\mu}}{d \tau} &= -\frac{1}{2} P^{\mu\nu}(g_\nu{}^\lambda - h^{\mathrm{R}\, \lambda}_\nu)\left(2 h_{\lambda \rho ; \sigma}^{\mathrm{R}}-h_{\rho \sigma ; \lambda}^{\mathrm{R}}\right) u^{\rho} u^{\sigma}\nonumber\\
&\quad -\frac{m_2}{2}R^{\mu}{}_{\alpha \beta \gamma}\left(1-\frac{1}{2}h^{\mathrm{R}}_{\rho\sigma}u^\rho u^\sigma\right)u^{\alpha}\hat S^{\beta \gamma} \nonumber\\
&\quad +\frac{m_2}{2} P^{\mu\nu}(2h^{\mathrm{R}}_{\nu(\alpha;\beta)\gamma}-h^{\mathrm{R}}_{\alpha\beta;\nu\gamma})u^{\alpha}\hat S^{\beta \gamma}\nonumber\\
&\quad+\mathcal{O}(\e^3,s^2),\label{eq:selfforceorbit}\\
\frac{D\hat S^{\mu\nu}}{d \tau} &= u^{(\rho}\hat S^{\sigma)[\mu}g^{\nu]\lambda}\left(2 h_{\lambda \rho ; \sigma}^{\mathrm{R}}-h_{\rho \sigma ; \lambda}^{\mathrm{R}}\right)  + \mathcal{O}(\e^2,s^2), \label{eq:selfforcespin}
\end{align}
\end{subequations}
where $D/d\tau\equiv u^\mu\nabla_\mu$, $u^\mu\equiv dz^\mu/d\tau$,  $P^{\mu\nu}\equiv g^{\mu\nu}+u^\mu u^\nu$, $\tau$ is the proper time as measured in $g_{\mu\nu}$, and $\nabla_\mu$ and semicolons both denote the covariant derivative compatible with $g_{\mu\nu}$. 

The effective four-velocity  and background four-velocity are both timelike normalized with respect to their corresponding metric and are related by $\hat u^{\alpha} =\frac{d\tau}{d\hat\tau}u^{\alpha}$ such that 
\begin{equation}
\frac{d\tau}{d\hat\tau}=\sqrt{1-h^{\rm R}_{\alpha\beta}u^\alpha u^\beta}.
\end{equation}

\subsection{Parameterization of the secondary spin}
\label{sec:spin}
In Paper I, we presented a useful parameterization of the secondary's spin degrees of freedom along with a complete description of the spin's precession and nutation. The net result is that we write the spin vector as
\begin{equation}
\hat S^{\alpha}=S^{\alpha}+\e \delta S^{\alpha}\trunctwo,
\end{equation}
where the two terms are
\begin{subequations}\label{eq:S and dS}
\begin{align}
S^{\alpha}&=\chi_{\parallel}\E_3^{\alpha}+\chi_{\perp}\cos\tilde\psi_s\,\E_1^{\alpha}+\chi_{\perp}\sin\tilde\psi_s\,\E_2^{\alpha},\\
\delta S^{\alpha}&=\chi_{\parallel}\delta\E_3^{\alpha}+\chi_\parallel \delta \vartheta_s \, \E^\alpha_1 - \chi_\parallel\delta \vartheta_c\, \E^\alpha_2\nonumber\\
&\quad +\chi_\parallel  (\delta\vartheta_c \sin \tilde\psi_s -\delta\vartheta_s \cos\tilde\psi_s)\sigma^\alpha_3 \nonumber\\
&\quad  +\chi_{\perp}\cos\tilde\psi_s\,\delta\E_1^{\alpha}+\chi_{\perp}\sin\tilde\psi_s\,\delta\E_2^{\alpha}.
\end{align}
\end{subequations}
Here $\chi_\parallel$ and $\chi_\perp$ are constant parameters, $\tilde\psi_s$ is a spin-precession angle, and $\delta\vartheta_s, \delta\vartheta_c$ are nutation angles. 

The parameterization~\eqref{eq:S and dS} involves the background orthonormal tetrad
 \begin{subequations}
  \label{eq:sigmatetrad}
  \begin{align}
 \E_0^\alpha &= u^\alpha,\\
  \E_1^\alpha &= \epsilon^{\alpha \beta \gamma \delta}\E^0_\beta \E^2_\gamma \E^3_\delta,\\
  \E_2^\alpha &= \frac{1}{N}P^{\alpha \beta}K_{\beta \gamma}u^{\gamma},\\
  \E_3^\alpha &= Y^\alpha_{\ \;\beta} u^{\beta}/\sqrt{\mathcal{K}},
  \end{align}
 \end{subequations}
 where $Y_{\nu\rho}$ is the Killing-Yano tensor, $K_{\mu\nu}=Y_{\mu}{}^\rho Y_{\nu\rho}$ is the Killing tensor, $\mathcal{K}\equiv K_{\alpha \beta}u^{\alpha}u^{\beta}$ is the Carter constant, and $N=-\sqrt{P^{\alpha \beta}K_{\beta \gamma}u^{\gamma}P_{\alpha}^{\, \delta}K_{\delta \lambda}u^{\lambda}}$. The tetrad is the same as an intermediary tetrad introduced in constructing the Marck tetrad \cite{Marck, Witzany:2019nml}. The triad perturbations, $\delta\E_A^{\alpha}$ ($A=1,2,3$), result from enforcing orthonomality of the tetrad $(\hat u^\alpha,\hat\E^\alpha_A=\E^\alpha_A+\delta\E^\alpha_A)$ with respect to the effective metric. It follows from orthonormality that
  \begin{equation}
 \label{eq:tetrad_shift}
  \delta \E_A^{\alpha} = u^{\alpha}  h^{\mathrm{R}}_{\beta \gamma}u^\beta \E_A^{\gamma} -\frac{1}{2}\E^{B\alpha}h^{\mathrm{R}}_{\beta \gamma}\E_B^{\beta}\E_A^{\gamma}.
  \end{equation}
As a consequence of the normalisations of the tetrad legs, we have
\begin{equation}
\hat S^{\alpha}\hat S_{\alpha}= S^{\alpha}S_{\alpha}=\chi_2^2=\chi_\parallel^2+\chi_\perp^2.
\end{equation}
Note that the secondary spin magnitude $S_2$ defined in the Introduction is related to $S^\alpha$ by $S_2 = (m_2)^2 S^\alpha S_\alpha$.

In Paper I, we also introduced the angular velocity   \begin{equation}\label{eq:omegaABhat}
      \hat\omega_A^{\ \,B} = \frac{\hat D\hat\E^\alpha_A}{d\hat\tau}\hat\E^B_\alpha = \omega_A^{\ \,B}+\delta\omega_A^{\ \,B},
  \end{equation}
such that
\begin{subequations}
\begin{align}
\omega_{AB}&=\frac{\hat D\E^\alpha_A}{d\tau}\E^B_\alpha,\\
 \delta\omega_{AB}&=-h^{\mathrm{R}}_{\alpha \beta;\gamma}u^\alpha\E_A^{[\beta}\E_B^{\gamma]}+{\cal O}(\e^2).
 \end{align}
\end{subequations}
The precession angle evolves according to
\begin{equation}
     \frac{d\tilde\psi_s}{d\tau} = \omega_{21} + \delta\omega_{21},\label{eq:dpsi_s eqn}
\end{equation}
and the nutation angles evolve according to
 \begin{align}
 \label{eq:dtheta_c eqn}
     \frac{d\delta\vartheta_c}{d\tau} - \omega_{12}\delta\vartheta_s &= -\delta\omega_{23},\\
 \label{eq:dtheta_s eqn}
     \frac{d\delta\vartheta_s}{d\tau} + \omega_{12}\delta\vartheta_c &= \delta\omega_{13}.   
 \end{align}

Finally expressing the spin tensor as $\hat S^{\alpha \beta} = S^{\alpha \beta} + \e \delta S^{\alpha \beta}+\mathcal{O}(\e^2)$,  by Eq.~\eqref{eq:SpinTensorFromVector} we have 
\begin{subequations}\label{eq:deltasab}
\begin{align}
S^{\mu \nu} &= -\epsilon^{\mu \nu \alpha \beta}S_{\alpha}u_{\beta},\\
 \delta S^{\alpha \beta} &= \epsilon^{\alpha \beta}{}_{\gamma \lambda}u^{\gamma}\delta S^{\lambda}+\frac{1}{2}P^{\gamma \lambda} h^{\mathrm{R}}_{\gamma \lambda} S^{\alpha \beta}-2 h^{\mathrm{R} \, [\beta}_{\lambda}S^{\alpha] \lambda}. 
\end{align}
\end{subequations}
The self-force equations of motion listed in Eq.~\eqref{eq:selfforceEOMs} are then easily re-expanded in terms of the background spin tensor or vector, but are more compact in their current form.

\section{Multiscale expansion for approximately equatorial orbits around a slowly spinning, slowly evolving primary}
\label{sec:tt}

We now specialise our binary configuration, following closely the two-timescale analysis in Sec.~IV of Ref.~\cite{Mathews:2021rod} (that extends from the scheme of Ref.~\cite{Miller:2020bft}) and making use of a specialization of the results of Paper I. The analysis of Ref.~\cite{Mathews:2021rod} was valid for a secondary in quasi-circular motion with spin (anti-)aligned with the orbital angular momentum, inspiraling into a non-spinning primary black hole. Keeping a similar structure, we adapt the scheme to allow for a generic precessing secondary spin and slow spin of the primary which is (anti-)aligned with the leading orbital angular momentum. 

For a slowly spinning primary, we may take the background metric $g_{\mu \nu}$ to be the Schwarzschild metric and encode the primary's spin in the metric perturbations. Due to radiation reaction, the primary's physical mass, $m_1$, and spin angular momentum, $S_1=(m_1)^2\chi_1$, are dynamical variables. The same is true for the secondary's mass and spin, but their evolution is beyond 1PA order~\cite{Poisson:2004cw}. We introduce the `background' masses and dimensionless spins ($m_i^{(0)}$ and $\chi_i^{(0)}$, respectively) such that the physical masses and dimensionless spins are
\begin{subequations}
\label{eq:massandspins}
\begin{align}
\label{eq:primary mass}
    m_1&=m_1^{(0)}+\ee\delta m_1,\\ 
\label{eq:primary spin}
    \chi_1&=\ee\delta\chi_1, 
\end{align}
\end{subequations}
where we have set the background mass as the initial mass such that $\delta m_1(t=0)=0$ and defined the \emph{initial} mass ratio $\ee\equiv m_2^{(0)}/m_1^{(0)}$. We continue to use $\e$ as a counting parameter which will now track powers of the initial mass ratio $\ee$. We make the slow spin condition explicit by setting $\chi_1^{(0)}=0$ and instead allowing for a non-zero initial condition in the sub-leading term, $\delta \chi_1(t=0)\neq0$. Note we have pulled out the explicit factor of $\ee$ so that $\delta\chi_1\sim \e^0$ and $\delta m_1\sim \e^0$. The mass of the secondary evolves according to $dm_2/dt={\cal O}(\e^5)$ if it is a Kerr black hole~\cite{Poisson:2004cw}, meaning we can treat it as a constant.

We work in Schwarzschild coordinates $x^\mu=(t,x^i)$, with $x^i=(r,\theta,\phi)$, and we use 
\begin{equation}\label{eq:trajectory}
z^i(t,\ee)=(r_p(t,\ee),\theta_p(t,\ee),\phi_p(t,\ee))    
\end{equation}
to denote the particle's orbital trajectory. For our specialized binary configuration, the particle's dynamics and the metric only depend on $t$ via the orbital and precession phases ($\phi_p$ and $\tilde\psi_s$ respectively) and the mechanical variables  $\varpi_I = (\Omega, \delta m_1, \delta\chi_1)$, where $\Omega\equiv d\phi_p/dt$.
The system is periodic in each of the phases, which evolve on the fast timescale $\sim 1/\Omega$.  The mechanical variables $\varpi_I$ evolve slowly, on the radiation-reaction timescale $\sim \Omega/\dot\Omega\sim 1/\e$. 

In addition to the phases and $\varpi_I$, the equations of motion depend linearly on the  secondary's conserved spin magnitudes, $\chi_{\parallel}$ and $\chi_\perp$. We will show that, for approximately equatorial inspirals, the nutation angles $\delta\vartheta_c$ and $\delta\vartheta_s$ are vanishing constants at 1PA order.

In Paper I, the list of independent phases that appear in the leading-order dynamics of generic inspirals is denoted as $\psi^i$. In the quasi-circular, approximately equatorial configuration we consider here, the set of phases has only one element, the orbital phase; $\psi^i=(\phi_p)$. The set of independent mechanical variables in the leading-order dynamics, denoted as $\pi_i$, likewise reduces down to a single element; $\pi_i=(r_p)$, where $r_p$ is the orbital radius. We work directly with the orbital frequency $\Omega$ instead of $r_p$ for reasons advocated in Ref.~\cite{Mathews:2021rod}, and the full set of mechanical variables we have continued to denote as $\varpi_I$. In Paper I, we perform an averaging transformation on the quantities $\psi^i$, $\tilde\psi_s$ and $\varpi_I$ to the variables $\mathring{\psi}^i$, $\mathring{\psi}_s$ and $\mathring{\varpi}_I$, such that the evolution of the ringed quantities depends only on $\mathring{\varpi}_I$. In this work, the ringed variables reduce to their non-averaged counterparts through first post-adiabatic order due to the simplified orbital configuration. Thus, we will not require an explicit averaging transformation. By using $\Omega$ as our orbital variable, we adopt the fixed-frequency gauge as described more generally in Sec.~IV~D of Paper~I.

\subsection{Orbital configuration}

We write the particle's trajectory as a function $z^i(t,\ee)=z^i(\phi_p(t,\ee),\tilde\psi_s(t,\ee),\varpi_I(t,\ee),\ee)$. Expanded in powers of $\ee$ at fixed $(\phi_p,\tilde\psi_s, \varpi_I)$, it then reads
\begin{equation}
\label{eqn: tt worldline}
z^i(\phi_p,\tilde\psi_s,\varpi_I, \ee) = z_{0}^i(\phi_p,\Omega)+\ee z_{1}^i(\tilde\psi_s, \varpi_I)+O(\e^2),
\end{equation}
where the leading-order trajectory corresponds to a circular equatorial inspiral,
\begin{equation}
z_0^i(\phi_p,\Omega) = (r_0(\Omega),\pi/2,\phi_p),
\end{equation}
while the first subleading term consists of a radial correction and an oscillatory, precession-induced polar correction,
\begin{equation}
z_1^i(\varpi_I) = (r_1(\varpi_I),\theta_1(\tilde\psi_s, \Omega),0).
\end{equation}
We define the frequencies via the rates of change of the orbital and precession phases,
\begin{subequations}
\label{Omegadef}
\begin{align}
 \frac{d\phi_p}{dt}& \equiv \Omega,\\
  \frac{d\tilde\psi_s}{dt}& \equiv \Omega_s.
 \end{align}
\end{subequations}
While $\Omega$ is an independent variable characterizing the system, the precession frequency $\Omega_s$ is a function of $\varpi_I$ (and of the system's non-evolving parameters).  

Altogether, the above ansatz represents a circular orbit with small, ${\cal O}(\e)$ polar oscillations about the equatorial plane and with a slowly evolving radius, orbital frequency and precession frequency. We take the convention that negative values of $\chi_1$ correspond to retrograde inspirals, while positive values correspond to prograde inspirals.

\subsection{Metric}

We expand the metric as
\begin{multline}
\textbf{g}_{\mu\nu} = g_{\mu\nu}(x^i) + \ee h^{1}_{\mu\nu}(x^i,\phi_p,\Omega,\delta m_1, \delta \chi_1)\\
+ \ee^2 h^{2}_{\mu\nu}(x^i,\phi_p, \tilde\psi_s,\varpi_I) + {\cal O}(\e^3),\label{g tt expansion}
\end{multline}
with the assumption that the dependence on the two phases $\phi_p, \tilde\psi_s$ is $2\pi$-periodic. In our ansatz~\eqref{g tt expansion}, there is no time dependence outside the dependence on the phase-space variables $(\phi_p, \tilde\psi_s,\varpi_I)$. To extend those variables away from the worldline, we adopt a hyperboloidal time $s$ that is equal to advanced time $v$ at the future horizon, Schwarzschild time $t$ at the particle, and retarded time $u$ at infinity. The phase-space variables $(\phi_p, \tilde\psi_s,\varpi_I)$ are then defined as constant on slices of constant $s$. In practice, we choose a ``sharp" slicing in which $s=t$ in a worldtube of diameter $4M$ (for $r_0 \le 20m^{(0)}_1$) or $8M$ (for $r_0>20m^{(0)}_1$) around the particle, $s=v$ between the horizon and the worldtube, and $s=u$ between the worldtube and future null infinity. Refer to Ref.~\cite{Miller:2023ers} for details, specifically Fig.~1 therein.  

The first-order perturbation $h^{1}_{\mu\nu}(x^i,\phi_p,\Omega,\delta m_1, \delta \chi_1)$ is linear in $\delta m_1$ and $\delta \chi_1$. We peel off the dependence on those parameters as
\begin{multline}
\label{eq:h1tt}
h^{1}_{\mu\nu} = h^{1(pp)}_{\mu\nu}(x^i,\phi_p,\Omega) + \delta m_1\, h^{1(\delta m_1)}_{\mu\nu}(x^i) \\
+ \delta{\chi}_1\, h^{1(\delta{\chi}_1)}_{\mu\nu}(x^i).
\end{multline}
If $\Omega$ were constant and $\phi_p$ took its geodesic value $\Omega t$, then $h^{1 (pp)}_{\mu\nu}(x^i,\phi_p,\Omega)\equiv h^{1}_{\mu\nu}(x^i,\phi_p,\Omega,0, 0)$ would be the standard linear perturbation to the Schwarzschild metric due to a point particle (hence ``$pp$") on a circular geodesic with frequency $\Omega$. The terms $\delta m_1\, h^{1(\delta m_1)}_{\mu\nu}(x^i)$ and $\delta\chi_1\, h^{1(\delta\chi_1)}_{\mu\nu}(x^i)$ are linear perturbations toward a Kerr black hole with mass described by Eq.~\eqref{eq:primary mass} and angular momentum $ \left(m_1^{(0)}\right)^2\chi_1$, with $\chi_1$ described by Eq.~\eqref{eq:primary spin}. 

In the same way, we  peel off the dependence on $\delta m_1$ and $\delta\chi_1$ as well as $\chi_\parallel$ and $\chi_\perp$ in the second-order perturbation:
\begin{align}
h^{2}_{\mu\nu} & = h^{2(pp)}_{\mu\nu}(x^i,\phi_p,\Omega) + \sum_{j+k=2}\delta m_1^j\delta \chi_1^k h^{2[j,k]}_{\mu\nu}(x^i)\nonumber\\
 &+\delta m_1\, h^{2(\delta m_1)}_{\mu\nu}(x^i,\phi_p,\Omega)+ \delta{\chi}_1\, h^{2(\delta{\chi}_1)}_{\mu\nu}(x^i,\phi_p,\Omega)\nonumber\\
& +\chi_\parallel h^{2(\chi_\parallel)}_{\mu\nu}(x^i,\phi_p,\Omega)
+\chi_\perp h^{2(\chi_\perp)}_{\mu\nu}(x^i,\phi_p, \tilde\psi_s,\Omega).\label{eq:h2tt}
\end{align}
Again, the first term is shorthand for $h^2_{\mu\nu}$ with parameters other than $m_2$ set to zero: $\delta m_1=\delta\chi_1=\chi_\parallel=\chi_\perp=0$. The second 
term in the first line is the quadratic perturbation toward a Kerr black hole with mass $m_1=m_1^{(0)}+\ee \delta m_1$ and spin $\chi_1=\ee \delta\chi_1$, which will not impact the waveform at 1PA order. The terms on the second line arise from the coupling of the mass and spin corrections with the linear point mass perturbation and do impact the waveform at 1PA order. The final two terms are the linear contributions of the secondary's spin for a spinning particle along a corresponding accelerated MPD trajectory. Note that the secondary's spin nutation parameters $\delta\vartheta_c$ and $\delta\vartheta_s$ first enter the metric at third order despite appearing in the second-order acceleration. 

\subsubsection{Up-down symmetry}

In general, we say a tensor field of arbitrary rank is up-down \emph{symmetric} if
\begin{equation}
    A_{\mu\nu ...}(\pi-\theta)=(-1)^{k} A_{\mu\nu...}(\theta),
\end{equation}
where $k$ is the integer number of polar indices labeling the tensor component. We have only included the field's explicit dependence on $\theta$ for simplicity, as its dependence on the other spacetime coordinates does not affect its up-down symmetry properties. On the equatorial plane, $\theta=\pi/2$, an up-down symmetric field therefore satisfies
\begin{equation}
A_{\mu \nu...}\left(\pi/2\right)=0\qquad  \text{$k$ is odd}.
\end{equation}
Conversely, we say the field is up-down \emph{antisymmetric} if
\begin{equation}
     A_{\mu\nu...}(\pi-\theta)=(-1)^{k+1} A_{\mu\nu...}(\theta),
\end{equation}
which implies that on the equatorial plane
\begin{equation}
A_{\mu \nu...}\left(\pi/2\right)=0\qquad  \text{$k$ is even}.
\end{equation}

Assuming a reasonable gauge choice such as the Lorenz gauge, every term in Eq.~\eqref{eq:h1tt} and Eq.~\eqref{eq:h2tt} is up-down symmetric \emph{except} for $h^{2(\chi_\perp)}_{\mu\nu}(x^i,\phi_p, \tilde\psi_s,\Omega)$, which is up-down antisymmetric. The same statement extends to the higher-rank tensor fields constructed exclusively from derivatives operating on those rank-two tensor fields. That this is true may be shown from examining the field equations with a given gauge choice sourced by the stress-energy of a spinning particle (approximately) in the equatorial plane. It is clear that the up-down symmetry/antisymmetry conditions extend to any inspiral of a spinning particle constrained to the equatorial plane at leading order, and may be used to vastly simplify the evolution equations for the trajectory and the secondary's spin.

\subsubsection{Fields on phase space}

The expansion~\eqref{g tt expansion} expresses the metric as a function on the mechanical phase space of the two-body system, replacing time dependence with dependence on mechanical variables. When substituting these expansions into the equations of motion and field equations, we apply the chain rule 
\begin{equation}
\frac{\partial}{\partial x^\alpha} = \eta^i_\alpha \frac{\partial}{\partial x^i} + s_\alpha\left(\frac{d\phi_p}{dt}\frac{\partial}{\partial\phi_p} +\frac{d\tilde\psi_s}{dt}\frac{\partial}{\partial\tilde\psi_s} +\frac{d\varpi_I}{dt}\frac{\partial}{\partial \varpi_I}\right),
\end{equation}
where $\eta^i_\alpha \equiv \frac{\partial x^i}{\partial x^\alpha}$ and $s_\alpha \equiv \partial_\alpha s$, and we have used the fact that $s=t$ along the particle's trajectory. We then treat the mechanical variables as independent of one another and re-expand while holding them fixed. 

This treatment implies the expansion
\begin{equation}\label{nabla tt}
\nabla_\alpha = \nabla^{0}_\alpha + \ee s_\alpha \left(\vec{\partial}_{\cal V} +\Omega_s^{(1)}\frac{\partial}{\partial\tilde\psi_s}\right) + {\cal O}(\e^2),
\end{equation}
where the zeroth-order covariant derivative is 
\begin{equation}
\nabla^{0}_\alpha = \eta^i_\alpha\frac{\partial}{\partial x^i} + s_\alpha \Omega\frac{\partial}{\partial\phi_p}+ s_\alpha \Omega_s^{(0)}\frac{\partial}{\partial\tilde\psi_s} + \text{Christoffel terms}.\label{nabla0}
\end{equation}
Here $\Omega_s^{(0)}$ and $\Omega_s^{(1)}$ are the leading-order precession frequency and its subleading correction, respectively, which we make precise later in Eq.~\eqref{eq:tt prec freq split}. ${\cal V}_I=(F^{(0)}_\Omega, F^{(1)}_{\delta m_1}, F^{(1)}_{\delta\chi_1})$ is the leading-order velocity through parameter space; i.e., $d\varpi_I/dt=\ee {\cal V}_I + {\cal O}(\e^2)$. The operator
\begin{align}
\vec{\partial}_{\cal V}\equiv{\cal V}_I\frac{\partial}{\partial \varpi_I} = F_{\Omega}^{(0)}\frac{\partial}{\partial\Omega} + F_{\delta m_1}^{(1)}\frac{\partial}{\partial\delta m_1} + F_{\delta\chi_1}^{(1)}\frac{\partial}{\partial\delta\chi_1}
\end{align}
is a derivative along this velocity. $\nabla^{0}_\alpha$, on the other hand, acts at fixed parameter values. Its action on $h^{1}_{\alpha\beta}(x^i,\phi_p,\Omega)$ is identical to the action of $\nabla_\alpha$ on the linear metric perturbation to the Schwarzschild metric from a point mass on a circular geodesic. The directional derivative $\vec{\partial}_{\cal V}$ then accounts for the system's slow movement through the parameter space. We note that since $\tilde\psi_s$ only enters the metric at second order, the correction term proportional to $\Omega^{(1)}_s$ in Eq.~\eqref{nabla tt} would only enter the Einstein equation at third order.

In the above expansions we have kept the same notation as Ref.~\cite{Mathews:2021rod}, in which the integer labels with parentheses indicate the post-adiabatic order at which the quantity enters, while the integer labels without parentheses correspond to the absolute order in $\ee$.

\subsection{Evolution equations}

As we have just explained, all functions are expanded in powers of $\ee$ at fixed values of the phase-space coordinates $(\phi_p, \tilde\psi_s,\varpi_I)$. Expanded in this way, the rates of change of the parameters $\varpi_I$ are
\begin{subequations}
\label{eq:ttinspiral}
\begin{align}
    \frac{d\Omega}{dt} &= \ee F_{\Omega}^{(0)}(\Omega) + \ee^2 F_{\Omega}^{(1)}(\varpi_I) + {\cal O}(\e^3),\label{Omegadot}\\        
    \frac{d\delta m_1}{dt} &= \ee F_{\delta m_1}^{(1)}(\Omega) + {\cal O}(\e^2),\label{Mdot}\\
    \frac{d\delta \chi_1}{dt} &= \ee F_{\delta\chi_1}^{(1)}(\Omega) + {\cal O}(\e^2).\label{Jdot}
\end{align}
\end{subequations}
From these expansions we obtain the expansion for the coordinate velocity, 
\begin{equation}
\dot z^\alpha \equiv \frac{dz^\alpha}{dt} = \dot z^\alpha_0(\Omega) + \ee \dot z^\alpha_1(\tilde\psi_s, \varpi_I) +O(\e^2),
\end{equation}
where
\begin{equation}
\dot z_0^\alpha \equiv \frac{dz_0^\alpha}{dt} = (1,0,0,\Omega),
\end{equation}
and
\begin{equation}\label{z1dot}
\dot z_1^\alpha = (0,\dot r_0,\dot \theta_1,0),
\end{equation}
with $\dot r_0(\Omega)=\frac{dr_0}{d\Omega}F_{\Omega}^{(0)}$ and $\dot \theta_1(\tilde\psi_s, \varpi_I)=\frac{d\theta_1}{d\tilde\psi_s}\Omega_s^{(0)}$; the proper four-velocity is
\begin{equation}
\label{eq:tt fourvel}
u^\alpha=u^\alpha_0(\Omega)+\ee u^t_0(\Omega)\dot z^\alpha_1(\tilde\psi_s, \varpi_I)+\mathcal{O}(\e^2),
\end{equation}
with $u^\alpha_0=u^t_0\dot z^\alpha_0$. From the timelike normalisation condition one obtains
\begin{equation}
    u_0^t(\Omega)=\frac{1}{\sqrt{1-\frac{3 m_1^{(0)}}{r_0(\Omega)}}},
\end{equation}
as for a circular geodesic in Schwarzschild spacetime. 

We determine the forcing functions that drive the evolution of the mechanical variables, $F_{J}^{(n)}$, from the expansion of the equations of motion~\eqref{eq:selfforceEOMs} and from the Einstein field equations.  Throughout this section, we have anticipated that 0PA quantities only depend on $(\phi_p,\Omega)$ and indicated that the slow evolution rates $d\varpi_I/dt$ are independent of the phases $\phi_p, \tilde\psi_s$. 
Finally, we reiterate that we have excluded $\Omega_s$ in the list of independent variables since it is uniquely determined by the other orbital parameters via the expansion of Eq.~\eqref{eq:dpsi_s eqn}. 

\subsubsection{Evolution of the secondary's spin}
\label{sec:precevolveeqn}

In the self-consistent formalism, the secondary's spin vector is completely described by Eq.~\eqref{eq:S and dS} along with the evolution equations~\eqref{eq:dpsi_s eqn}--\eqref{eq:dtheta_s eqn}. We now make these explicit in our multiscale analysis of quasi-circular and approximately equatorial inspirals. 

Substituting Eq.~\eqref{eqn: tt worldline} and Eq.~\eqref{eq:tt fourvel} into Eq.~\eqref{eq:sigmatetrad}, we construct the $\sigma_A^{\alpha}$ triad evaluated along the worldline:
\begin{subequations}
\label{tt marck triad}
\begin{align}
\E^{\alpha}_{1}(z^i,\Omega)&=(0,f_0^{-1/2},0,0)+\mathcal{O}(\e),\\
\E^{\alpha}_{2}(z^i,\Omega)&=-r_0 \sqrt{f_0} u_0^t(\Omega f_0^{-1},0,0,r_0^{-2})+\mathcal{O}(\e),\\
\E^{\alpha}_{3}(z^i,\Omega)&= \left(0,0,-r_0^{-1},0 \right)+\mathcal{O}(\e),
\end{align}
\end{subequations}
with $f_0\equiv 1-\frac{2m_1^{(0)}}{r_0}$. We do not display the sub-leading terms for compactness. 

We next substitute our expansion of the worldline, the four-velocity and Eq.~\eqref{tt marck triad} into Eq.~\eqref{eq:dtheta_c eqn} and Eq.~\eqref{eq:dtheta_s eqn}. After imposing the up-down symmetry of $h_{\mu\nu}^1$, we find that we are free to pick the trivial solution $\delta\vartheta_c=0=\delta\vartheta_s$. We are then left with only one dynamical quantity in the evolution of the secondary's spin: the precession phase. Expanding Eq.~\eqref{eq:dpsi_s eqn} and recalling that $\frac{d \tilde\psi_s}{d t} = \Omega_s$, we obtain
\begin{equation}
\label{eq:tt prec freq}
\Omega_s=\Omega_s^{(0)}+\ee \Omega_s^{(1)},
\end{equation}
with
\begin{subequations}\label{eq:tt prec freq split}
\begin{align}
\Omega_s^{(0)} &= \Omega/u_0^t(\Omega),\\
\Omega_s^{(1)} &= h^{ \rm R1}_{\beta \gamma; \delta}(z_0^i,\phi_p, \varpi_I)\E_1^{[\beta}(z_0^i,\Omega)\E_2^{\delta]}(z_0^i,\Omega)k^{\gamma}(\Omega) \nonumber\\
&\qquad + 3u_0^t \Omega^2 \sqrt{y(\Omega)}r_1(\varpi_I) ,\nonumber\\
&=\Omega_s^{(1,0)}+ \delta\chi_1\Omega_s^{(1,\delta\chi_1)}+ \delta m_1\Omega_s^{(1,\delta m_1)}+ \chi_\parallel \Omega_s^{(1,\chi_\parallel)}.
\end{align}
\end{subequations}
Here, we have introduced the helical killing vector 
\begin{equation}
k^{\alpha}\equiv u_0^{\alpha}/u^t_0,
\end{equation}
and the inverse-separation parameter 
\begin{equation}\label{eq:ydef}
y\equiv m_1^{(0)}/r_0 = (m_1^{(0)}\Omega)^{2/3},
\end{equation}
and we have denoted the first-order contribution to the regularized metric perturbation, $h^{\rm R}_{\alpha \beta}$, as $h^{\rm R1}_{\alpha \beta}$.
Note that $\Omega_s^{(1,0)}$ is simply related to the spin-precession invariant first computed in Ref.~\cite{Dolan:2013roa}. The other terms are test-body terms and may be derived analytically from the MPD equations in Kerr spacetime. Their expressions are
\begin{subequations}
\label{eq:precfreqterms}
\begin{align}
    m_1^{(0)}\Omega_s^{(1,\delta\chi_1)}&=\frac{(1-y) y^3}{\sqrt{1-3 y}},\\
    m_1^{(0)}\Omega_s^{(1,\delta m_1)}&=\frac{(1-4 y) y^{3/2}}{\sqrt{1-3 y}},\\
    m_1^{(0)}\Omega_s^{(1,\chi_\parallel)}&=-\frac{3 y^4}{\sqrt{1-3 y}}.
\end{align}
\end{subequations}

We discuss the relevance of $\Omega_s^{(1)}$ at the end of Sec.~\ref{sec:SummaryTT}.

\subsubsection{Expansion of the equations of motion}
\label{sec:EOM expand}
In expanding the equations of motion~\eqref{eq:selfforceEOMs}, we expand the force as
\begin{equation}
a^\alpha = \ee a^\alpha_1(\tilde\psi_s, \varpi_I) + \ee^2 a^\alpha_2(\tilde\psi_s, \varpi_I)+{\cal O}(\e^3),
\end{equation}
where the numeric labels correspond to the explicit powers of $\ee$.  Recall that $\phi_p$ does not appear in the forces. We make the secondary's spin contribution explicit by writing
\begin{align}
a^\alpha_{1}(\tilde\psi_s,\varpi_I)&=a^\alpha_{1}(\varpi_I)+\chi_\parallel a^\alpha_{1(\chi_\parallel)}(\varpi_I)\nonumber \\
& \qquad +\chi_\perp a^\alpha_{1(\chi_\perp)}(\tilde\psi_s,\varpi_I),\\
a^\alpha_{2}(\tilde\psi_s,\varpi_I)&=a^\alpha_{2}(\varpi_I)+\chi_\parallel a^\alpha_{2(\chi_\parallel)}(\varpi_I)\nonumber \\
& \qquad +\chi_\perp a^\alpha_{2(\chi_\perp)}(\tilde\psi_s,\varpi_I),
\end{align}
where $a^\alpha_{1}(\varpi_I)$ is the leading, ``MiSaTaQuWa''~\cite{Mino:1996nk,Quinn:1996am} force generated by the first-order metric perturbation and $a^\alpha_{1(\chi_\parallel)}(\varpi_I)$ and $a^\alpha_{1(\chi_\perp)}(\tilde\psi_s,\varpi_I)$ are the test-body MPD force terms induced by $ S^{\alpha}$.

$a^\alpha_{2}(\varpi_I)$ is the second-order
self-force generated by the first- and second-order metric perturbations, excluding the linear secondary spin terms. The two secondary spin terms $a^\alpha_{2(\chi_\parallel)}(\varpi_I)$ and $a^\alpha_{2(\chi_\perp)}(\tilde\psi_s,\varpi_I)$ contain:
\begin{enumerate}
    \item The MPD force correction generated by $\delta S^{\alpha\beta}$.\footnote{It is in this term that generic, off-equatorial inspirals would depend on the nutation angles $\delta\vartheta_c$ and $\delta\vartheta_s$.}
    \item The self-force generated by the linear secondary spin's metric perturbation.
    \item Additional self-force terms that go as $h^1\cdot S^\alpha$.
    \item Subleading secondary spin corrections to the first-order MiSaTaQuWa force via the dependence on the expanded worldine.
\end{enumerate}
We refer to Sec. IVC of Paper I for a more detailed description of these various terms.

At this point we flag that $a^\alpha_{1(\chi_\perp)}(\tilde\psi_s,\varpi_I)$ and $a^\alpha_{2(\chi_\perp)}(\tilde\psi_s,\varpi_I)$ depend on the precession phase $\tilde\psi_s$, but this dependence does not enter into $d\varpi_I/dt$ (which is always constructed from an average over phases as described in Paper~I). Moreover, $\chi_\perp$ itself does not contribute to $d\varpi_I/dt$ until 2PA order because $\chi_\perp$ terms are purely oscillatory at lower orders. In the case of the first-order MPD force, $a^\alpha_{1(\chi_\perp)}(\tilde\psi_s,\varpi_I)$ is conservative,\footnote{Refer to Appendix B of Paper~I for a discussion of conservative versus dissipative forces.} and it has only a $\theta$ component. Its sole effect in our configuration is to produce the small conservative polar oscillation $\theta_1(\tilde\psi_s,\varpi_I)$. The same idea extends to $a^\alpha_{2(\chi_\perp)}(\tilde\psi_s,\varpi_I)$ by invoking the up-down (anti)symmetries of the metric perturbations and inspecting the equations of motion. In doing so, we find it is straightforward (but tedious) to show that $a^\alpha_{2(\chi_\perp)}(\tilde\psi_s,\varpi_I)$, like $a^\alpha_{1(\chi_\perp)}(\tilde\psi_s,\varpi_I)$, has only a $\theta$ component in our configuration. This induces a second-order polar oscillation of the worldline and does not contribute to the inspiral dynamics at 1PA order. All of the other force terms, which are independent of $\chi_\perp$, have a vanishing $\theta$ component.

\subsubsection{Evolution of the worldline}
\label{sec:TT evolve worldline}

Substituting all of the above expansions into Eq.~\eqref{eq:selfforceorbit}, we can straightforwardly solve order by order in $\e$, equating coefficients of powers of $\e$ at fixed $\varpi_I$ rather than at fixed $t$. We obtain from the radial component
\begin{align}\label{radius}
r_0(\Omega) = \left(m_1^{(0)}\right)^{\frac{1}{3}}\Omega^{-\frac{2}{3}}, \quad r_1(\varpi_I) = -\frac{a_1^r(\varpi_I)}{3(u_0^t)^2 f_0\Omega^2}.
\end{align}
Meanwhile the polar equation simplifies to
\begin{equation}\label{polarangleeqn}
(\Omega^{(0)}_s)^2 \partial_{\tilde\psi_s}^2\theta_1+\Omega^2 \theta_1 = -\frac{a_1^\theta(\tilde\psi_s, \varpi_I)}{(u_0^t)^2},
\end{equation}
where
\begin{equation}
    a_1^\theta(\tilde\psi_s, \varpi_I)=-3 \chi_\perp \sqrt{f_0}\left( m_1^{(0)}u^t_0\right)^2 \Omega \cos \tilde\psi_s,
\end{equation}
and where we have used Eq.~\eqref{eq:tt prec freq split} to rewrite $\Omega_s^{(0)}$ in terms of $\Omega$.
The solution is\footnote{In the test body limit, this formula recovers equivalent expressions from Refs.~\cite{Piovano:2025aro, Drummond:2022xej}.}
\begin{equation}
\theta_1(\tilde\psi_s ,\varpi_I) =  - \chi_\perp \sqrt{f_0} r_0\Omega\cos\tilde\psi_s,
\end{equation}
where we have discarded homogeneous solutions that would correspond to a geodesic, nonprecessing tilt of the orbital plane. 

From the dissipative sector [$t$ or $\phi$ component of~\eqref{eq:selfforceorbit}], we obtain
\begin{align}
F_\Omega^{(0)} &= -\frac{3 f_0\Omega a^t_1(\Omega)}{y(u_0^t)^4(1-6y)},\label{Omegadot0}\\
F_\Omega^{(1)} &= -\frac{3 f_0\Omega a^t_2(\varpi_I)}{y(u_0^t)^4(1-6y)} 
 - \frac{2 \vec{\partial}_{\cal V}a_1^r(\varpi_I)}{\sqrt{y}(u_0^t)^4 f_0(1-6y)}\nonumber \\
& \qquad \qquad  - \frac{4(1-6y+12y^2)a_1^r(\varpi_I) a_1^t(\Omega)}{y^{3/2} (u_0^t)^6f_0(1-6y)^2}\label{Omegadot1}
\end{align}
in terms of the parameter $y$ defined in Eq.~\eqref{eq:ydef}. The forcing functions $F_{\delta m_1}^{(1)}(\Omega)$ and $F_{\delta \chi_1}^{(1)}(\Omega)$ are determined from the Einstein equations~\cite{Miller:2020bft} and are identical to the fluxes of energy and angular momentum through the horizon due to a point mass on a circular geodesic orbit of frequency $\Omega$.

If we rewrite Eqs.~\eqref{Omegadef} and \eqref{Omegadot}--\eqref{Jdot} in terms of  a ``slow time'' variable $\tilde t\equiv\e t$, then it is clear they have asymptotic solutions with the following form:
\begin{align}
    \phi_p &= \ee^{-1}\phi^{(0)}_p(\tilde t)+\phi^{(1)}_p(\tilde t)+{\cal O}(\e),\label{phip - slow t}\\
    \tilde\psi_s &= \ee^{-1}\psi^{(0)}_s(\tilde t)+\psi^{(1)}_s(\tilde t)+{\cal O}(\e),\label{psis - slow t}\\
    \Omega &= \Omega^{(0)}(\tilde t)+\ee\,\Omega^{(1)}(\tilde t)+{\cal O}(\e^2),\\
    \Omega_s &= \Omega_s^{(0)}(\tilde t)+\ee\,\Omega_s^{(1)}(\tilde t)+{\cal O}(\e^2)\\
    \delta m_1 &= \delta m_1^{(1)}(\tilde t)+{\cal O}(\e),\\
    \delta\chi_1 &= \delta\chi_1^{(1)}(\tilde t)+{\cal O}(\e),\label{dJ - slow t}
\end{align}
with constant $\chi_\parallel$ and $\chi_\perp$, $d\phi^{(n)}_p/d\tilde t=\Omega^{(n)}(\tilde t)$, and with easily worked out equations for $d\Omega^{(n)}/d\tilde t$. As shown in Ref.~\cite{Wardell:2021fyy}, directly using such asymptotic solutions leads to lost accuracy in waveforms, but they are a useful guide for the behavior of each variable over the course of an inspiral.

\subsection{Stress-energy tensor}

After substituting our expansions for the wordline~ \eqref{eqn: tt worldline}, the four-velocity~\eqref{eq:tt fourvel} and the spin vector parameterisation of Eq.~\eqref{eq:S and dS} with the triad~\eqref{tt marck triad}  into the stress-energy~\eqref{eq:pole-dipole}, we obtain
\begin{align}
\!\!T^{\mu\nu} = \ee T_{1}^{\mu\nu}(x^i,\phi_p,\Omega) + \ee^2 T_{2}^{\mu\nu}(x^i,\phi_p, \tilde\psi_s,\varpi_I) + {\cal O}(\e^3).\!\label{T tt expansion}
\end{align}

We have at leading order
\begin{equation}
 \label{T1 multiscale}
  T^{\mu\nu}_{1}  = \frac{m_1^{(0)}u_0^t}{r_0^2}  \dot z_0^{\mu}\dot z_0^{\nu} \delta_{\phi}\delta_{r_0}\delta_{\theta_0},
 \end{equation}
 in which we have used the shorthand expressions for the Dirac delta distributions $\delta_\phi\equiv\delta(\phi-\phi_p)$, $\delta_{\theta_0}\equiv\delta(\theta-\pi/2)$ and $\delta_{r_0}\equiv\delta(r-r_0)$. 
 
Meanwhile the sub-leading stress-energy takes contributions from the sub-leading corrections to the monopole, $T_{(m)}^{\mu \nu}$, and the leading terms from the dipole, $T_{(d)}^{\mu \nu}$:
\begin{align}
T^{\mu\nu}_{2} &=\frac{m_1^{(0)}u_0^t}{r_0^2}\bigg[2 \Big(\dot{z}_0^{(\mu}\delta_{\theta}^{\nu)} \dot \theta_1+\dot{z}_0^{(\mu}\delta_{r}^{\nu)} \dot r_0\Big)\delta_{r_0}\delta_{\theta_0} \delta_{\phi} \nonumber \\
& +   \dot z_0^{\mu}\dot z_0^{\nu} \delta_{\phi}\Big(\delta_{r_0}\delta_{\theta_0}\Big(1-\frac{2r_1}{r_0}\Big) - r_1 \delta'_{r_0}\delta_{\theta_0}  - \theta_1 \delta_{r_0}\delta'_{\theta_0}\Big) \bigg] \nonumber \\
& +\frac{1}{2}h^{{\rm R}1 }_{\alpha \beta}\left(u_0^\alpha u_0^\beta-g^{\alpha\beta}\right)T^{\mu\nu}_{1}+ T_{(d)}^{\mu \nu}.\label{T2 multiscale}
\end{align}
The leading dipole term is
\begin{align}
\label{eq:ttdipoleSE}
    T_{(d)}^{\mu \nu} &= \frac{1}{r^{2} \sin \theta} \Bigl( K_{1}^{\mu \nu} \delta_{r_0} \delta_{\theta_0} \delta_{\phi} +K_{2}^{\mu \nu} \delta_{r_0} \delta_{\theta_0} \delta_{\phi}' \nonumber \\
    & \qquad +K_{3}^{\mu \nu} \delta_{r_0}' \delta_{\theta_0} \delta_{\phi}+K_{4}^{\mu \nu} \delta_{r_0} \delta_{\theta_0}' \delta_{\phi} \Bigr),
\end{align}
with $\delta_\phi'\equiv \partial_\phi\delta(\phi-\phi_p)$, $\delta_{\theta_0}'\equiv\partial_\theta\delta(\theta-\pi/2)$ and $\delta_{r_0}'\equiv\partial_r\delta(r-r_0)$. The components of the symmetric tensors $K_1^{\mu \nu}$, $K_2^{\mu \nu}$, $K_3^{\mu \nu}$ and $K_4^{\mu \nu}$ are given in Appendix~\ref{app:Kcomps}.

\subsection{Field equations and Fourier expansions}

We next consider a Fourier expansion of the field equations. Our procedure is a simple modification of the one described in Sec.~IV~D of Ref.~\cite{Mathews:2021rod}, with the major change being  the metric's periodic dependence on the additional phase, $\tilde\psi_s$. We begin at the same point, substituting the expansions~\eqref{g tt expansion} and~\eqref{T tt expansion} into the field equations \eqref{eq:FE} and equating coefficients of powers of $\ee$ (which we stress is the initial, rather than evolving, mass ratio) at fixed $(\phi_p, \tilde\psi_s,\varpi_I)$. We then obtain
\begin{subequations}
\label{tt EFE}
\begin{align}
 G_{\mu\nu}[g] &= 0,\\
 G^{(1,0)}_{\mu\nu}[h^{1}] &= 8\pi T^{1}_{\mu\nu},\label{tt EFE1}\\
 G^{(1,0)}_{\mu\nu}[h^{2}] &= 8\pi T^{2}_{\mu\nu} - G^{(2,0)}_{\mu\nu}[h^{1},h^{1}] -G^{(1,1)}_{\mu\nu}[h^{1}].\label{tt EFE2}
\end{align}
\end{subequations}
The operators $G^{(n,j)}_{\mu\nu}$ are found by first expanding the spacetime's full Einstein tensor in powers of $h_{\mu\nu}$,
\begin{equation}
G_{\mu\nu}[g+h]=G_{\mu\nu}+G^{1}_{\mu\nu}[h]+G^{2}_{\mu\nu}[h,h]+\mathcal{O}(h^3),
\end{equation}
 and then replacing covariant derivatives in $G^{n}_{\mu\nu}$ via Eq.~\eqref{nabla tt}. Thus $G^{(1,0)}_{\mu\nu}$ is the linearized Einstein tensor with $\nabla_\alpha$ replaced by $\nabla^{0}_\alpha$, and $G^{(n,1)}_{\mu\nu}$ derives from terms linear in ${\cal V}_I$ and $\Omega_s^{(1)}$ in $G^{n}_{\mu\nu}$. Because $h^1_{\alpha \beta}$ is independent of $\tilde\psi_s$, the linear $\Omega_s^{(1)}$ term in $G^{(n,1)}_{\mu\nu}$ does not appear until third order in the actual field equations via terms proportional to $\partial_{\tilde{\psi}_s}h^2_{\alpha \beta}$. For a more thorough description, refer to Refs.~\cite{Miller:2020bft,Miller:2023ers,Mathews:2025nyb}. 

To take advantage of the periodic dependence on each phase, $\phi_p, \tilde\psi_s$, we represent the metric as a Fourier series,
\begin{subequations}\label{two-time modes}
\begin{align}
& h^{1}_{\alpha\beta}(x^i,\phi_p,\varpi_I) = \sum_{m=-\infty}^\infty  h^{(1,m)}_{\alpha\beta}(x^i,\varpi_I)e^{-im\phi_p},\\    
& h^{2}_{\alpha\beta}(x^i,\phi_p, \tilde\psi_s,\varpi_I) = \nonumber \\
& \qquad \sum_{m=-\infty}^\infty \sum_{k=-1}^{1} h^{(2,m,k)}_{\alpha\beta}(x^i,\varpi_I)e^{-i(m\phi_p+k\tilde\psi_s)},
\end{align}
\end{subequations}
where $k=\pm1$ modes only appear for the term $h^{2(\chi_\perp)}_{\mu\nu}$ in Eq.~\eqref{eq:h2tt}. 
We then have $\frac{\partial}{\partial\phi_p} \to -im$ and $\frac{\partial}{\partial\tilde\psi_s} \to -ik$ when acting on individual modes, implying Eq.~\eqref{nabla0} becomes
\begin{equation}\label{del0 modes}
\nabla^{0}_\alpha \to \eta^i_\alpha\frac{\partial}{\partial x^i} -i s_\alpha \omega_{m k}+ \text{Christoffel terms},
\end{equation}
with 
\begin{equation}
\label{eq:spectrum}
    \omega_{m k} \equiv m \Omega + k \Omega_s^{(0)}.
\end{equation}

Because the precession phase only appears as a linear function of $\cos\tilde\psi_s$ and $\sin\tilde\psi_s$ in the metric (through second order), there are only three modes of precession ($k=0,\pm1$) in the spectrum of frequencies in Eq.~\eqref{eq:spectrum}. Inspecting Eq.~\eqref{eq:tt prec freq} and Eq.~\eqref{eq:tt prec freq split}, we find that there are no possible resonances between $\Omega$ and $\Omega_s$,\footnote{To see this, note that $1/u_0^{t}\rightarrow 1$ as $r_0 \rightarrow \infty$ and $1/u_0^{t}\rightarrow 0$ as $r_0 \rightarrow 3 m_1^{(0)}$ (and takes no integer values in between). Thus inspecting $\Omega_s^{(0)}$ (and arguing that $\Omega_s^{(1)}$ should remain small) reveals there are no possible resonances in the frequency modes $\omega_{m\pm1}=m \Omega \pm \Omega_s$ given $m$ takes integer values.} justifying our treatment of the two phases $\phi_p$ and $\tilde\psi_s$ as independent. As a side remark, we also note that the spectrum of frequencies contain harmonics of frequencies that are too low to resonate with the primary's quasi-normal modes \cite{Cardoso:2021vjq, Yin:2024nyz}.

Finally, we highlight that the metric perturbation only depends on $\phi$ and $\phi_p$ in the combination $(\phi-\phi_p)$; this implies that the mode number $m$ in Eq.~\eqref{two-time modes} is the same azimuthal mode number that appears in a (tensor) spherical harmonic expansion of the metric perturbation.

By expanding all quantities in the field equations in these discrete Fourier series, we reduce the field equations to  partial differential equations, in $(r,\theta,\phi)$, for the mode coefficients $h^{(n,m,k)}_{\alpha\beta}$. A crucial aspect of our expansion is that the left-hand side of these field equations is identical to the frequency-domain equations one obtains for the linearized Einstein equations with Fourier modes $e^{-i\omega_{mk} s}$ (where, recall, $s$ is our hyperboloidal time), even though here $\phi_p\neq\Omega s$ and $\tilde\psi_s\neq\Omega^{(0)}_s s$. This makes the equations amenable to standard methods of solving frequency-domain equations. We refer to Ref.~\cite{Miller:2023ers} for more details on the formulation of the field equations in a multiscale expansion.

\subsection{Summary: two-timescale evolution with precessing spin}
\label{sec:SummaryTT}

In the preceding sections we have extended the multiscale framework of Refs.~\cite{Miller:2020bft,Mathews:2021rod}, which were limited to (anti-)aligned secondary spin, to allow for a precessing secondary spin.
Or, equivalently, we have specialized the generic treatment of Paper~I to the case of a slowly spinning primary whose spin is at most slightly misaligned with the orbital angular momentum. The resulting waveform-generation scheme divides into an offline and an online step:
\begin{enumerate}
\item {\em Offline computations}. Offline computations are performed on a grid of $\Omega$ values. Since the other dynamical mechanical variables ($\delta m_1$ and $\delta \chi_1$) and the secondary parameters ($m_2$, $\chi_\parallel$, and $\chi_\perp$) appear only linearly (and quadratically in the case of $m_2$), we calculate their coefficients and leave their values unspecified until the online step. We solve the field equations~\eqref{tt EFE1} and~\eqref{tt EFE2} on this $\Omega$ grid for the mode amplitudes $h^{(n,m,k)}_{\alpha\beta}$. From these, we compute the forcing functions $F^{(0)}_\Omega(\Omega)$, $F^{(1)}_{\delta m_1}(\Omega)$, $F^{(1)}_{\delta \chi_1}(\Omega)$, and $F^{(1)}_\Omega(\Omega,\delta m_1,\delta \chi_1)$ and $\Omega_s^{(1)}$ on the grid.\footnote{$\Omega_s^{(1,0)}$ in Eq.~\eqref{eq:tt prec freq split} can be extracted from the directly related calculations in Ref.~\cite{Dolan:2013roa} while the other contributions to $\Omega_s^{(1)}$ are given analytically in Eq.~\eqref{eq:precfreqterms}.} Our use of Fourier expansions means the orbital phases factor out of these offline computations; only the Fourier mode coefficients enter.
\item {\em Online simulation}. The online stage comprises a fast evolution through phase space, using the pre-computed forcing functions, and a summation of waveform modes, using the pre-computed waveform amplitudes. We choose values of $m_1^{(0)}$, $\delta\chi_1(t=0)$, $m_2$, $\chi_\parallel$, and $\chi_\perp$, recalling $\delta m_1(t=0)\equiv0$. We then solve Eqs.~\eqref{Omegadef} and \eqref{eq:ttinspiral} for the phase-space trajectory$$(\phi_p, \tilde\psi_s(t,\e),\Omega(t,\e),\delta m_1(t,\e), \delta \chi_1(t,\e)).$$
From the trajectory and the pre-computed mode amplitudes, $h^{(n,m,k)}_{\alpha\beta}$, we then generate the waveform:\footnote{In actuality, the second-order terms in this expressions are ill defined at infinity due to the asymptotic irregularity of the Lorenz gauge~\cite{Cunningham:2024dog}. In practice, we use a matching procedure, subtract a ``puncture'' at large $r$ to obtain a regular residual field, and then transform the total metric perturbation to the Bondi-Sachs gauge. See Refs.~\cite{Miller:2023ers,Cunningham:2024dog} for details.}
\begin{multline}\label{eq:waveform}
    \hspace{25pt}\lim_{r\to\infty}r\sum_{n,m,k} \e^n h^{(n,m,k)}_{\alpha\beta}(x^i,\varpi_I(t,\e))\\[-5pt]
    \times e^{-i[m\phi_p(t,\e)+k\tilde\psi_s(t,\e)]}.
\end{multline}
Here $t$ refers to Schwarzschild time along the particle trajectory; equivalently, since the phase-space variables are constant on slices of constant $s$, we can replace $t$ with hyperboloidal time $s$ in the waveform.

\end{enumerate}

We note that solving Eqs.~\eqref{Omegadef} and \eqref{eq:ttinspiral} in the online step does not actually involve $\chi_\perp$ because, as we have explained, the evolution equations do not depend on $\chi_\perp$ at 1PA order. The intuitive reason for this is that the $k\neq0$ modes of the second-order metric perturbation cannot interact with any first-order modes to generate a nonzero flux of energy (or any other secular effect); such products will always be oscillatory functions of the two independent phases $\phi_p$ and $\tilde\psi_s$, which then average to zero. A nonzero secular effect first enters at 2PA, when $k=\pm1$ modes can multiply against each other to generate a non-oscillatory ${\cal O}(\e^4)$ flux. The spin precession hence only enters our waveform through the direct contribution of the $n=2$, $k\neq0$ modes in Eq.~\eqref{eq:waveform}. We include this contribution, even though it is itself a 2PA effect, because it is distinct from other 2PA effects: it modulates the waveform by introducing additional frequencies, while the 2PA effects we neglect would only contribute to the evolution of $\phi_p$ through a forcing function $F^{(2)}_\Omega$.  We refer again to Paper I for a more detailed discussion. 

Given that the spin precession is already a 2PA effect in the waveform, one might suspect that the correction $\Omega_s^{(1)}$ is an entirely negligible 3PA effect. However, its impact is somewhat subtler. To see this, we follow Paper I in writing each $m$ mode of the waveform as $h_m = A_m e^{-i\Phi_m}$, with $A_m=|h_m|$ and $\Phi_m = -{\rm arg}(h_m)$. Both $A_m$ and $\Phi_m$ here are functions of angles on the sky; $m$ corresponds to a Fourier mode number in Eq.~\eqref{eq:waveform}, not an azimuthal mode number in a spherical-harmonic decomposition (though the two mode numbers become equal if the Fourier coefficients $h^{(n,m,k)}_{\alpha\beta}$ are decomposed in spin-weighted spherical harmonics). Decomposing Eq.~\eqref{eq:waveform} in this way, we find
\begin{multline}\label{eq:waveform real phase}
    \Phi_{m} = m\phi_p -\e^0\arg\bigl(h^{(1)}_m\bigr)
    +\frac{\e}{\bigl|h^{(1)}_m\bigr|^2}\Bigl(d^{\,0}_m
  +\chi_\perp d^{\,\rm s}_m\sin\tilde\psi_s\\
 +\chi_\perp d^{\,\rm c}_m\cos\tilde\psi_s\Bigr)
 + {\cal O}(\e^2),
\end{multline}
with (sky-angle-dependent) coefficients $d^{\,0}_m$, $d^{\,\rm s}_m$, and $d^{\,\rm c}_m$ given by Eq.~(197) of Paper I. Recalling that an order-$\e$ term in the phase is 2PA, we see that $d^{\,0}_m$ represents a 2PA phase correction without precession, and the other two order-$\e$ terms represent the precession's modulation of the 2PA phase. A 3PA term in the waveform phase would be proportional to $\e^2$ in the above expression, which is \emph{not} the effect of $\Omega_s^{(1)}$. Instead, as implied in Eq.~\eqref{psis - slow t}, the correction $\Omega^{(1)}_s$ affects $\tilde\psi_s$ at order $\e^0$; hence, rather than being a 3PA term, it is an order-unity effect in the modulation of the 2PA term. The impact of this modulation on parameter estimation bears further investigation, which we defer to future work.

\section{Flux balance and the first law of binary black hole mechanics}
\label{sec:fluxbal}

In Sec.~\ref{sec:tt}, we provided formulae for the rate of change of the orbital frequency, $F^{(n)}_{\Omega}$, in terms of the local self-force at the worldline: Eqs.~\eqref{Omegadot0} and~\eqref{Omegadot1}. Current state-of-the-art self-force waveform models do not use these directly, and instead approximate $F_\Omega^{(n)}$ by employing an energy flux balance law which only partially involves local self-force calculations~\cite{Wardell:2021fyy, Albertini:2022rfe}. In this section, we clarify this approach along with the approximation of the binary's binding energy that the models presently rely upon. Later, in Sec.~\ref{sec:1PAT1R}, we describe how the energy balance law approach informs an effective re-summation of $F_\Omega$ that significantly enhances the waveform model accuracy.

\subsection{Flux balance law}
\label{sec:energy balance law}
Computing the local self-force at the wordline either involves first computing the retarded metric perturbation and using a regularisation procedure~\cite{Heffernan:2022zgc, vandeMeent:2017bcc} to account for its singular behavior, or directly computing the effective metric via an effective-source/puncture method~\cite{PanossoMacedo:2022fdi, Leather:2023dzj, PanossoMacedo:2024pox, Upton:2023tcv} and then computing the self-force. The former technique has only been developed for the first-order self-force, and we require the inclusion of the second-order self-force in Eq.~\eqref{Omegadot1}. The latter technique is favored though significant effort is front-loaded into the calculation of the effective source. Local calculations of the second-order self-force require highly accurate computations of a sufficiently smooth effective source. Meanwhile, the calculation of the second-order gravitational energy flux at future null infinity~\cite{Warburton:2021kwk} is less sensitive to the smoothness and accuracy of the effective source. Motivated by such practicalities, we outline the balance law and the determination of a well-motivated approximation to  $F^{(n)}_{\Omega}$ that enables the 1PA inspiral to be determined from the asymptotic calculations of the second-order energy flux encoded in $h^2_{\alpha \beta}$, combined with local calculations of $h^1_{\alpha \beta}$. 

First, we define the binary's binding energy as a function of the hyperboloidal time coordinate, $s$,
\begin{equation}
\label{eq:bindingenergy}
    E(s)=M_{\text{Bondi}}(s)-m_1(s)-m_2,
\end{equation}
following the  analysis of non-spinning binaries in Sec.~II~B of Ref.~\cite{Albertini:2022rfe}. While we have indicated that the primary's mass is time dependent, we have neglected any $s$ dependence in $m_2$ since its evolution is a 3PA effect~\cite{Poisson:2004cw}. The primary mass $m_1(s)$ and Bondi mass $M_{\text{Bondi}}(s)$ can be measured directly from the quasistationary ($m=k=0$), spherically symmetric modes of the metric on surfaces of constant $s$ where they intersect the primary's horizon and future null infinity (${\cal I}^+$). Through this procedure, as explained in Refs.~\cite{Pound:2019lzj,Bonetto:2021exn,Cunningham:2024dog}, the binding energy can be computed as a function of $\varpi_I$. Differentiating Eq.~\eqref{eq:bindingenergy} with respect to $s$ then yields
\begin{equation}
\label{eq:dEds}
    \frac{\partial E}{\partial \varpi_I}\frac{d \varpi_I}{ds}=\frac{d M_{\text{Bondi}}}{d s}-\frac{d m_{1}}{d s}.
\end{equation}
The Bondi-Sachs mass-loss equation states \cite{Bondi:1962px,Sachs:1962wk}
\begin{equation}
\label{eq:BSmassloss}
 \frac{d M_{\text{Bondi}}}{d s}=-{ \mathcal{F}}^\infty,
\end{equation}
in which ${\mathcal{F}}^\infty$ is the gravitational wave energy flux through the $s=\text{constant}$ cut of $\mathcal{I}^+$. Meanwhile, 
\begin{equation}\label{eq:dm1ds}
\frac{d m_{1}}{d s}={\mathcal{F}}^{\mathcal{H}},
\end{equation}
where ${\mathcal{F}}^{\mathcal{H}}$ is the gravitational wave energy flux through the $s=\text{constant}$ cut of the primary's 
horizon \cite{Ashtekar:2004cn}. 
Substituting these flux-balance equations into Eq.~\eqref{eq:dEds} and rearranging for $d\Omega/ds$, we obtain an evolution equation for the orbital frequency:
\begin{equation}
  \label{eq:dOmegads}
    \frac{d\Omega}{ds}= -\frac{{\cal F}+\displaystyle\frac{\partial E}{\partial \delta m_1}\frac{d\delta m_1}{ds}+\frac{\partial E}{\partial \delta \chi_1}\frac{d\delta \chi_1}{ds}}{\partial E/\partial\Omega}.  
\end{equation}
Here for convenience we have defined the total gravitational wave energy flux ${\mathcal{F}}={\mathcal{F}}^{\mathcal{H}}+{\mathcal{F}}^{\infty}$. 

Recall that $s$ reduces to $t$ at the particle, such that Eq.~\eqref{eq:dOmegads} is equivalent to the evolution equation~\eqref{Omegadot}. We next expand each of the quantities on the right-hand side of Eq.~\eqref{eq:dOmegads} in powers of $\ee$, expressing them as functions of $\varpi_I$ (and of the system's constant parameters) in the process. This then allows us to read off the forcing functions $F^{(n)}_\Omega$ in Eq.~\eqref{Omegadot}.

First we note that Eq.~\eqref{eq:dm1ds} implies
\begin{equation}
\epsilon \frac{d \delta m_{1}}{d s}={\mathcal{F}}^{\mathcal{H}}.
\end{equation}
Additionally, if we label the gravitational angular momentum flux at the horizon as ${\mathcal{L}}_{\mathcal{H}}$, then
\begin{equation}
    \frac{d}{ds}\left(m_1^2\chi_1\right)={\mathcal{L}}^{\mathcal{H}},
\end{equation}
which in turn implies
\begin{equation}
\ee \frac{d \delta \chi_1}{ds} =\left({\mathcal{L}}^{\mathcal{H}}-2 m_1\chi_1{\mathcal{F}}^{\mathcal{H}}\right)m_1^{-2}.
\end{equation}

Next, for our quasi-circular, approximately equatorial inspiral, we expand the binding energy as 
\begin{multline}
\label{eq:bindingenergyexpand}
E= \ee E_1(\Omega) +\ee^2 E_2(\Omega) +\ee^2 \delta \chi_1 E_{(\delta \chi_1)}(\Omega)\\+\ee^2 \chi_\parallel E_{(\chi_\parallel)}(\Omega)+\ee^2 \frac{\delta m_1}{m_1^{(0)}} E_{(\delta m_1)}(\Omega) +\mathcal{O}(\e^3),
\end{multline}
noting that $\chi_\perp$ does not contribute to the binding energy at this order.\footnote{The dynamical effects linearly proportional to $\chi_\perp$ of a spinning test-body are purely oscillatory and thus cannot contribute to the system's constants of motion. This definition of the binding energy reduces to the equivalent conserved quantity in the test-body limit, corresponding to an eternal circular bound orbit of frequency $\Omega$ in a Schwarzschild background spacetime.} We likewise expand the fluxes in a similar form,
\begin{multline}
    \mathcal{F}=\ee \bigl[\ee \mathcal{F}_1(\Omega) +\ee^2  \mathcal{F}_{2}(\Omega)+\ee^2 \delta \chi_1 \mathcal{F}_{(\delta \chi_1)}(\Omega)\\+\ee^2 \chi_\parallel \mathcal{F}_{(\chi_\parallel)}(\Omega)+\ee^2 \frac{\delta m_1}{m_1^{(0)}} \mathcal{F}_{(\delta m_1)}(\Omega)\bigr] +\mathcal{O}(\e^4),
\end{multline}
with an identical form for the individual horizon and $\mathcal{I}^+$ terms (and for the angular momentum flux).\footnote{Note that we are working with the full binding energy and angular momentum ($\sim \ee$) and their fluxes ($\sim \ee^2$), not the flux of the \emph{specific} energy and angular momentum ($\sim \ee$).} Since $\mathcal{F}_1$ is dimensionless (with units of mass divided by time), it can only depend on $m^{(0)}_1$ and $\Omega$ in the dimensionless combination $m^{(0)}_1\Omega$; and since ${\cal F}_{(\delta m_1)}=m_1^{(0)}\partial {\cal F}_1/\partial m_1$, it follows that ${\cal F}_{(\delta m_1)}=\Omega\,\partial{\cal F}_1/\partial\Omega$. Again, $\chi_\perp$ does not contribute to the fluxes at this order. The other flux terms must be computed independently.

After making these substitutions in Eq.~\eqref{eq:dOmegads}, we read off the following coefficients of $\ee$ (0PA) and $\ee^2$ (1PA):
\begin{subequations}
\label{eq:FOmegaBalanceLaw}
\begin{align}
F^{(0)}_\Omega&=-\mathcal{F}_1\left(\frac{\partial E_1}{\partial \Omega}\right)^{-1},\\
F^{(1)}_\Omega&=F^{(1,0)}_\Omega+\delta \chi_1F^{(1,\delta \chi_1)}_\Omega+\frac{\delta m_1}{m_1^{(0)}} F^{(1,\delta m_1)}_\Omega+\chi_\parallel F^{(1,\chi_\parallel)}_\Omega,
\end{align}
\end{subequations}
in which we have defined
\begin{subequations}
\begin{align}
F^{(1,0)}_\Omega =& -\left(\mathcal{F}_2\frac{\partial E_1}{\partial \Omega}- \mathcal{F}_1 \frac{\partial E_2}{\partial \Omega}\right)\left(\frac{\partial E_1}{\partial \Omega}\right)^{-2}\nonumber\\
&-\left(\frac{\partial E_1}{\partial \Omega}\right)^{-1}\left(\mathcal{F}_1^{\mathcal{H}}E_{(\delta m_1)} +\frac{\mathcal{L}_1^{\mathcal{H}}E_{(\delta \chi_1)}}{(m_1^{(0)})^2}\right),\\
F^{(1,\delta\chi_1)}_\Omega =& -\left(\mathcal{F}_{(\delta\chi_1)}\frac{\partial E_1}{\partial \Omega}- \mathcal{F}_1 \frac{\partial E_{(\delta\chi_1)}}{\partial \Omega}\right)\left(\frac{\partial E_1}{\partial \Omega}\right)^{-2},\\
F^{(1,\delta m_1)}_\Omega =& -\left(\mathcal{F}_{(\delta m_1)}\frac{\partial E_1}{\partial \Omega}- \mathcal{F}_1 \frac{\partial E_{(\delta m_1)}}{\partial \Omega}\right)\left(\frac{\partial E_1}{\partial \Omega}\right)^{-2},\\
F^{(1,\chi_\parallel)}_\Omega =& -\left(\mathcal{F}_{(\chi_\parallel)}\frac{\partial E_1}{\partial \Omega}- \mathcal{F}_1 \frac{\partial E_{(\chi_\parallel)}}{\partial \Omega}\right)\left(\frac{\partial E_1}{\partial \Omega}\right)^{-2}.
\end{align}
\end{subequations}

Given the fluxes, the only additional knowledge required to explicitly compute the quantities in Eq.~\eqref{eq:FOmegaBalanceLaw} is the expansion of the binding energy in Eq.~\eqref{eq:bindingenergyexpand}. While this may be computed directly from the asymptotic metric as in Ref.~\cite{Pound:2019lzj,Cunningham:2024dog}, extant 1PA self-force models have not used $E_2$ as defined from the Bondi mass in Eq.~\eqref{eq:bindingenergy} because Ref.~\cite{Pound:2019lzj}'s computation of the Bondi mass used a different choice of time slicing than the second-order flux computations in Ref.~\cite{Warburton:2021kwk}. Instead, the 1PA waveform models in Ref.~\cite{Wardell:2021fyy} used a notion of binding energy derived from the first law of binary black hole mechanics~\cite{LeTiec:2011ab}, which Ref.~\cite{Pound:2019lzj} showed to be a reasonably accurate approximation to $E$. 
See Sec.~IID of Ref.~\cite{Albertini:2022rfe} for an assessment of the impact of this approximation upon the accuracy of the waveform phase. We return to this point in Sec.~\ref{sec:evolution with FL}. 

\subsection{Binding energy from the first law}
\label{sec: FLBE}

For a binary of spinning point particles in an eternal circular orbit, the first law of binary black hole mechanics through linear order in each spin is the variational relationship~\cite{Blanchet:2012at,Ramond:2022ctc}
\begin{equation}\label{eq:FL variational}
    \delta M_{\rm FL} - \Omega\, \delta L_{\rm FL} =\sum_{i=1,2}\left(z^{[i]} \delta m_i + \omega^{[i]}\delta S_i\right).
\end{equation}
Here $M_{\text{FL}}$ and $L_{\text{FL}}$ are the Arnowitt-Deser-Misner (ADM) mass and angular momentum \cite{Arnowitt:1962hi} of the eternally circular system. $\Omega$ still denotes the orbital frequency, and $z^{[i]}$ and $\omega^{[i]}$ are the Detweiler `redshift'~\cite{Detweiler:2008ft} and precession frequency\footnote{Here, $\omega^{[i]}$ are defined in a Hamiltonian sense. See Equation 4.5 of Ref.~\cite{Blanchet:2012at}.} of body $i$, respectively. Since $M_{\rm FL}$ and $L_{\rm FL}$ are functions of the independent variables $(\Omega,m_i,S_i)$, the variational relationship implies a set of partial differential equations:
\begin{subequations}
\label{eq:FL}
\begin{align}
\label{eq:FLa}
   \frac{\partial{M_{\text{FL}}}}{\partial \Omega}-\Omega\frac{\partial{L_{\text{FL}}}}{\partial \Omega}&=0,\\
\label{eq:FLb}
   \frac{\partial{M_{\text{FL}}}}{\partial m_i}-\Omega\frac{\partial{L_{\text{FL}}}}{\partial m_i}&=z^{[i]},\\
\label{eq:FLc}
    \frac{\partial{M_{\text{FL}}}}{\partial S_i}-\Omega\frac{\partial{L_{\text{FL}}}}{\partial S_i}&=\omega^{[i]}.
\end{align}
\end{subequations}
Note that the partial derivatives are with respect to the full physical mass and spins of each black hole (as opposed to $\chi_i$, for example). 

In our context, the redshift for the secondary is defined as 
\begin{equation}
\label{eq:FLredshift}
z^{[2]}\equiv \frac{d\hat\tau}{dt} = \frac{1}{\hat u^t}.    
\end{equation}
The secondary's spin precession frequency is 
\begin{equation}
\omega^{[2]}=-\Omega_s, 
\end{equation}
with a sign difference due to a choice of convention. Since in black hole perturbation theory the primary is an extended object rather than a point particle, quantities such as its redshift and spin precession frequency would have to be defined as appropriate integrals over its horizon.\footnote{Though we note that on scales large compared to the primary's size, its redshift $z^{[1]}$ can be identified with its surface gravity~\cite{Zimmerman:2016ajr,LeTiec:2017ebm,Albalat:2022lfz}.} However, in what follows, we will not make use of the equations involving those quantities [\eqref{eq:FLb} and \eqref{eq:FLc} with $i=1$], and so our calculations will be insensitive to those definitions.

A subtler point is that in these relationships we should not interpret $M_{\rm FL}$ and $L_{\rm FL}$ as the ADM mass and angular momentum of our physical, inspiraling system. Those quantities would be constants, as they would be for any asymptotically flat spacetime. Instead, for any set of values of $(\Omega,m_i,S_i)$, we construct an effective, asymptotically flat spacetime for an eternally circular binary with those parameters. Such a system is unphysical because no exact solution to the Einstein equations can be both helically symmetric and asymptotically flat~\cite{Ramond:2022ctc}. When using the first law, we work under the (unproved) assumption that the same relationships~\eqref{eq:FL variational} and \eqref{eq:FL} hold true for our physical, inspiraling system at each value of hyperboloidal time $s$. One can think of this heuristically as $M_{\rm FL}$ and $L_{\rm FL}$ being good proxies for the Bondi mass and angular momentum at time $s$. (As discussed below, that surrogacy breaks down at 4PN order and at order $\e^2$ in the Bondi mass~\cite{Trestini:2025nzr,Grant:2025InPrep}, but it remains a reasonably good approximation.)

We return to the applicability of the first law in our setting at the end of the next section. For now, we blithely proceed to use it to derive a formula for the binding energy~\eqref{eq:bindingenergyexpand}. 

To do so, we introduce the first law binding energy,
\begin{equation}
\label{eq:FLBE}
E_{\text{FL}}=M_{\text{FL}}-m_1-m_2.
\end{equation}

We expand the binding energy and angular momentum at fixed values of the dimensionless frequency $m_1\Omega$ 
to $\mathcal{O}(\e^3)$ as
\begin{subequations}\label{eq:EFL and LFL sums}
\begin{align}
E_{\text{FL}}&=m_2\sum_{a,b,c=0}^{a+b+c=1} E^{\text{FL}}_{(a,b,c)}\left(\frac{m_2}{m_1}\right)^a\left(\frac{S_1}{m_1^2}\right)^b\left(\frac{S_2}{m_1 m_2}\right)^c,\label{eq:EFL sum}\\
L_{\text{FL}}&=m_1 m_2\sum_{a,b,c=0}^{a+b+c=1} L^{\text{FL}}_{(a,b,c)}\left(\frac{m_2}{m_1}\right)^a\left(\frac{S_1}{m_1^2}\right)^b\left(\frac{S_2}{m_1 m_2}\right)^c,\label{eq:LFL sum}
\end{align}
\end{subequations}
where the coefficients $E^{\text{FL}}_{(a,b,c)}$ and $L^{\text{FL}}_{(a,b,c)}$ are dimensionless functions of $m_1\Omega$. 
The leading terms ($a=b=c=0$) represent the binding energy and angular momentum for a test mass $m_2$ in a Schwarzschild spacetime of mass $m_1$. For our slowly spinning primary and compact secondary, $m_2/m_1 \sim \ee$, $S_1/m_1\sim \ee$, and $S_2/(m_1 m_2)\sim \ee$. However, we avoid expanding directly in powers of $\ee$ here because it is defined in terms of $m_1^{(0)}$ instead of the physical mass $m_1$. 

We likewise expand the redshift and precession frequency to $\mathcal{O}(\e^3)$;
\begin{subequations}
\begin{align}
    z^{[2]}&=\sum_{a,b,c=0}^{a+b+c=1} z^{[2]}_{(a,b,c)}\left(\frac{m_2}{m_1}\right)^a\left(\frac{S_1}{m_1^2}\right)^b\left(\frac{S_2}{m_1 m_2}\right)^c,\\
    m_1\omega^{[2]}&=\sum_{a,b,c=0}^{a+b+c=1}\omega^{[2]}_{(a,b,c)}\left(\frac{m_2}{m_1}\right)^a\left(\frac{S_1}{m_1^2}\right)^b\left(\frac{S_2}{m_1 m_2}\right)^c,
\end{align}
\end{subequations}
where again the coefficients are dimensionless functions of $m_1\Omega$. 
Our arguments that follow will be based on the fact that $m_2/m_1$, $S_1/m_1^2$, and $S_2/(m_1 m_2)$ are independent, arbitrarily specified small quantities. This implies that from an equation of the form 
\begin{equation}
\sum_{abc}f_{abc}(m_1\Omega)\left(\frac{m_2}{m_1}\right)^a \left(\frac{S_1}{m_1^2}\right)^b \left(\frac{S_2}{m_1 m_2}\right)^c=0, 
\end{equation}
we can infer $f_{abc}(m_1\Omega)=0$.

Given this reasoning and the above expansions, Eq.~\eqref{eq:FLa} immediately implies
\begin{equation}
    \partial_\Omega E^{\text{FL}}_{(a,b,c)}= m_1 \Omega\,\partial_\Omega L^{\text{FL}}_{(a,b,c)},
\end{equation}
which we can recast as
\begin{equation}
\label{eq:FLaexpand}
     m_1 L^{\text{FL}}_{(a,b,c)}=-\partial_\Omega\left( E^{\text{FL}}_{(a,b,c)}- m_1\Omega L^{\text{FL}}_{(a,b,c)}\right).
\end{equation}
Next, from Eq.~\eqref{eq:FLb} with $i=2$ we have 
\begin{subequations}
\label{eq:FLbexpandv1}
    \begin{align} 
    E^{\text{FL}}_{(0,0,0)}&=z^{[2]}_{(0,0,0)}-1+m_1\Omega L^{\text{FL}}_{(0,0,0)},\\
    E^{\text{FL}}_{(0,1,0)}&=z^{[2]}_{(0,1,0)}+m_1\Omega L^{\text{FL}}_{(0,1,0)},\\
    E^{\text{FL}}_{(1,0,0)}&=\frac{1}{2}z^{[2]}_{(1,0,0)}+m_1\Omega L^{\text{FL}}_{(1,0,0)}.
    \end{align}
\end{subequations}
Equation~\eqref{eq:FLb} also implies
\begin{equation}
 z^{[2]}_{(0,0,1)}=0,
\end{equation}
since the factors of $m_2$ cancel for $a=0,c=1$ in Eq.~\eqref{eq:EFL sum}, meaning the derivative with respect to $m_2$ vanishes on the left-hand side of \eqref{eq:FLb} for these terms. 

Combining Eq.~\eqref{eq:FLaexpand} and Eq.~\eqref{eq:FLbexpandv1}, we can write most terms in the binding energy completely in terms of the redshift:
\begin{subequations}
\label{eq:FLbexpand}
    \begin{align} 
    E^{\text{FL}}_{(0,0,0)}&=z^{[2]}_{(0,0,0)}-1-\Omega \partial_\Omega z^{[2]}_{(0,0,0)},\\
    E^{\text{FL}}_{(0,1,0)}&=z^{[2]}_{(0,1,0)}-\Omega \partial_\Omega z^{[2]}_{(0,1,0)},\\
    E^{\text{FL}}_{(1,0,0)}&=\frac{1}{2}z^{[2]}_{(1,0,0)}-\frac{1}{2}\Omega \partial_\Omega z^{[2]}_{(1,0,0)}.
    \end{align}
\end{subequations}
These leave us with $E^{\text{FL}}_{(0,0,1)}$ as the only undetermined coefficient. Equation~\eqref{eq:FLc} with $i=2$  provides the additional relationship
\begin{equation}
\label{eq:FLcexpand}
     E^{\text{FL}}_{(0,0,1)}-\Omega m_1 L^{\text{FL}}_{(0,0,1)}=\omega^{[2]}_{(0,0,0)}.
\end{equation}
We can combine the relation with Eq.~\eqref{eq:FLaexpand} to obtain the final coefficient:
\begin{equation}
\label{eq:FLbexpandv1s2}
E^{\text{FL}}_{(0,0,1)}=\omega^{[2]}_{(0,0,0)}-\Omega \partial_\Omega\omega^{[2]}_{(0,0,0)}. 
\end{equation}

\subsection{Evolution equation with the first law}\label{sec:evolution with FL}

Suppose now we assume the two different binding energies are equal: $E_{\text{FL}}=E$, recalling that $E$ is defined in Eq.~\eqref{eq:bindingenergy} and $E_{\rm FL}$ in Eq.~\eqref{eq:FLBE}. Then we can compute each term in Eq.~\eqref{eq:bindingenergyexpand} by substituting Eq.~\eqref{eq:massandspins} into our first-law result and re-expanding in powers of the initial mass ratio, $\ee$. The outcome is%
\begingroup\allowdisplaybreaks
\begin{subequations}%
\label{eq:FLBEexpressions}
\begin{align}%
E_1(\Omega)&=m_1^{(0)}E^{\text{FL}}_{(0,0,0)}\left(m_1^{(0)}\Omega\right),\\
E_2(\Omega)&=m_1^{(0)}E^{\text{FL}}_{(1,0,0)}\left(m_1^{(0)}\Omega\right),\\
E_{(\delta\chi_1)}(\Omega)&=m_1^{(0)}E^{\text{FL}}_{(0,1,0)}\left(m_1^{(0)}\Omega\right),\\
E_{(\delta m_1)}(\Omega)&=m_1^{(0)}\Omega \partial_{\Omega}E^{\text{FL}}_{(0,0,0)}\left(m_1^{(0)}\Omega\right),\\
E_{(\chi_\parallel)}(\Omega)&=m_1^{(0)}E^{\text{FL}}_{(0,0,1)}\left(m_1^{(0)}\Omega\right).
\end{align}%
\end{subequations}%
\endgroup%
By combining the definition of the redshift in Eq.~\eqref{eq:FLredshift} with the timelike normalization condition $\hat{g}_{\mu \nu}\hat u^\mu \hat u^\nu =-1$ in our quasi-circular, approximately equatorial configuration, we obtain expressions for the coefficients $z^{[2]}_{(a,b,c)}$. Taking the redshift coefficients and $\omega^{[2]}_{(0,0,0)}=-\Omega_s^{(0)}$ as inputs to Eq.~\eqref{eq:FLbexpandv1} and Eq.~\eqref{eq:FLbexpandv1s2} respectively, we arrive at
\begin{subequations}\label{eq:Evals}%
\begin{align}%
E_1(\Omega)&=m_1^{(0)}\left(\frac{1-2y}{\sqrt{1-3y}}-1\right),\\
E_2(\Omega)&=\frac{1}{2}m_1^{(0)}\left(z_{\rm 1SF}-\Omega \partial_\Omega z_{\rm 1SF}\right),\label{eq:E2val}\\
E_{(\delta\chi_1)}(\Omega)&=m_1^{(0)}\left( \frac{-2y^{5/2}(2-3y)}{3(1-3y)^{3/2}}\right),\\
E_{(\delta m_1)}(\Omega)&=-m_1^{(0)}\frac{y}{3}\frac{1-6y}{(1-3y)^{3/2}},\\
E_{(\chi_\parallel)}(\Omega)&=m_1^{(0)}\left(-\frac{y^{5/2}}{\sqrt{1-3y}}\right),
\end{align}
\end{subequations}%
having defined $z_{\rm 1SF}\equiv z^{[2]}_{(1,0,0)}=\frac{1}{2}h^{(pp)R}_{\alpha\beta}u_0^\alpha k^\beta$, which is the standard first-order self-force redshift~\cite{Detweiler:2008ft}.

With the first-law binding energy at hand, we can now express Eq.~\eqref{eq:FOmegaBalanceLaw} in terms of the frequency, the redshift and the asymptotic fluxes. %
Expressing the $\Omega$ dependence in terms of the quantity $y$ defined in Eq.~\eqref{eq:ydef}, and defining $\bar F_\Omega^{(...)}\equiv \left(m_1^{(0)}\right)^2F_\Omega^{(...)}$, we obtain the following:
\begin{widetext}%
\begingroup\allowdisplaybreaks
\begin{subequations}%
\label{eq:OmegaDotFL}
\begin{align}
    \bar F_\Omega^{(0)}&=\frac{3 \sqrt{y} (1-3y)^{3/2} \mathcal{F}_1}{(1-6y)},\\
    \bar F_\Omega^{(1,\delta\chi_1)}&=\frac{3 \sqrt{y} (1-3y)^{3/2} \mathcal{F}_{(\delta\chi_1)}}{(1-6y)}-\frac{2 \sqrt{1-3 y}
   y^2 \left(36 y^2-33 y+10\right) \mathcal{F}_1}{ (1-6 y)^2},\\
    \bar F_\Omega^{(1,\delta m_1)}&=\frac{3 \sqrt{y} (1-3y)^{3/2}}{(1-6y)} \left(\frac{2}{3}y{\cal F}_1'(y)\right)+\frac{\sqrt{1-3 y} \sqrt{y} (3 (7-6 y) y-2) \mathcal{F}_1}{ (1-6 y)^2},\\
    \bar F_\Omega^{(1,\chi_\parallel)}&=\frac{3 \sqrt{y} (1-3y)^{3/2} \mathcal{F}_{(\chi_\parallel)}}{(1-6y)}-\frac{3 (1-3 y)^{3/2}
   y^2 (12 y-5) \mathcal{F}_1}{(1-6 y)^2},\\
    \bar F_\Omega^{(1,0)}&=\frac{(1-3y)^3 y^{1/2} (z_{\rm 1SF}'(y)-2 y z_{\rm 1SF}''(y))\mathcal{F}_1}{(1-6
   y)^2}+\frac{3 \sqrt{y} (1-3y)^{3/2} \mathcal{F}_2}{(1-6y)}-\left(y^{3/2}\frac{(5-12
   y) }{ (1-6y)}\right)\mathcal{F}_1^{\mathcal{H}}\label{eq:FOmega1PA}.
\end{align}%
\end{subequations}%
To group together the explicit horizon flux terms above, we have used the fact that $\mathcal{L}_{1}^{\mathcal{H}}(y)=m_1^{(0)}y^{-3/2}\mathcal{F}_1^{\mathcal{H}}$, for quasi-circular inspirals.
Thus by computing the relevant flux and redshift terms on a grid of $y$ values, we can integrate the coupled system in Eq.~\eqref{eq:ttinspiral} to obtain the parameter-space trajectory $\varpi_I(t)$. Equation~\eqref{eq:ttinspiral} reduces to
\begin{subequations}
\label{eq:ttinspiralFlux}
\begin{align}
    \left(m_1^{(0)}\right)^2\frac{d\Omega}{ds} &= \ee \bar F_{\Omega}^{(0)}(y) + \ee^2 \Bigl[\bar F^{(1,0)}_\Omega(y)+\delta \chi_1 \bar F^{(1,\delta \chi_1)}_\Omega(y) +\frac{\delta m_1}{m_1^{(0)}} \bar F^{(1,\delta m_1)}_\Omega(y)+\chi_\parallel \bar F^{(1,\chi_\parallel)}_\Omega(y)\Bigr]+\mathcal{O}(\e^3),\\      
    \frac{d\delta m_1}{ds} &= \ee \mathcal{F}_1^{\mathcal{H}}(y)+\mathcal{O}(\e^2),\\
    \left(m_1^{(0)}\right)^{2}\frac{d\delta \chi_1}{ds} &= \ee \mathcal{L}_{1}^{\mathcal{H}}(y) + {\cal O}(\e^2),
\end{align}%
\end{subequations}%
\endgroup%
with $y=y(\Omega(t))$.
\end{widetext}%

We now return to the question of whether our use of the first law is valid. An important facet of that question is that our use of the first law is actually exact and fully justified for all terms in Eq.~\eqref{eq:ttinspiralFlux} except the first term in Eq.~\eqref{eq:FOmega1PA} for $\bar F_\Omega^{(1,0)}$. The second term in $\bar F_\Omega^{(1,0)}$ accounts for the dissipation of the $E_1$ contribution to the binding energy due to the second-order energy flux, and its derivation relies solely upon the multiscale expansion of Eqs.~\eqref{eq:dEds}--\eqref{eq:dm1ds}. The last term in $F_\Omega^{(1,0)}$ derives from the time dependence of the primary's mass and spin corrections, which are explicit in Eq.~\eqref{eq:dOmegads}.

To explain why all the other terms are exact, we first remark that in the binding energy, $E_1$,  $E_{(\delta m_1)}$, $E_{(\delta \chi_1)}$, and $E_{(\chi_\parallel)}$ are all test-body terms that may be derived by linearizing (in $\delta m_1$, $\delta\chi_1$, and $\chi_\parallel$) the energy of a spinning test body in Kerr around the energy $E_1$ of a test mass in Schwarzschild spacetime. Specifically, in the test-body limit, we have the equality 
\begin{equation}
E\doteq m_2\left(\mathcal{E}-1\right),
\end{equation}
where the dot is to stress that the equality holds only in this limit, and we have defined the specific mechanical energy
\begin{equation}
\mathcal{E}\equiv -u^\alpha t_\alpha-\frac{m_2}{2}S^{\alpha\beta}\nabla_\alpha t_\beta.
\end{equation}
For these terms, the binding energy can be defined from the mechanical energy of a test body on a circular orbit. One can directly verify that the first law is valid when $E_{\rm FL}$ and $L_{\rm FL}$ are the energy and angular momentum of a spinning test body, and the first law's formulas for $E_1$,  $E_{(\delta m_1)}$, $E_{(\delta \chi_1)}$, and $E_{(\chi_\parallel)}$ can be directly verified from the test-body energy. Moreover, regardless of the first law, the test-body energy satisfies the flux-balance law~\eqref{eq:dEds} \emph{exactly} at the orders in spin and mass ratio that we use here~\cite{Akcay:2019bvk,Mathews:2021rod,Mathews:2025nyb}. In that sense, we did not need to make explicit use of the first law to calculate these contributions to $F^{(1)}_\Omega$. However, it provides a convenient, unified method of deriving the binding energy. 

Finally, we address our use of the first law in deriving $\bar F_\Omega^{(1,0)}$. This forcing function only pertains to the non-spinning, quasi-circular sector of the problem; all other effects have been stripped off and isolated as linear corrections. More specifically, in our use of the first law in Eq.~\eqref{eq:E2val}, we have replaced $E_2$---the leading non-test-body term in the binding energy~\eqref{eq:bindingenergy}---with 
\begin{equation}\label{eq:E2 FL}
E_2^{\rm FL}\equiv \frac{1}{2}m_1^{(0)}\left(z_{\rm 1SF}-\Omega\partial_\Omega z_{\rm 1SF}\right).
\end{equation}
Reference~\cite{Pound:2019lzj} found a numerically small, but nonzero difference between $E_2$ (as measured directly from the metric) and this $E_2^{\rm FL}$, and Ref.~\cite{Albertini:2022rfe} estimated that this probably leads (via $F^{(1)}_\Omega$) to a small but non-negligible difference in the waveform phase. However, the true impact on the waveform is unknown because $E_2(\Omega)$ and ${\cal F}_2(\Omega)$ are both dependent on the choice of hyperboloidal slicing. $E_2(\Omega)$ in Ref.~\cite{Pound:2019lzj} was calculated with a \emph{different} choice of slicing than ${\cal F}_2(\Omega)$ in Ref.~\cite{Warburton:2021kwk}, meaning it cannot be used in place of $E^{\rm FL}_2$. Here we highlight forthcoming work to better understand the relationship between $E_2$ and $E^{\rm FL}_2$:
\begin{enumerate}
    \item In Ref.~\cite{Lewis:2025ydo}, we will clarify that $E_{\rm FL}$ is the on-shell value of the local Hamiltonian governing the conservative sector of the secondary's orbital dynamics, even in the presence of dissipation (at least at the orders we consider here, for a nonspinning secondary). See also Refs.~\cite{Isoyama:2014mja,Fujita:2016igj,Blanco:2022mgd}. A disagreement between $E_2$ and $E_2^{\rm FL}$ therefore indicates that the rate of energy emission is \emph{not} equal to the rate of change of the secondary's mechanical Hamiltonian energy.
    \item In Ref.~\cite{Grant:2025InPrep}, we show analytically that $E_2$ differs from $E^{\rm FL}_2$ by an explicit, computable, nonzero correction term. This extends recent work by Trestini~\cite{Trestini:2025nzr}, who established an analogous result at 4PN order.
\end{enumerate}
Future work will compare the results of Ref.~\cite{Grant:2025InPrep} to the numerically computed $E_2$ in Ref.~\cite{Pound:2021qin} and incorporate the correction to $E_2^{\rm FL}$ into our waveform model.

However, we also point out that, in addition to the uncertainty surrounding the binding energy, there are known omissions in the flux ${\cal F}_2$:
\begin{enumerate}
    \item Existing ${\cal F}_2$ data omits the second-order flux through the primary's horizon. The effect of this omission, which should be numerically small, is discussed in Ref.~\cite{Albertini:2022rfe}.
    \item In Ref.~\cite{Cunningham:2024dog} we showed that existing data for ${\cal F}_2$ \emph{also} omits a `memory distortion' contribution arising from coupling between gravitational-wave memory and oscillatory modes. This conclusion will be further solidified in other forthcoming work~\cite{Spiers:InPrep}. However,  the impact of this contribution is expected to be small due to the small magnitude of memory modes.
\end{enumerate}

\subsection{Summary: balance law, binding energy and inspiral evolution}
In Sec.~\ref{sec:energy balance law}, we outlined how an energy balance law can be employed to express the rate of change of the orbital frequency in terms of a binding energy, $E$, and the gravitational wave energy flux at $\mathcal{I}^+$ and the primary's horizon (see Eq.~\eqref{eq:FOmegaBalanceLaw}). The expressions can be used in inspiral evolution schemes in place of the equivalent expressions derived in terms of the local self-force, which we initially derived in Sec.~\ref{sec:TT evolve worldline} --- see Eq.~\eqref{Omegadot0} and Eq.~\eqref{Omegadot1}. The intermediary equation, Eq.~\eqref{eq:dOmegads}, inspires a physically motivated effective re-summation that we implement in the 1PAT1R waveform model in Sec.~\ref{sec:1PAT1R}.

In Sec.~\ref{sec: FLBE}, we defined a different binding energy by making use of the first law of binary black hole mechanics. The expression for the first law binding energy, $E_{\text{FL}}$, is provided in Eq.~\eqref{eq:FLBEexpressions}.

In Sec.~\ref{sec:evolution with FL}, we derived explicit expressions for the rate of change of the orbital frequency by substituting $E_{\text{FL}}$ as an approximation to $E$ in Eq.~\eqref{eq:FOmegaBalanceLaw}, which we utilize in the waveform models of Sec.~\ref{sec:waveform}. We highlighted the specific terms impacted by this approximation and commented on other minor approximations employed in computing the second order energy flux, $\mathcal{F}_2$.

\section{Offline computations}
\label{sec:implementation}

\subsection{Overview of offline data}

In this section, we outline the general strategy we employ to obtain the offline data for our waveform model. 

We largely use data computed previously in Refs.~\cite{Warburton:2021kwk,Mathews:2021rod}: specifically, the asymptotic amplitudes of $h^{1(pp)}_{\mu\nu}$, $h^{2(pp)}_{\mu\nu}$, $h^{2(\delta{\chi}_1)}_{\mu\nu}$, and $h^{2(\chi_\parallel)}_{\mu\nu}$ that enter the waveform~\eqref{eq:waveform}; and the corresponding fluxes $\mathcal{F}_1$, $\mathcal{F}_{2}$, $\mathcal{F}_{(\delta \chi_1)}$, and $\mathcal{F}_{(\chi_\parallel)}$ and redshift $z_{1\rm SF}$ entering the evolution equation~\eqref{eq:ttinspiralFlux}. The new ingredients, not previously computed, are the amplitudes $h^{2(\delta m_1)}_{\mu\nu}$ and $h^{2(\chi_\perp)}_{\mu\nu}$. However, for completeness we summarize the calculations of all ingredients.

\subsubsection{Field equations, puncture fields and residual fields}

The data are computed from solutions to the field equations~\eqref{tt EFE}. In these equations we substitute the first- and second-order metric perturbations in the forms~\eqref{eq:h1tt} and~\eqref{eq:h2tt} along with the stress-energy tensor~\eqref{T tt expansion}. We thus obtain equations that determine each term in the metric perturbation, separated by peeling off the coefficients of the powers of $\ee$, $\delta m_1$, $\delta \chi_1$, $\chi_\parallel$ and $\chi_\perp$. We then decompose these equations into Fourier modes, using the expansion~\eqref{two-time modes}, and into tensor-spherical-harmonic $\ell m$ modes, as described in Refs.~\cite{Miller:2020bft,Mathews:2021rod}, to obtain radial ordinary differential equations for each $\ell m k$ mode coefficient.

As inputs in the second-order field equations (and to compute the first-law binding energy), we also need to split several of the metric perturbations into their singular and regular pieces:
\begin{equation}
\label{eq:singularregular}
h^{...}_{\mu \nu}=h^{ \mathrm{R}...}_{\mu \nu}+h^{ \mathrm{S}...}_{\mu \nu}\,;
\end{equation}
refer to the comments above Eq.~\eqref{eq:effectivemet} and the references therein. In practice, the singular fields are not known exactly but are approximated with \emph{puncture} fields $h_{\alpha\beta}^{\mathcal{P}...}\approx h_{\alpha\beta}^{\mathrm{S}...}$ near the worldline, from which we define the residual fields 
\begin{equation}\label{eq:punctureresidual}
h_{\alpha\beta}^{\mathcal{R}...}\equiv h_{\alpha\beta}^{...}-h_{\alpha\beta}^{\mathcal{P}...}.
\end{equation}
If the puncture is expanded to sufficiently high order (in distance from the particle), then the values of $h_{\alpha\beta}^{\mathcal{R}...}$ and its first derivatives on the worldline are identical to those of $h_{\alpha\beta}^{\mathrm{R}...}$, such that the residual field can be used in place of $h_{\alpha\beta}^{\mathrm{R}...}$ in the particle's equations of motion.
The puncture field for a spinning object, through second order in its mass and to sufficient order in distance, is given in local, co-moving coordinates in Ref.~\cite{Pound:2012dk} and in covariant form in Refs.~\cite{Pound:2014xva,Mathews:2021rod}.

\subsection{First-order fields}

The first-order point-mass field $h^{1(pp)}_{\mu\nu}$ satisfies
\begin{equation}
 G^{(1,0)}_{\mu\nu}[h^{1(pp)}] = 8\pi T^{1}_{\mu\nu}.
\end{equation}
Methods of solving this equation, decomposed in tensor-harmonic and Fourier modes, are standard in the literature~\cite{Barack:2018yvs}. 
As input for the second-order field equations, we specifically solve the mode-decomposed equation in the Lorenz gauge using the \texttt{h1Lorenz} code~\cite{h1Lorenz}, developed in Refs.~\cite{Akcay:2010dx,Akcay:2013wfa} and available on the Black Hole Perturbation Toolkit \cite{BHPToolkit}. 

The second-order field equations and binding energy also require the first-order puncture and residual field. We calculate the tensor-harmonic modes of the Lorenz-gauge puncture as described in Refs.~\cite{Wardell:2015ada,Bourg:2024vre}; the modes of the residual field are then obtained by subtracting the puncture modes from the retarded-field modes. We refer to Ref.~\cite{Upton:2025bja} for full details.

In order to more efficiently populate the grid of $\Omega$ values, we also separately generate data for the first-order waveform amplitudes and fluxes using the \texttt{Teukolsky} package~\cite{BHPT_Teukolsky} from the Black Hole Perturbation Toolkit. This data is used in our online waveform generation, while the Lorenz-gauge data is only used in offline calculations in the interior of the spacetime, specifically in the construction of second-order source terms and in the calculation of $z_{\rm 1SF}$ from the first-order regular field at the particle. 

Finally, we require the second two terms in Eq.~\eqref{eq:h1tt}, $h^{1(\delta m_1)}_{\mu\nu}(x^i)$ and $h^{1(\delta{\chi}_1)}_{\mu\nu}(x^i)$, which are smooth vacuum perturbations. Their corresponding sources vanish, and they contribute only to the first-order regular field. In principle, they are easily obtained analytically by expanding the Kerr metric with mass $m_1=m_1^{(0)}+\ee \delta m_1$ and dimensionless spin $\chi_1=\ee \delta\chi_1$ about the (non-spinning) Schwarzschild metric with mass $m_1^{(0)}$. In practice, we require them in the Lorenz gauge, as given analytically in Appendix~D of Ref.~\cite{Miller:2020bft}.

\subsection{Second-order fields}

\subsubsection{Linear secondary spin terms}

The secondary's spin contributes to the second-order metric perturbation via the fields $h^{2(\chi_\parallel)}_{\mu\nu}(x^i,\phi_p,\Omega)$ and $h^{2(\chi_\perp)}_{\mu\nu}(x^i,\phi_p, \tilde\psi_s,\Omega)$. These fields are straightforward to compute since their governing equations contain only linear terms, as there is no secondary spin dependence in $h^1_{\alpha\beta}$;
\begin{subequations}
    \begin{align}
        G^{(1,0)}_{\mu\nu}[h^{2(\chi_\parallel)}] &= 8\pi T^{2(\chi_\parallel)}_{\mu\nu},\label{EFE chi par}\\
        G^{(1,0)}_{\mu\nu}[h^{2(\chi_\perp)}] &= 8\pi T^{2(\chi_\perp)}_{\mu\nu}.
    \end{align}
\end{subequations}
Unlike for $h^1_{\alpha\beta}$, we do not require their corresponding punctures and residual fields --- we need only compute the retarded fields. 

In Eq.~\eqref{eq:ttinspiralFlux}, we provide the exact contribution of these fields (although $h^{2(\chi_\perp)}_{\mu\nu}$ does not contribute here) to the 1PA inspiral evolution in terms of asymptotic energy fluxes that are easily computed directly from the retarded metric perturbations. The relevant flux data were computed in Ref.~\cite{Mathews:2021rod}, which we have incorporated into this work. Rather than directly solving the mode-decomposed version of Eq.~\eqref{EFE chi par}, Ref.~\cite{Mathews:2021rod} computed this data by solving the corresponding Regge-Wheeler-Zerilli equations. 

Likewise, we obtain the waveform-amplitude contributions of both $h^{2(\chi_\parallel)}_{\mu\nu}$ and $h^{2(\chi_\perp)}_{\mu\nu}$  from the retarded solutions to the Regge-Wheeler-Zerilli equations. For the calculation of $h^{2(\chi_\perp)}_{\mu\nu}$ we adapted the method of Ref.~\cite{Mathews:2021rod}, where  $h^{2(\chi_\parallel)}_{\mu\nu}$ was already calculated. We detail the calculation of $h^{2(\chi_\perp)}_{\mu\nu}$ in Appendix~\ref{app:chiperp}.

\subsubsection{Nonlinear point-mass terms}

With the secondary spin terms accounted for, we next focus on the point-mass term $h^{2(pp)}_{\mu\nu}$ and its associated flux ${\cal F}_2$.

Due to the second-order source's strong singularity at the particle, the field equation is reformulated using a puncture scheme~\cite{Rosenthal:2006iy,Detweiler:2011tt,Pound:2012nt,Gralla:2012db}, adapting methods developed at first order~\cite{Barack:2007jh,Vega:2007mc,Wardell:2015ada}. In this approach, we solve for $h_{\alpha\beta}^{\mathcal{R}2(pp)}$ directly by re-arranging the point-mass terms in Eq.~\eqref{tt EFE} after substituting Eq.~\eqref{eq:punctureresidual}:
\begin{align}
 G^{(1,0)}_{\mu\nu}[h^{\mathcal {R}2(pp)}] &= 8\pi T^{2(pp)}_{\mu\nu} - G^{(2,0)}_{\mu\nu}[h^{1(pp)},h^{1(pp)}]\nonumber \\
 & \quad -G^{(1,1)}_{\mu\nu}[h^{1(pp)}]- G^{(1,0)}_{\mu\nu}[h^{\mathcal{P}2(pp)}].\label{tt EFE2 R pp}
\end{align}
The puncture term on the right cancels the singularities in the other source terms, leaving a more regular, effective source. Equation~\eqref{tt EFE2 R pp} is solved in a worldtube surrounding the particle's worldline. Outside the worldtube, we solve for the retarded field $h^{2(pp)}_{\mu\nu}$.

Reference~\cite{Upton:2025bja} describes our method of computing the source terms in the above equation, building on technology from Refs.~\cite{Wardell:2015ada,Miller:2016hjv,Miller:2020bft,Spiers:2023mor}. This source construction requires not only the first-order retarded field but also the first-order residual field (which enters into $T^{2(pp)}_{\mu\nu}$ and $h^{2{\cal P}(pp)}_{\mu\nu}$) and the first-order puncture field. The latter is required because, to accurately calculate~$G^{(2,0)}_{\mu\nu}[h^{1(pp)},h^{1(pp)}]$ in the worldtube around the worldline, we decompose it as~\cite{Miller:2016hjv,Upton:2025bja},
\begin{align}
    G_{\mu\nu}^{(2,0)}[h^{1(pp)},h^{1(pp)}] &= G_{\mu\nu}^{(2,0)}[h^{\mathcal{P}1},h^{\mathcal{P}1}] \nonumber\\
    &\quad + G_{\mu\nu}^{(2,0)}[h^{\mathcal{P}1},h^{{\cal R}1(pp)}] \nonumber\\
    &\quad + G_{\mu\nu}^{(2,0)}[h^{\mathcal{R}1(pp)},h^{{\cal R}1(pp)}].
\end{align}
The $\ell m$ modes of the first term are computed from the analytically known, four-dimensional $h^{\mathcal{P}1}_{\mu\nu}$, while the other two terms are computed from $\ell m$ modes of the first-order fields.

Our method of solving the (mode-decomposed) equations in Lorenz gauge is described in Ref.~\cite{Miller:2023ers}. The ultimate output is the asymptotic amplitude and flux data computed in Ref.~\cite{Warburton:2021kwk}, which we use here. We refer readers to Refs.~\cite{Miller:2020bft,Miller:2023ers,Upton:2025bja,Cunningham:2024dog} for further details on computing $h^{2(pp)}_{\mu\nu}$ and extracting its contribution to the waveform (and flux) at future null infinity.

\subsubsection{Primary mass terms}

In principle, $h^{2(\delta m_1)}_{\mu\nu}$ can be computed with the full second-order infrastructure described for the point-mass terms, constructing a nonlinear source involving products of $h^{1(pp)}_{\mu\nu}$ and $h^{1(\delta m_1)}_{\mu\nu}$. However, we can also calculate $h^{2(\delta m_1)}_{\mu\nu}$ far more easily, as it is necessarily equivalent to the field that would be produced by slightly shifting the background mass parameter in $h^{1(pp)}_{\mu\nu}$: 
\begin{equation}
h^{1(pp)}_{\mu\nu}(m_1) = h^{1(pp)}_{\mu\nu}(m^{(0)}_1) + \ee\delta m_1 \frac{\partial h^{1(pp)}_{\mu\nu}}{\partial m_1^{(0)}} + \ldots 
\end{equation}
From this, we read off $h^{2(\delta m_1)}_{\mu\nu} = \partial h^{1(pp)}_{\mu\nu}/\partial m_1^{(0)}$. 
We refer to the discussion around Eq.~(31) of Ref.~\cite{Warburton:2024xnr} for more information.

The associated flux is not explicitly required because, in formulating Eq.~\eqref{eq:OmegaDotFL}, we already made use of the equality ${\cal F}_{(\delta m_1)} = m_1^{(0)} \partial {\cal F}_1/\partial m_1^{(0)}$.

\subsubsection{Primary spin terms}

The final term we require in $h^{2}_{\mu\nu}$ is $h^{2(\delta{\chi}_1)}_{\mu\nu}$\footnote{We do not require the $h_{\mu \nu}^{2[j,k]}$ mass and spin corrections to the second-order metric as described in Eq.~\eqref{eq:h2tt}, since they couple to the waveform at second post-adiabatic order.}. Like $h^{2(\delta m_1)}_{\mu\nu}$, this term, and its associated flux, can be calculated in two ways: directly, using our full second-order infrastructure; or by linearizing first-order data. We explain each method in turn and then compare their results as a consistency check.

In the first approach, we follow the same procedure we used to compute $h^{2(pp)}_{\mu\nu}$, solving a field equation analogous to Eq.~\eqref{tt EFE2 R pp}. However, the source in the field equation for $h^{2(\delta{\chi}_1)}_{\mu\nu}$ has a more straightforward structure. More precisely, there are no terms quadratic in the first-order puncture (as $h^{1(\delta\chi_1)}_{\mu\nu}$ only contributes to $h^{{\rm R}1}_{\mu\nu}$), and there is no contribution from the slow-evolution term $G^{(1,1)}_{\mu\nu}[h^{1}]$ in the source~\eqref{tt EFE2} (as $h^{1(\delta\chi_1)}_{\mu\nu}$ is independent of $\Omega$, and the time derivative of $\delta\chi_1$ is proportional to a flux that is independent of $\delta\chi_1$). 
Hence, we can write the field equation for $h^{2(\delta{\chi}_1)}_{\mu\nu}$ as
\begin{align}\label{eq:delta_chi1_source}
 G^{(1,0)}_{\mu\nu}[h^{\mathcal {R}2(\delta\chi_1)}] &= 8\pi T^{2(\delta \chi_1)}_{\mu \nu}-G^{(2,0)}_{\mu \nu}[h^{1 (\delta \chi_1)}, h^{1(pp)}]\nonumber\\
 &\quad - G^{(1,0)}_{\mu \nu}[h^{\mathcal{P}2(\delta \chi_1) }].
\end{align}
We solve this equation in Lorenz gauge using the method in Ref.~\cite{Miller:2023ers}.
The flux, $\mathcal{F}^{2(\delta\chi_1)}$, is then computed from the asymptotic amplitudes of the metric perturbation.

In the second, simpler approach, the linear-in-$\chi_1$ amplitudes and flux are computed using the first-order Teukolsky formalism for a particle in Kerr spacetime. As in the discussion of primary-mass perturbation, we can write 
\begin{equation}
h^{2(\delta{\chi}_1)}_{\mu\nu} = \left.\frac{\partial h^{1(pp)}_{\mu\nu}}{\partial\chi_1}\right|_{\chi_1=0}
\end{equation}
and
\begin{equation}
\mathcal{F}_{(\delta\chi_1)} = \left.\frac{\partial {\cal F}_1}{\partial\chi_1}\right|_{\chi_1=0},
\end{equation}
where $h^{1(pp)}_{\mu\nu}$ and ${\cal F}_1$ on the right are calculated on a Kerr background with spin parameter $a=m_1^{(0)}\chi_1$. 
We can use these results as a check on the complicated machinery used to construct the source in Eq.~\eqref{eq:delta_chi1_source} and the numerical methods used to integrate the source to get the second-order field.

There is a slight subtlety in the comparison as, for $\delta\chi_1\neq0$, standard frequency-domain Teukolsky codes compute the fluxes on a spin-weighted spheroidal harmonic and not a spherical harmonic basis.
Thus to make the comparison we first expand the spin-weight -2 spheroidal harmonics,${}_{-2}S_{lm}$, onto a basis of spin-weight -2 spherical harmonics,  ${}_{-2}Y_{\ell m}$, such that
\begin{align}
    {}_{-2}S_{lm} = \sum_{l=l_{\rm min}}^\infty b^\ell_{lm} {}_{-2}Y_{\ell m},
\end{align}
where $l_{\rm min} = {\rm max}(|s|,|m|)$.
The spheroidal-to-spherical expansion coefficients, $b^\ell_{lm}$, can be computed using methods found in, e.g., Ref.~\cite{Hughes:1999bq}. In practice we evaluate the coefficients using the implementation in the \texttt{SpinWeightedSpheroidalHarmonics} Mathematica package \cite{BHPT_SWSH} of the Black Hole Perturbation Toolkit.

Defining $Z^\infty_{l m}$ as the amplitude of the Teukolsky radial function at infinity, we can write the flux for a given \textit{spherical harmonic} mode as 
\begin{equation}\label{eq:Teuk_flux_spherical}
    \mathcal{F}^{\rm Teuk}_{\ell m} = \sum_{l=l_{\rm min}}^{\infty}\frac{\epsilon^2}{4\pi m^2 \Omega^2_{\rm Kerr}} |b_{lm}^\ell Z_{lm}|^2,
\end{equation}
where $\Omega_{\rm Kerr} = \sqrt{m_1^{(0)}}/\left(r_0^{3/2}+ \chi_1 \left(m_1^{(0)}\right)^{3/2}\right)$.
Note that as we are only interested in the linear-in-$\chi_1$ flux, the limits on the summand in Eq.~\eqref{eq:Teuk_flux_spherical} can be reduced to ${\rm max}(l_{\rm min},l-1)<l<\ell+1$.
In order to evaluate Eq.~\eqref{eq:Teuk_flux_spherical} we use the \texttt{Teukolsky} Mathematica package~\cite{BHPT_Teukolsky} from the Black Hole Perturbation Toolkit to compute the $Z_{lm}$'s.
To extract the linear-in-$\chi_1$ piece of the flux we evaluate Eq.~\eqref{eq:Teuk_flux_spherical} on a grid of 25 evenly spaced values of $\chi_1 \in [0, 0.25]$. 
We then fit the data to a 5th-order polynomial in $\chi_1$ and extract the linear coefficient.

As a check on our results we can with compare against known PN series.
The leading PN term of the linear-in-$\chi_1$ flux can be extracted from the PN series in Refs.~\cite{Fujita:2014eta,Messina:2018ghh}. This leading term is given by
\begin{align}\label{eq:spin_flux_PN}
    \mathcal{F}^{2(\delta\chi_1)}_{\rm PN} = -\frac{256}{15}y^{13/2} + \mathcal{O}(y^{15/2}).
\end{align}
where $y=(m_1^{(0)}\Omega)^{2/3}$.

\begin{figure}
    \includegraphics[width=0.48\textwidth]{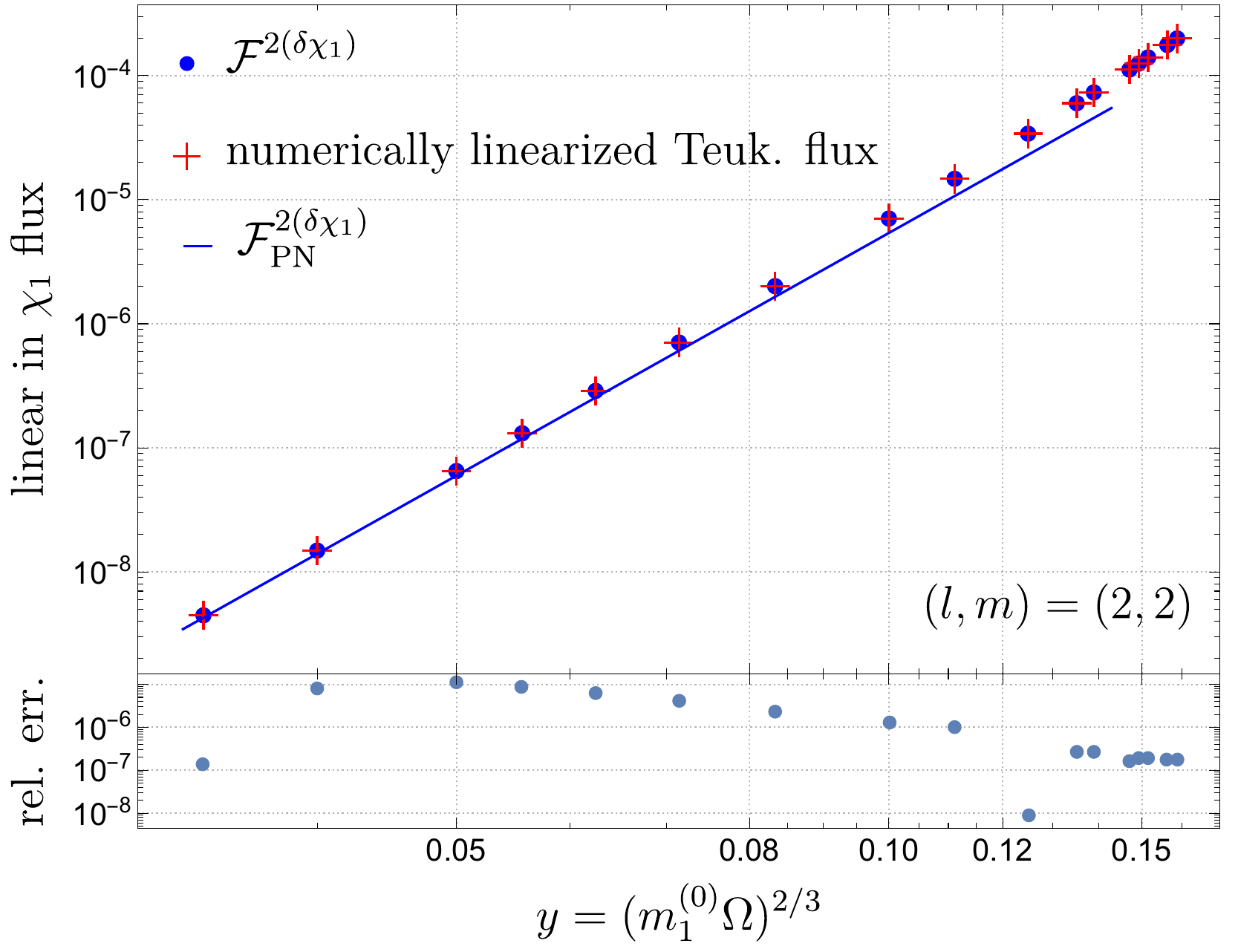}
    \caption{(Top) Comparison of the flux computed from the $\delta\chi_1$ perturbation and the linear-in-$\chi_1$ Teukolsky flux. 
    We also plot the leading PN behaviour given in Eq.~\eqref{eq:spin_flux_PN}. 
    (Bottom) The relative difference between the $\delta\chi_1$ and the linearized-in-$\chi_1$ Teukolsky flux. For orbital radii in the range $ 6 < r_0/m^{(0)}_1 \le 30$, the relative error is always less than $1.5\times10^{-5}$.}\label{fig:deltachi1_flux}
\end{figure}

We compare the results of the linearized Teukolsky flux to $\mathcal{F}^{2(\delta\chi_1)}$ in Fig.~\ref{fig:deltachi1_flux}, where we find for the (2,2) mode the relative error is always less than $1.2\times 10^{-5}$.
As discussed above, the structure of the $\delta\chi_1$ source in Eq.~\eqref{eq:delta_chi1_source} is the same as for the full second-order metric perturbation but without the contributions from $G_{\mu\nu}^{(2,0)}[h^{\mathcal{P}1},h^{\mathcal{P}1}]$ and $G^{(1,1)}_{\mu\nu}[h^1]$.
As such, the excellent agreement we find between the flux computed by integrating the full $\delta\chi_1$ source and the linearized Teukolsky flux lends confidence to the computational infrastructure that we use to compute the point particle second-order source and to integrate it to find $h^{{\cal R} 2(pp)}_{\mu\nu}$.

\section{Waveform models}
\label{sec:waveform}
We now present our final waveform model(s). Reference~\cite{Mathews:2021rod} outlined how the secondary's spin may be modularly added to the quasi-circular non-spinning 1PA waveform models of Ref.~\cite{Wardell:2021fyy}, when the spin is (anti-)aligned with the orbital angular momentum. Since then, Ref.~\cite{Burke:2023lno} assessed the impact of the (anti-)aligned secondary's spin in a parameter estimation study using 1PA waveforms, and Ref.~\cite{Albertini:2024rrs} combined the results of Ref.~\cite{Mathews:2021rod} with Ref.~\cite{Wardell:2021fyy} to inform the modelling of the (anti-)aligned secondary's spin terms for asymmetric mass systems in the EOB model \texttt{TEOBResumS}.

In addition to comparing the spinning 1PA waveforms with NR, we present several key improvements on these existing waveform models:
\begin{enumerate}
    \item We allow for a slowly spinning primary, with a spin of order $\e$ and a small, ${\cal O}(\e)$ misalignment with the orbital angular momentum.
    \item We remove the requirement that the secondary's spin be (anti-)aligned with the orbital angular momentum, allowing for a generic precessing spin. See Sec.~\ref{sec:SummaryTT} above and Sec.~V~C of Paper I for a discussion of the (2PA) contribution of the secondary's precession to the waveform.
    \item We no longer neglect the evolution of the primary's mass and spin.
    \item We compare slight variations of the 1PA model that hold different quantities fixed while expanding in powers of the mass ratio.
    \item We define a `re-summed' model and demonstrate its increased accuracy.
\end{enumerate}

In Sec.~\ref{sec:SFwaveform}, we summarise the `native' self-force model  that follows immediately from our multiscale anaylsis. In Sec.~\ref{sec:Rexpand}, we outline straightforward re-expansions and re-summations of the initial model and define several model variations based upon these. In Sec.~\ref{sec:results}, we compare the different 1PA waveform models with NR waveforms.

\subsection{Native self-force model}
\label{sec:SFwaveform}

The waveform strain is extracted in multiscale form from Eq.~\eqref{eq:waveform}. We specifically require the two polarizations of the transverse-tracefree piece of that expression, which we decompose in the usual way in terms of spin-weighted spherical harmonics:
\begin{equation}
\label{eq:waveformstrain}
h_+ - i h_\times = \frac{m_1^{(0)}}{D_L}\sum_{\ell=2}^\infty \sum_{m=-\ell}^{\ell}h_{\ell m} \,_{-2}Y_{\ell m}(\theta,\phi).
\end{equation}
Here $D_L$ is the luminosity distance, which we have adimensionalised with the factor of the initial primary mass. Accounting for the multiscale form~\eqref{eq:waveform}, we may write each waveform mode in terms of \emph{complex} mode amplitudes and phase factors:
\begin{equation}
\label{eq:waveformmodes}
 h_{\ell m}(t)=\sum_{k=-1}^1 R_{\ell m k}(\varpi_I(t))e^{-i \left[m \phi_p(t)+k\tilde\psi_s(t)\right]},
\end{equation}
suppressing dependence on $\ee$ and other constant parameters.

Given the expansions~\eqref{eq:h1tt} and \eqref{eq:h2tt} of the metric perturbation, the waveform amplitudes take the form
\begin{align}
R_{\ell m k}&= \ee R_{\ell m k}^{1(pp)}(y) + \ee^2 R_{\ell m k}^{2(pp)}(y)+ \ee^2\delta m_1 R_{\ell m k}^{2(\delta m_1)}(y)\nonumber \\
&\quad +\ee^2\delta \chi_1 R_{\ell m k}^{2(\delta \chi_1)}(y)+\ee^2\chi_\parallel R_{\ell m k}^{2(\chi_\parallel)}(y) \nonumber\\
&\quad+\ee^2\chi_\perp R_{\ell m k}^{2(\chi_\perp)}(y),\label{eq:ampexpand}
\end{align}
where $y$ is defined in Eq.~\eqref{eq:ydef}.

\begin{figure*}[t]
 \center
 \includegraphics[width=0.48\textwidth]{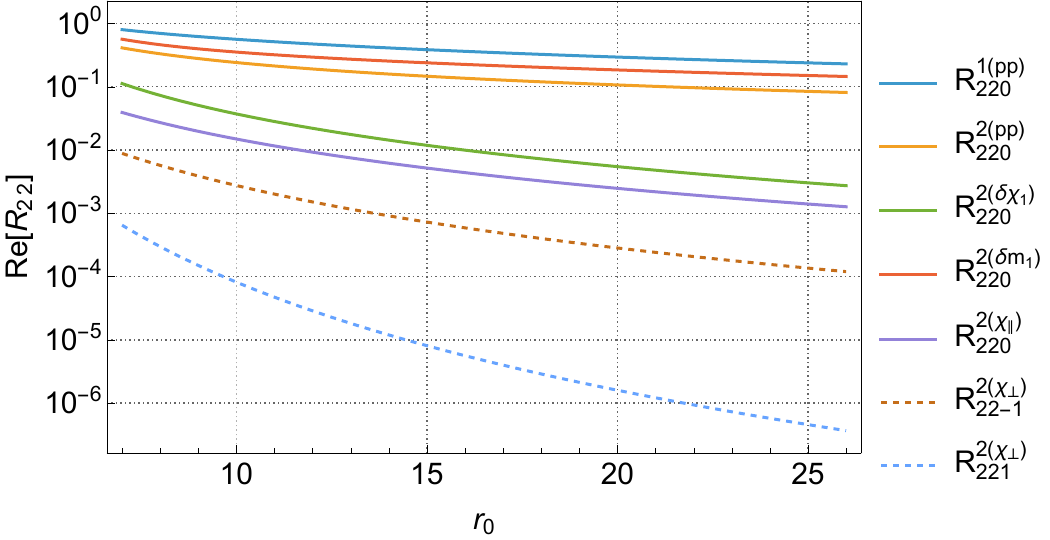}
 \hspace{5 mm}
 \includegraphics[width=0.48\textwidth]{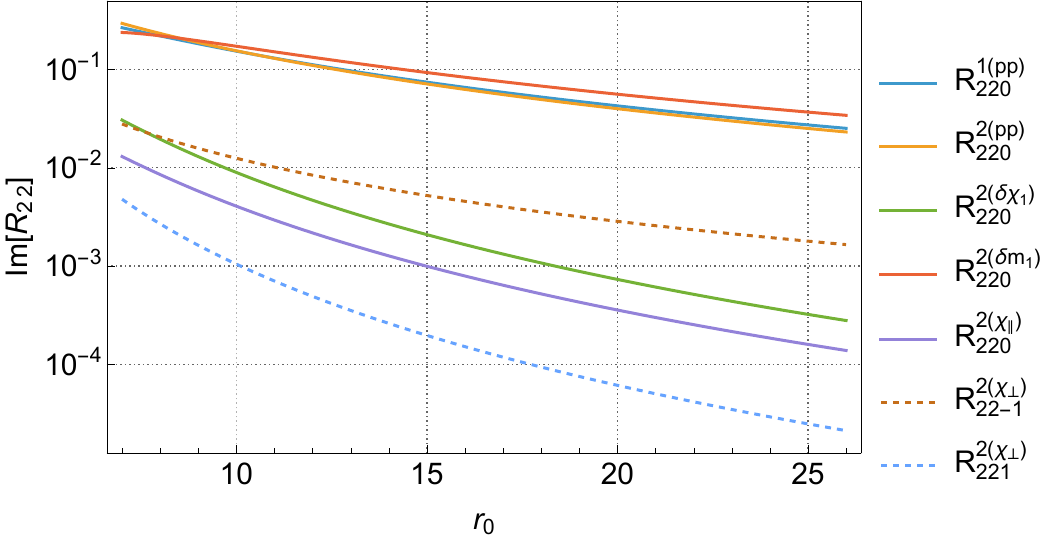}
 \hspace{5 mm}
 \includegraphics[width=0.48\textwidth]{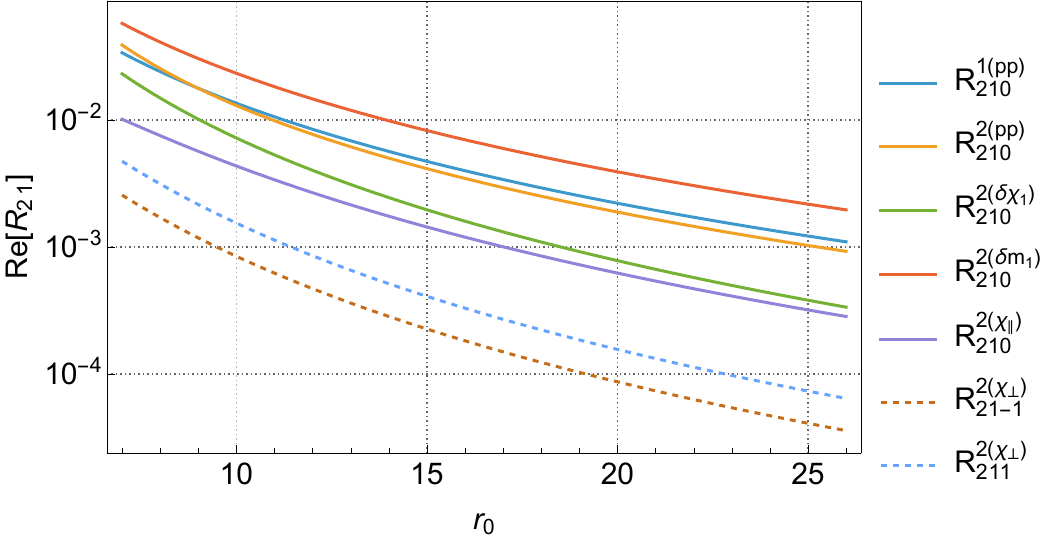} 
 \hspace{5 mm}
 \includegraphics[width=0.48\textwidth]{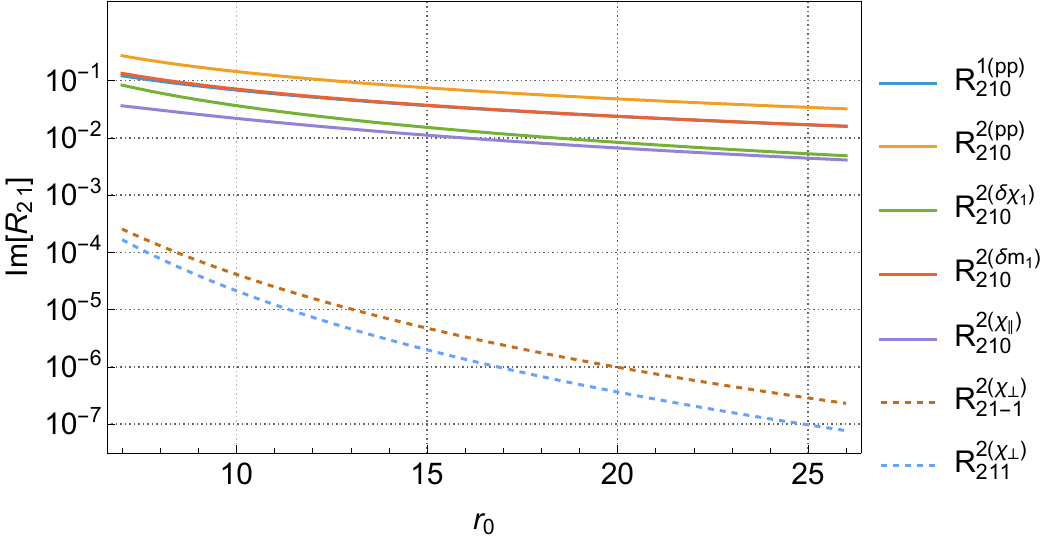} 
 \caption{\label{fig:RelAmps} The amplitude terms in Eq.~\eqref{eq:ampexpand} as functions of the separation, $r_0$. {\bf Top:} the various contributions to the $(\ell, m)=(2,2)$ mode (real and imaginary part respectively). {\bf Bottom:} the same but with $(\ell, m)=(2,1)$. Notice that the precessing modes ($k\neq0$) are represented by the dashed lines and are generally at least an order of magnitude smaller than the leading amplitude even before being suppressed by the additional factor of $\ee$.} 
 \end{figure*}

Importantly, $R_{\ell m 0}^{2(\chi_\perp)}=0$ and in all the other terms there are only modes with $k=0$; the phase factor $e^{\pm i\tilde\psi_s}$ \emph{only multiplies the $\chi_\perp$ term}. While the secondary's precession did not couple to the 1PA evolution of $\phi_p, \tilde\psi_s$, it does contribute to the sub-leading complex amplitude via the linear-in-$\chi_\perp$ term. We find that term is numerically very small (see Fig.~\ref{fig:RelAmps}). 

\subsection{Re-expanded and re-summed models}
\label{sec:Rexpand}
In Ref.~\cite{Warburton:2021kwk}, the second-order fluxes for a quasi-circular binary with aligned-spins and a slowly spinning primary were compared with NR results and found to agree well even at moderate mass ratios. Reference~\cite{Wardell:2021fyy} found similar agreement with NR at  the level of the waveform for non-spinning binaries. 

A critical step in achieving good accuracy away from the extreme-mass-ratio regime was the re-expansion in terms of the symmetric mass ratio, $\nu=m_1 m_2/M^2$, at fixed values of the total mass, $M=m_1+m_2$. When we add the spins into the picture, we also have freedom to chose which spin parameters we hold fixed in the $\nu$~expansion. As in Ref.~\cite{Albertini:2024rrs}, we consider two different choices: the dimensionless spins $\chi_i$ which appear naturally from our MPD-Harte framework and the reduced (also dimensionless) spins 
\begin{equation}
\tilde a_i\equiv S_i/(M m_i).    
\end{equation}
Defining the mass ratios 
\begin{equation}
\mu_i\equiv m_i/M,    
\end{equation}
we have $\mu_i=\frac{1}{2}(1\pm\Delta)$, taking the positive (negative) sign for $i=1$ ($i=2$) given the convention $m_1\geq m_2$, with $\Delta\equiv \sqrt{1-4\nu}$ and $\tilde a_i=\mu_i \chi_i$. The reason that the variables $\tilde a_i$ are also `obvious' choices of parameters to fix in the expansion is that in weak-field formalisms which do not rely upon a small-mass-ratio expansion (such as PN theory~\cite{Messina:2018ghh}), the spin dependence of the expressions describing the inspiral and waveform are more compactly expressed in terms of $\tilde a_i$ and it restores some of the symmetry under the interchange of each body. For example, in expansions in $\nu$ at fixed $\tilde a_i$, both spins contribute at each order in the series while re-expanding at fixed $\chi_i$,  the secondary's spin first enters at subleading order. This suggests we may more accurately capture the spin effects for generic mass ratios by holding these parameters fixed. In the flux comparisons in Ref.~\cite{Warburton:2021kwk}, the authors also opted to hold $\tilde a_i$ fixed for this reason.

In total, we define five models: 1PAT1-$a$, 1PAT1-$\chi$, 1PAT1e-a, 1PAT1e-$\chi$ and 1PAT1R. The models whose label include `e' and the 1PAT1R model account for the evolution of the primary's mass and spin, while the other models neglect these effects. The models labelled with `$\chi$' (`$a$') hold the spin variables $\chi_i$ ($\tilde a_i$) fixed in the symmetric mass-ratio expansion. The `1PAT1' prefix refers to the terminology of Ref.~\cite{Wardell:2021fyy}, where the authors defined several variants of 1PA waveforms: 1PAT1, 1PAT2 and 1PAF1. Here, we focus only on models of the 1PAT1 variety. All of the 1PAT1 model varieties rely on the following approximations:
\begin{enumerate}
    \item In computing the second order energy flux we make the approximation that $\mathcal{F}_2\simeq \mathcal{F}_2^{\infty}$ and neglect the second-order energy flux through the horizon. This is justified since the horizon flux is known to be numerically subdominant (see Ref.~\cite{Albertini:2022rfe}). However, as mentioned in Sec.~\ref{sec:evolution with FL}, work is ongoing to compute $\mathcal{F}_2^{\mathcal{H}}$ so that it can be included in future models.
    \item As we have  highlighted in detail in Sec.~\ref{sec:fluxbal}, we approximate the leading self-force correction to the binding energy with the binding energy predicted by the first law. We additionally neglect ``memory distortion'' terms in the waveform amplitudes and flux~\cite{Cunningham:2024dog}.
    \item The models are limited to the inspiral and do not yet include the transition to plunge and merger-ringdown parts of the waveform.
\end{enumerate}

Next, we summarise each model and the re-expansion in $\nu$. In the 1PAT1e models we expand in powers of the \emph{initial} symmetric mass-ratio at fixed powers of the \emph{initial} total mass (while in the other two models this distinction is redundant). To reduce notation-related jargon, we abuse the notation from our previous paragraph and recycle the labels `$\nu$' and `$M$' as the initial symmetric mass ratio and initial total mass respectively; 
\begin{equation}
\nu \equiv m_1^{(0)} m_2/M \qquad \text{and}\qquad M \equiv m_1^{(0)} + m_2.    
\end{equation}
Since we re-expand at fixed values of $M$, we also define the dimensionless variable $x\equiv (M \Omega)^{2/3}$ to replace our use of the variable $y$.

In all models we first re-write the waveform strain as
\begin{equation}
\label{eq:waveformstrainM}
h_+ - i h_\times = \frac{M}{D_L}\sum_\ell \sum_{m=-\ell}^{\ell}h_{\ell m} \,_{-2}Y_{\ell m}(\theta,\phi),
\end{equation}
so that the luminosity distance may be cleanly given units of $M$; this will be essential for comparison with NR, where units with $M=1$ are used. Note that this step is a \emph{redefinition} of $h_{\ell m}$ rather than a re-expansion:
\begin{equation}\label{eq:hlm redef}
    h^{\rm new}_{\ell m}= \frac{m_1^{(0)}}{M}h^{\rm old}_{\ell m}.
\end{equation}

\subsubsection{1PAT1e-$\chi$}

In 1PAT1e-$\chi$, we re-expand our baseline model in powers of $\nu$ at fixed $M$.

The expression for the individual waveform modes still takes the form in Eq.~\eqref{eq:waveformmodes}, but the re-expansion of Eq.~\eqref{eq:ampexpand}  yields
\begin{align}
R_{\ell m k}&= \nu R_{\ell m k}^{1(pp)}(x) + \nu^2 \biggl[R_{\ell m k}^{1(pp)}(x) -\frac{2}{3}x \partial_x R_{\ell m k}^{1(pp)}(x)\biggr] \nonumber\\
&\quad +\nu^2 R_{\ell m k}^{2(pp)}(x)+\nu^2\delta \chi_1 R_{\ell m k}^{2(\delta \chi_1)}(x)+\nu^2\delta m_1 R_{\ell m k}^{2(\delta m_1)}(x) \nonumber \\
&\quad + \nu^2\chi_\parallel R_{\ell m k}^{2(\chi_\parallel)}(x) +\nu^2\chi_\perp R_{\ell m k}^{2(\chi_\perp)}(x),\label{eq:ampexpandechi}
\end{align}
where here $R_{\ell m k}=\frac{m_1^{(0)}}{M}R^{\rm old}_{\ell mk}$. The terms in square brackets come from the redefinition and re-expansion. The $\nu^2 R_{\ell m k}^{1(pp)}(x)$ comes from the redefinition~\eqref{eq:hlm redef} and from re-writing $\ee$ in terms of $\nu$. The $\nu^2\partial_x R_{\ell m k}^{1(pp)}(x)$ term comes from rewriting the leading amplitude in terms of $x$ as opposed to $y$. The rest of the re-expansion in the subleading term is trivial since $\ee=\nu +\mathcal{O}(\nu^2).$

We also re-expand the inspiral evolution equations from Eq.~\eqref{eq:ttinspiralFlux}:
\begin{subequations}
\label{eq:ttinspiralFluxechi}
\begin{align}
    M^2\frac{d\Omega}{dt} &= \nu \bar G_{\Omega}^{(0)}(x) + \nu^2 \left[\bar G_{\Omega}^{(1,0)}(x)+\delta m_1 \bar G_{\Omega}^{(1,\delta m_1)}(x)\right]\nonumber \\
    &\quad +\nu^2\left[ \delta\chi_1 \bar G_{\Omega}^{(1, \chi_1)}(x)+ \chi_\parallel \bar G_{\Omega}^{(1, \chi_\parallel)}(x)\right]\nonumber\\ &\quad+\mathcal{O}(\nu^3),\\      
    \frac{d\delta m_1}{dt} &= \nu \mathcal{F}_1^{\mathcal{H}}(x)+\mathcal{O}(\nu^2),\\
    M^2\frac{d\delta \chi_1}{dt} &= \nu \mathcal{L}_{1}^{\mathcal{H}}(x) + {\cal O}(\nu^2),
\end{align}
\end{subequations}
with
\begin{subequations}
\label{eq:omegadotechi}
\begin{align}
\bar G_{\Omega}^{(0)}(x)&=\bar F_{\Omega}^{(0)}(x),\\
\bar G_{\Omega}^{(1,0)}(x)&=\bar F_{\Omega}^{(1,0)}(x)+4 \bar F_{\Omega}^{(0)}(x)-\frac{2}{3}x \partial_x\bar F_{\Omega}^{(0)}(x),\\
\bar G_{\Omega}^{(1,\chi_\parallel)}(x)&=\bar F_{\Omega}^{(1,\chi_\parallel)}(x),\\
\bar G_{\Omega}^{(1,\delta m_1)}(x)&= \frac{m_1^{(0)}}{M}\bar F_{\Omega}^{(1,\delta m_1)}(x),\\
\bar G_{\Omega}^{(1,\delta \chi_1)}(x)&=\bar F_{\Omega}^{(1,\delta \chi_1)}(x),
\end{align}
\end{subequations}
noting $\bar F_{\Omega}^{(1,\delta m_1)}\propto \left(m_1^{(0)}\right)^{-1}$ such that $\bar G_{\Omega}^{(1,\delta m_1)}$ is independent of $m_1^{(0)}$. 

The phases are then computed via  numerical integration of Eq.~\eqref{Omegadef}, having re-expanded the precession frequency in Eq.~\eqref{eq:tt prec freq} to linear order in $\nu$ at fixed values of $M$, $\chi_i$ and $x$. We do not expand the phases in their asymptotic forms implied by Eq.~\eqref{phip - slow t} and Eq.~\eqref{psis - slow t} due to the large loss of waveform accuracy this incurs~\cite{Wardell:2021fyy}.

\subsubsection{1PAT1e-$a$}

In 1PAT1e-$a$, we re-expand 1PAT1e-$\chi$ in powers of $\nu$ at fixed $\tilde a_1$. The re-expansion of the amplitudes is a trivial modification to Eq.~\eqref{eq:ampexpandechi},
\begin{align}
R_{\ell m k}&= \nu R_{\ell m k}^{1(pp)}(x) + \nu^2 \left[R_{\ell m k}^{1(pp)}(x)-\frac{2}{3}x\partial_x R_{\ell m k}^{1(pp)}(x)\right]\nonumber \\
&\quad +\nu^2R_{\ell m k}^{2(pp)}(x)+\nu \tilde a_1 R_{\ell m k}^{2(\delta \chi_1)}(x)+\nu^2\delta m_1 R_{\ell m k}^{2(\delta m_1)}(x) \nonumber\\
&\quad +\nu \tilde a_\parallel R_{\ell m k}^{2(\chi_\parallel)}(x) +\nu \tilde a_\perp R_{\ell m k}^{2(\chi_\perp)}(x),\label{eq:ampexpandea}
  \end{align}
where $\tilde a_{\parallel/\perp}\equiv \mu_2 \chi_{\parallel/\perp}$ and the slow primary spin condition is $\tilde a_1(t=0)\sim \nu$. The inspiral evolution equations are also a trivial modification of Eq.~\eqref{eq:ttinspiralFluxechi},
\begin{subequations}
\label{eq:ttinspiralFluxea}
\begin{align}
    \label{eq:1PATeafreqdot}
    M^2\frac{d\Omega}{dt} &= \nu \bar G_{\Omega}^{(0)}(x) + \nu^2 \left[\bar G_{\Omega}^{(1,0)}(x)+\delta m_1 \bar G_{\Omega}^{(1,\delta m_1)}(x)\right]\nonumber \\
    &\quad +\nu\left[ \tilde a_1 \bar G_{\Omega}^{(1, \chi_1)}(x)+ \tilde a_\parallel \bar G_{\Omega}^{(1, \chi_\parallel)}(x)\right] +\mathcal{O}(\nu^3),\\      
    \frac{d\delta m_1}{dt} &= \nu \mathcal{F}_1^{\mathcal{H}}(x)+\mathcal{O}(\nu^2),\\
    M^2\frac{d\tilde a_1}{dt} &= \nu^2 \mathcal{L}_{1}^{\mathcal{H}}(x) + {\cal O}(\nu^3),
\end{align}
\end{subequations}
with the same definitions for $\bar G_\Omega^{(...)}$ as in Eq.~\eqref{eq:omegadotechi}.

\subsubsection{1PAT1-$\chi$ (1PAT1-$a$)}

The 1PAT1-$\chi$ (1PAT1-$a$) model neglecting the evolution of the primary's mass and spin is easiest to describe as two simple changes to the 1PAT1e-$\chi$ (1PAT1e-$a$) model:
\begin{enumerate}
    \item Fix the primary's spin to its initial value; $\delta \chi_1=\delta \chi_1(t=0)$ [$\tilde a_1=\tilde a_1(t=0)$]. Fix $\delta m_1=0$. Discard their evolution equations.
    \item In $\bar G_{\Omega}^{(1,0)}(x)$, neglect the term $\propto \mathcal{F}_1 ^{\mathcal{H}}$ that enters via the last term of $\bar F_{\Omega}^{(1,0)}(x)$ in Eq.~\eqref{eq:FOmega1PA}.
\end{enumerate}

\subsubsection{1PAT1R}
\label{sec:1PAT1R}
The 1PAT1R model (`R' for `re-summed') is an effective re-summation of the 1PATe-$a$ model. Rather than fully expanding the rearranged balance law~\eqref{eq:dOmegads}, we expand the numerator and denominator in powers of $\nu$ at fixed values of $M,x,\tilde a_1$, but we leave the fraction unexpanded. We call this an \emph{effective} re-summation because Eq.~\eqref{eq:dOmegads} is an exact implication of the fully nonlinear Einstein equations. The only assumption is the one underlying our multiscale expansion: that the metric's time dependence can be reduced to a dependence on mechanical phase-space variables. That ``assumption'' can itself be derived from the more generic self-consistent formulation of self-force theory~\cite{Lewis:2025ydo}.

To glean why this evolution model might be more accurate than a fully expanded one, note that all of our models break down at the innermost stable circular orbit (ISCO), where $\partial E/\partial\Omega$ vanishes. As the secondary approaches the ISCO, $\partial E/\partial\Omega$ becomes small and $d\Omega/ds$ grows large. Our assumption that the frequency is slowly evolving then breaks down, signaling the failure of our multiscale expansion.  In the fully expanded evolution equation, the singularity occurs at the test-mass ISCO, $y=1/6$ (or $x=1/6$ following our re-expansions), where $\partial E_1/\partial\Omega=0$. As we can see from Eq.~\eqref{eq:OmegaDotFL}, this introduces the pole structure $1/(1-6y)^n$ in the $n$PA forcing function (for $n=0,1$; see Table~I of Ref.~\cite{Albertini:2022rfe}for the structure at higher orders). If we avoid expanding the fraction, then the singularity is partially displaced to the physical ISCO, where $\frac{d E}{d\Omega}=0$ in Eq.~\eqref{eq:dOmegads}. A singularity remains at $x=1/6$ because the second-order field $h^{2(pp)}_{\mu\nu}$, and therefore the flux ${\cal F}_2$, diverges there; this divergence is caused by the $G^{(1,1)}_{\mu\nu}$ source term in Eq.~\eqref{tt EFE2}, which is proportional to the 0PA approximation to $\dot\Omega$, which diverges at the zeroth-order ISCO. However, the pole structure is more mild at both singular points. The $\dot\Omega$ appearing in $h^{2(pp)}_{\mu\nu}$ and ${\cal F}_2$ can also be re-expanded to move its divergence to the physical ISCO.

The location of the physical ISCO is
\begin{equation}
M\Omega_{\rm ISCO}=\frac{1}{6 \sqrt6}-\frac{\tilde{a}_1}{216}+\frac{\tilde{a}_2}{48}+\nu\frac{(C_\Omega(0)-1)}{6 \sqrt6}+\mathcal{O}(\e^2),
\end{equation}
which we have checked term-by-term against the existing literature~\cite{Jefremov:2015gza,Barack:2009ey, Akcay:2012ea, Isoyama:2014mja}. Note $C_\Omega(0)$ is a numerical coefficient defined in Ref.~\cite{Isoyama:2014mja} in terms of the redshift as a function of $\chi_1$ with $C_\Omega(0)=1.251 015 39\pm4\times10^{-8}$.
In terms of $x$, the ISCO lies at
\begin{equation}\label{eq:x ISCO}
    x_{\rm ISCO}=\frac{1}{6}-\frac{\tilde{a}_1}{54\sqrt6}+\frac{\tilde{a}_\parallel}{12\sqrt6}+\frac{\nu}{9}(C_\Omega(0)-1)+\mathcal{O}(\e^2).
\end{equation}
When we expand the fraction in Eq.~\eqref{eq:dOmegads}, the corrections to the ISCO location in Eq.~\eqref{eq:x ISCO} manifest themselves as the higher-order poles at $x=1/6$ in the fully expanded models.

\subsection{Results}
\label{sec:results}

\begin{figure*}[htb!]
\center
\includegraphics[width=.95\textwidth]{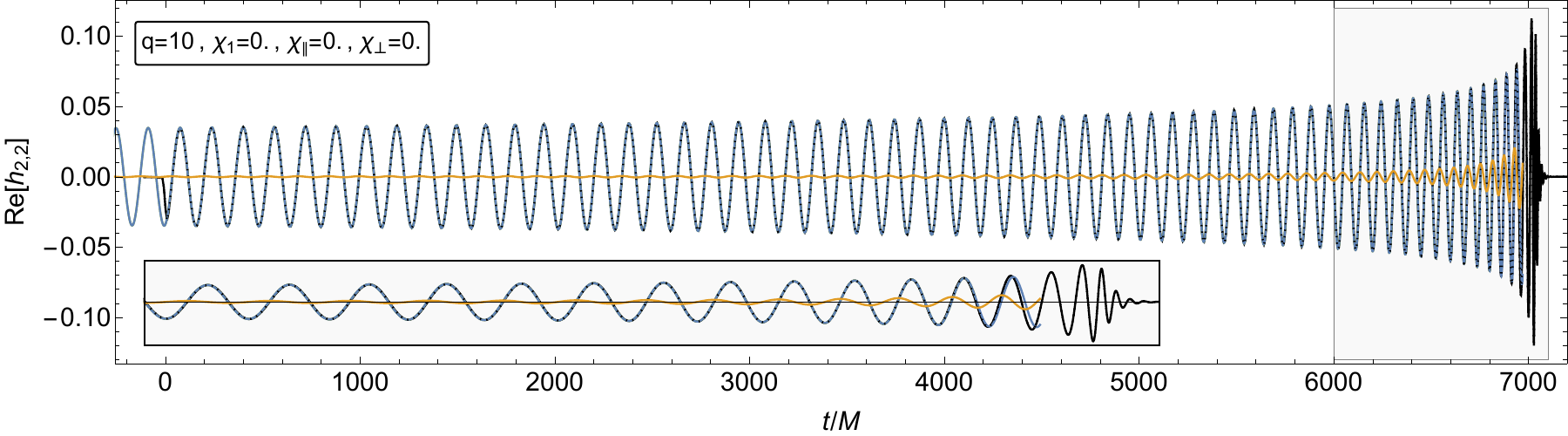}

\vspace{10pt}

\includegraphics[width=0.475\textwidth]{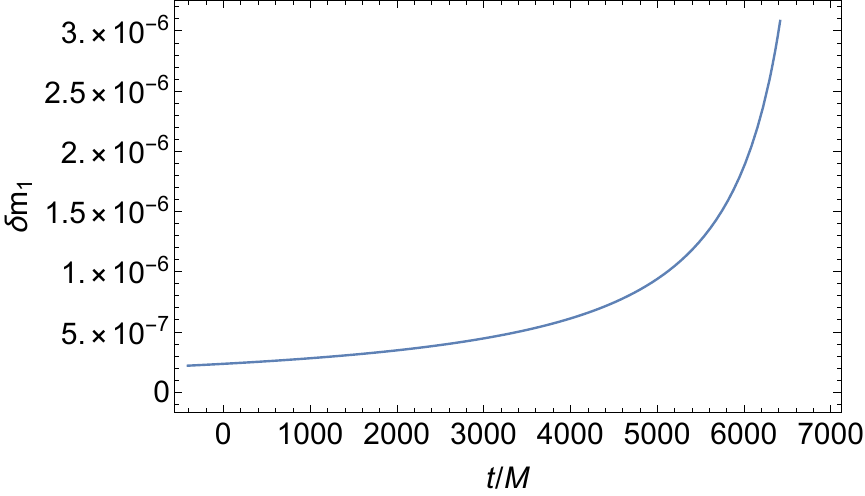}
\hspace{5 mm}
\includegraphics[width=0.475\textwidth]{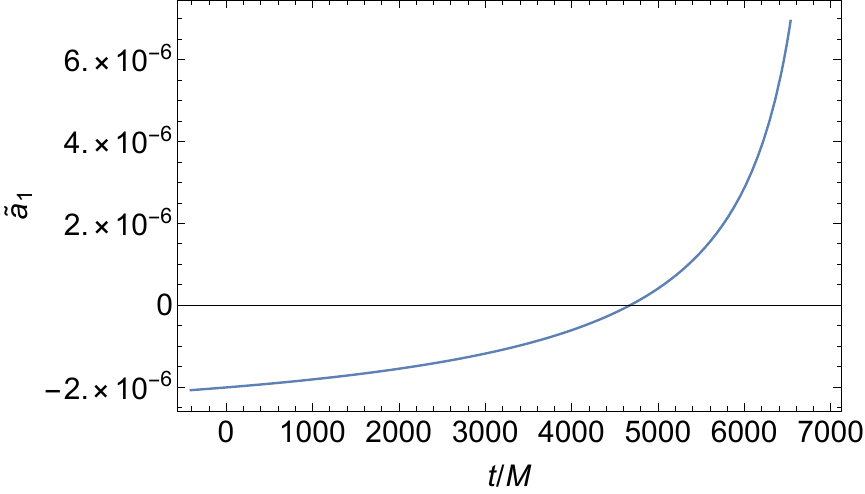}
\caption{\label{fig:1PAT1vs1PAT1e} \textbf{Top}: Waveform comparison for the same nonspinning binary configuration considered in Figure 1 of Ref.~\cite{Wardell:2021fyy}. The NR simulation~\cite{sxs_collaboration_2019_3302023} is in black. In blue is the 1PAT1e-$a$ waveform. In green is the original 1PAT1 waveform, which is not visible in the plot due to excellent overlap with the 1PAT1e-$a$ waveform. In orange is the difference between the 1PAT1e-$a$ waveform and the 1PAT1 waveform \emph{scaled by a factor of 100}. \textbf{Bottom left (right)}: The evolution of the primary's mass (spin) correction. Their values remain small over the entire inspiral, hence the very small difference between the 1PAT1 and 1PATe-$a$ waveforms.} 

\vspace{10pt}

\includegraphics[width=.95\textwidth]{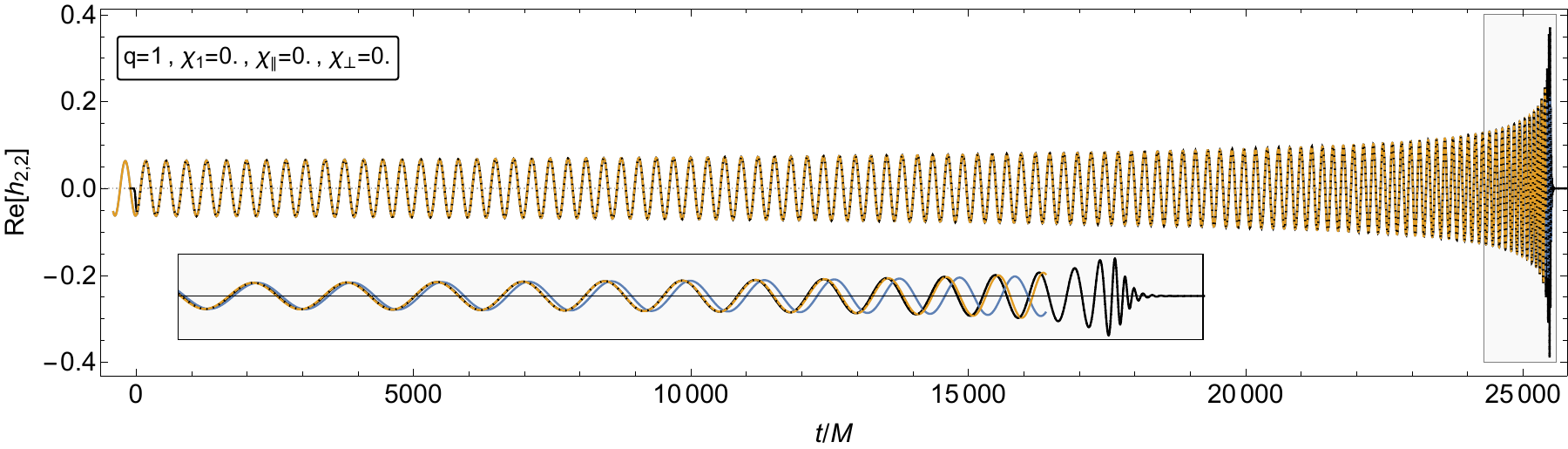}
\caption{\label{fig:1PAT1vs1PAT1R} Waveform comparison for the same equal-mass, nonspinning binary configuration shown in the top panel of Figure 1 in the supplemental material of Ref.~\cite{Wardell:2021fyy}. The NR simulation \cite{sxs_collaboration_2019_2649511} is in black. The original 1PAT1 model is in blue. The 1PAT1R model is in orange. We have slightly reduced the dephasing of the non-spinning 1PAT1 model against the NR simulation compared with the original plot in Ref.~\cite{Wardell:2021fyy} by more carefully matching the frequencies of the two waveforms at the reference time. The 1PAT1R waveform exhibits less dephasing against the NR waveform towards the transition-to-plunge region than the 1PAT1 waveform.} 
\end{figure*}

\begin{figure*}[htb!]
\center
\includegraphics[width=.8625\textwidth]{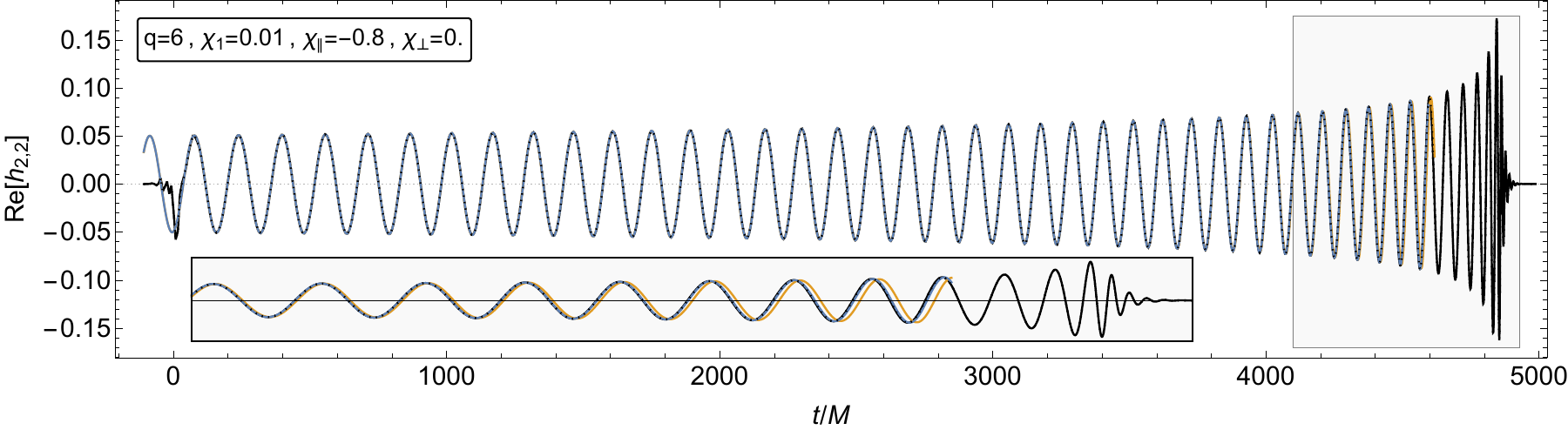}

\vspace{10pt}

\includegraphics[width=.8625\textwidth]{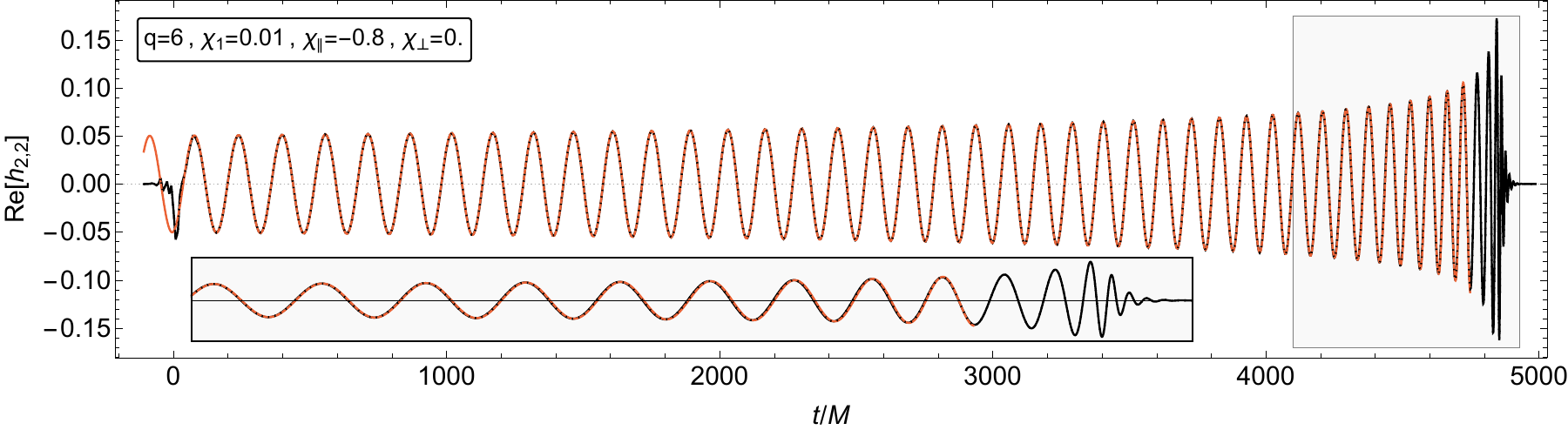}
\caption{\label{fig:spinaligned1} Waveform comparisons for a binary configuration with a rapid anti-aligned spin on the secondary. \textbf{Top:} Comparison of the NR simulation \cite{sxs_collaboration_2019_3272906} (in black) with the 1PAT1e-$a$ model (in blue) and the 1PAT1e-$\chi$ model (in orange). The binary has anti-aligned spins with a very small spin on the primary. \textbf{Bottom:} The same comparison but against the 1PAT1R model (in red).} 

\vspace{10pt}
\includegraphics[width=.8625\textwidth]{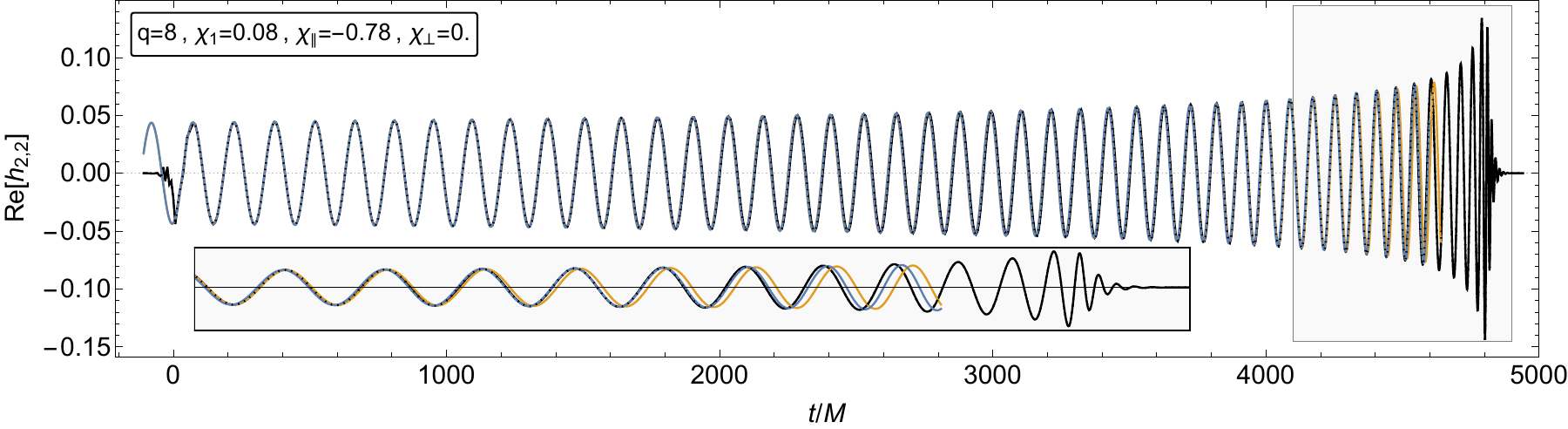}

\vspace{10pt}

\includegraphics[width=.8625\textwidth]{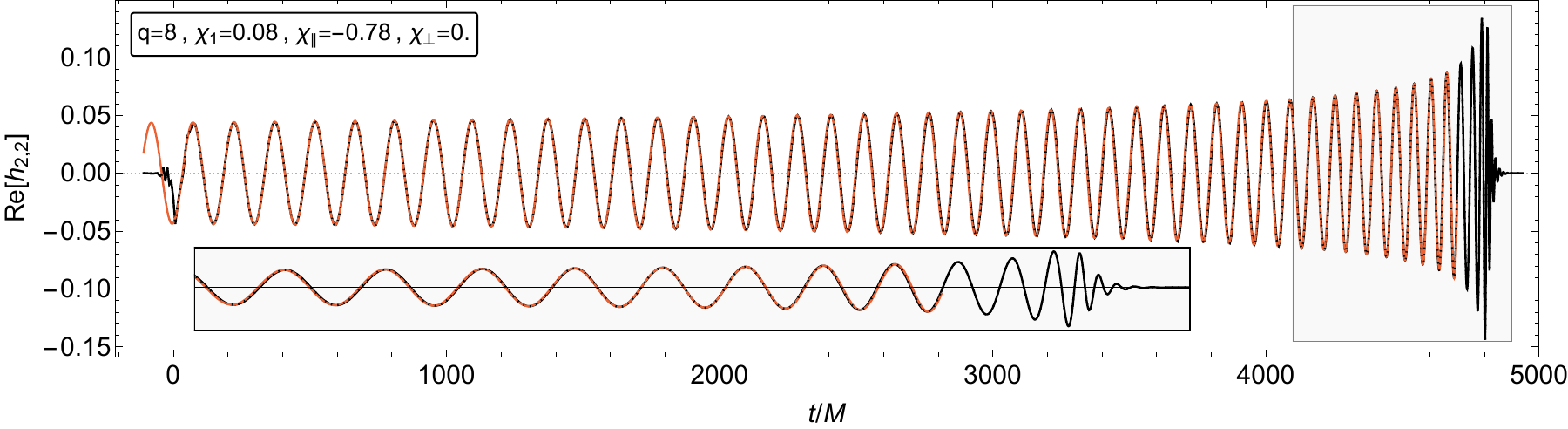}
\caption{\label{fig:spinaligned2}Waveform comparisons for a binary with anti-aligned spins and a small spin on the primary. \textbf{Top:} Comparison of an NR simulation \cite{sxs_collaboration_2019_3302671} (in black) with the 1PAT1e-$a$ model (in blue) and the 1PAT1e-$\chi$ model (in orange). \textbf{Bottom:} The same comparison but against the 1PAT1R model (in red). We note that the 1PAT1R's accuracy is significantly better than the 1PAT1e models for higher values of primary spin.} 
\end{figure*}

\begin{figure*}[htb!]
\center
\includegraphics[width=.8625\textwidth]{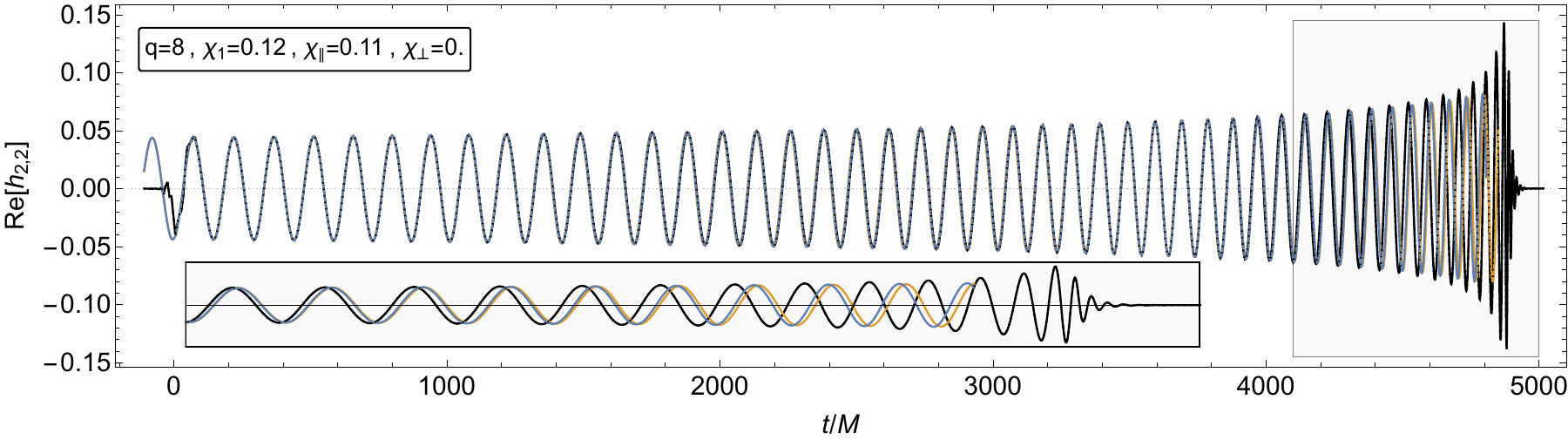}

\includegraphics[width=.8625\textwidth]{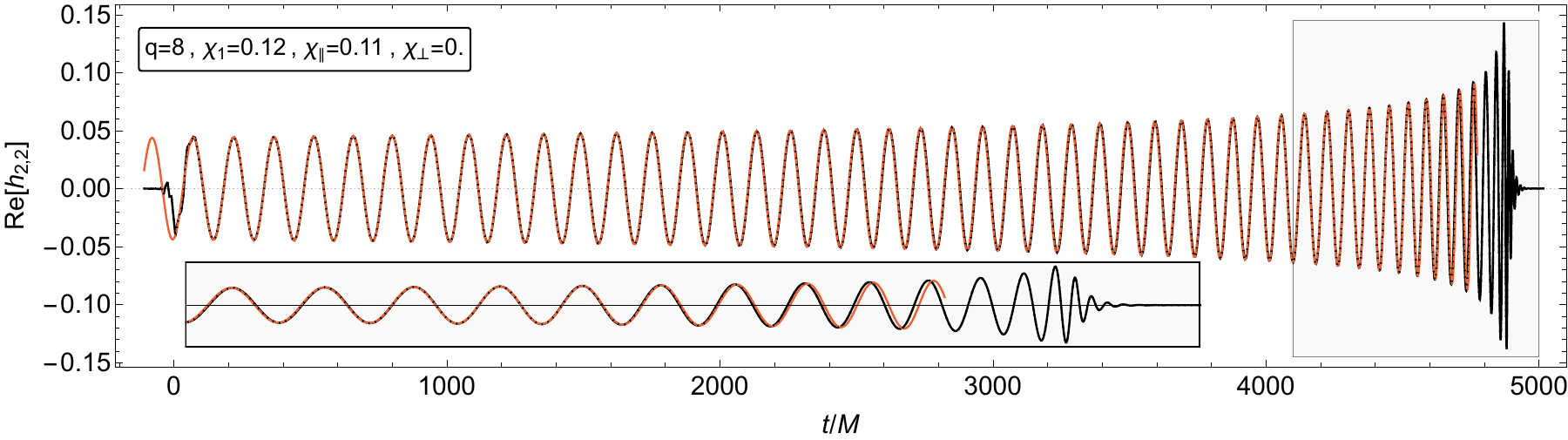}
\caption{\label{fig:spinaligned3}Waveform comparisons for a binary with aligned spins and a  small spin on the primary.
\textbf{Top:} Comparison of an NR simulation \cite{jonathan_blackman_2019_2644087} (in black) with the 1PAT1e-$a$ model (in blue) and the 1PAT1e-$\chi$ model (in orange).  \textbf{Bottom:} The same comparison but against the 1PAT1R model (in red).} 
\end{figure*}

\begin{figure*}[htb!]
\center
\includegraphics[width=.8625\textwidth]{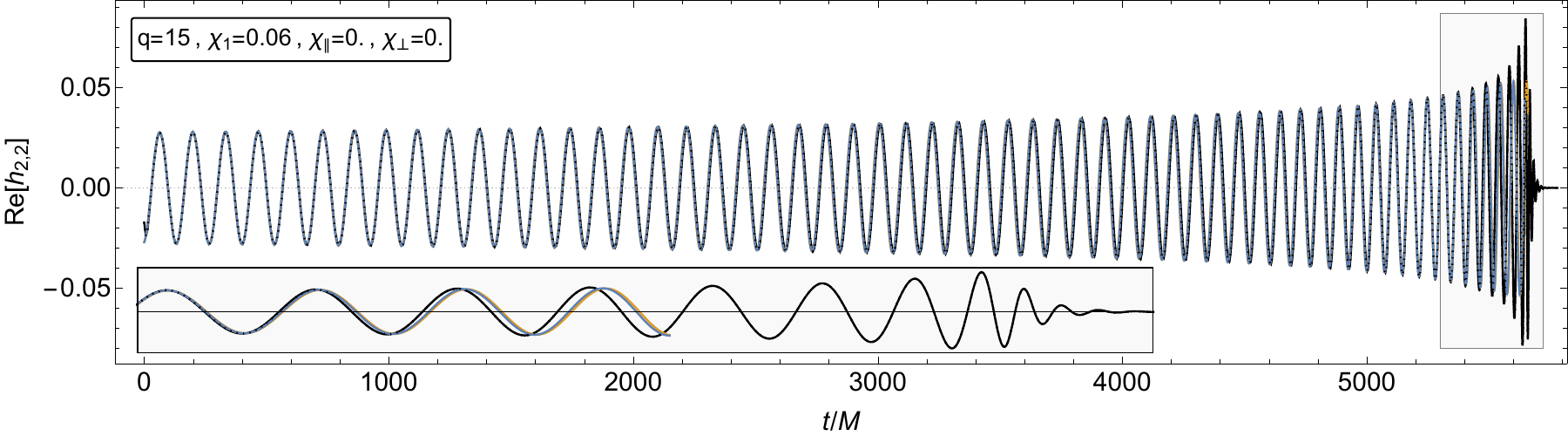}

\includegraphics[width=.8625\textwidth]{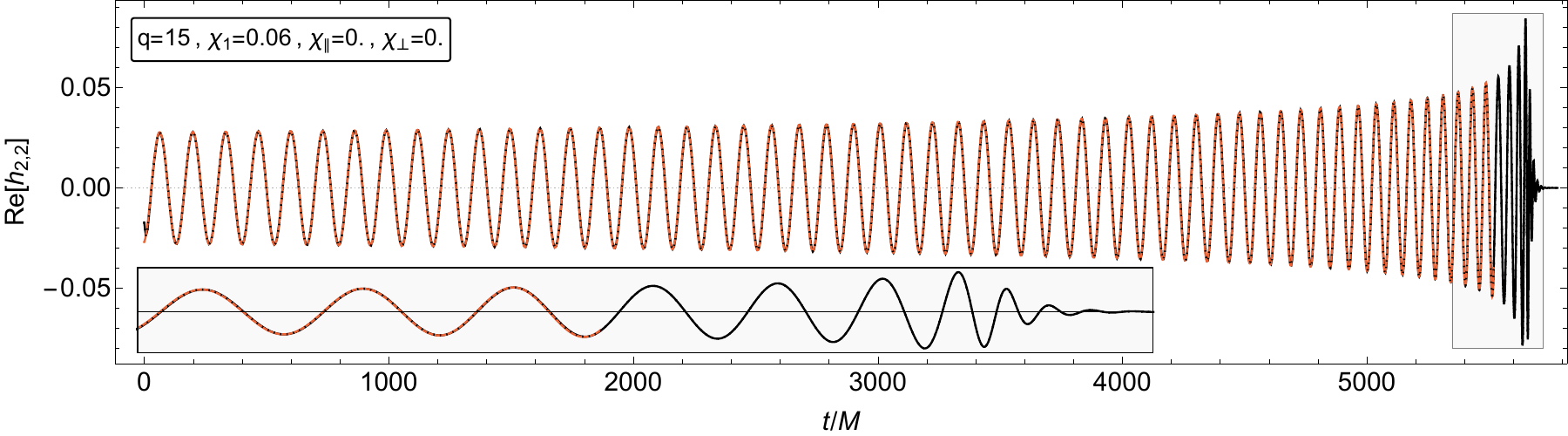}
\caption{\label{fig:spinaligned4}Higher $q$ waveform comparisons for a binary with a small aligned spin on the primary.
\textbf{Top:} Comparison of an NR simulation \cite{sxs_collaboration_2024_13147624} (in black) with the 1PAT1e-$a$ model (in blue) and the 1PAT1e-$\chi$ model (in orange).  \textbf{Bottom:} The same comparison but against the 1PAT1R model (in red).} 
\end{figure*}

\begin{figure*}[htb!]
\center
\includegraphics[width=.8625\textwidth]{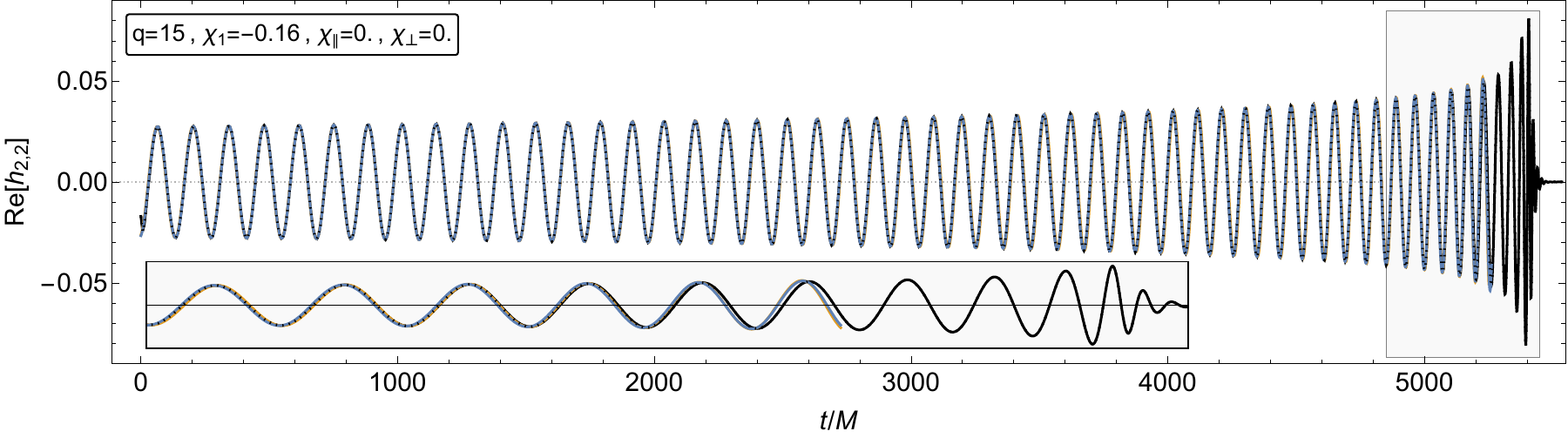}

\includegraphics[width=.8625\textwidth]{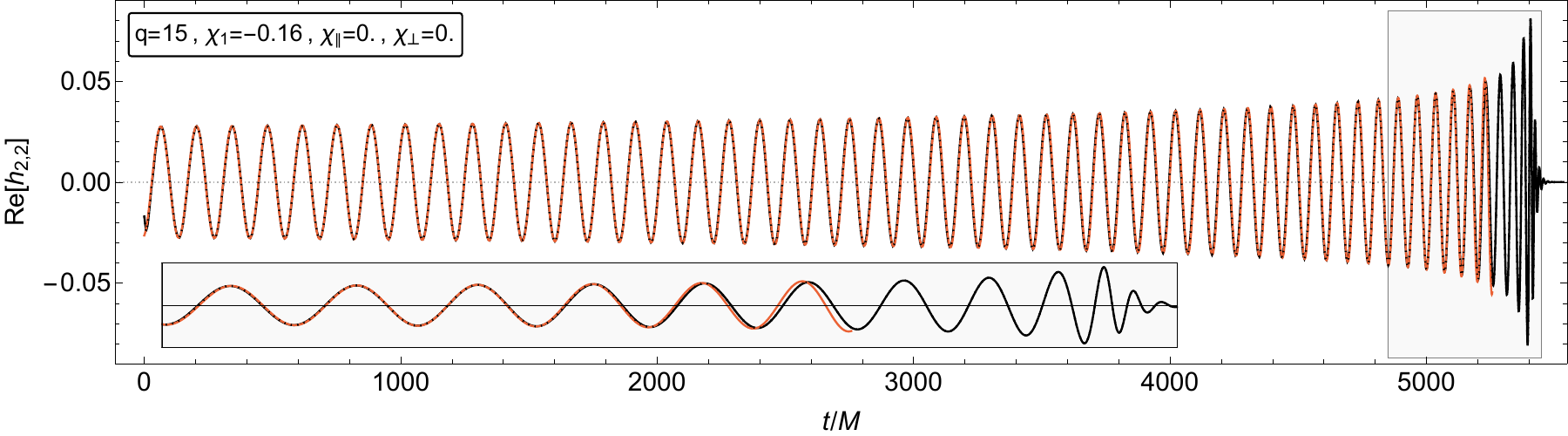}
\caption{\label{fig:spinaligned5}Similar to Fig.~\ref{fig:spinaligned4} but with a slightly more rapid, retrograde spin.
\textbf{Top:} Comparison of an NR simulation \cite{sxs_collaboration_2024_13147573} (in black) with the 1PAT1e-$a$ model (in blue) and the 1PAT1e-$\chi$ model (in orange but barely visible beneath blue).  \textbf{Bottom:} The same comparison but against the 1PAT1R model (in red).} 
\end{figure*}

\begin{figure*}[htb!]
 \center
 \includegraphics[width=.8625\textwidth]{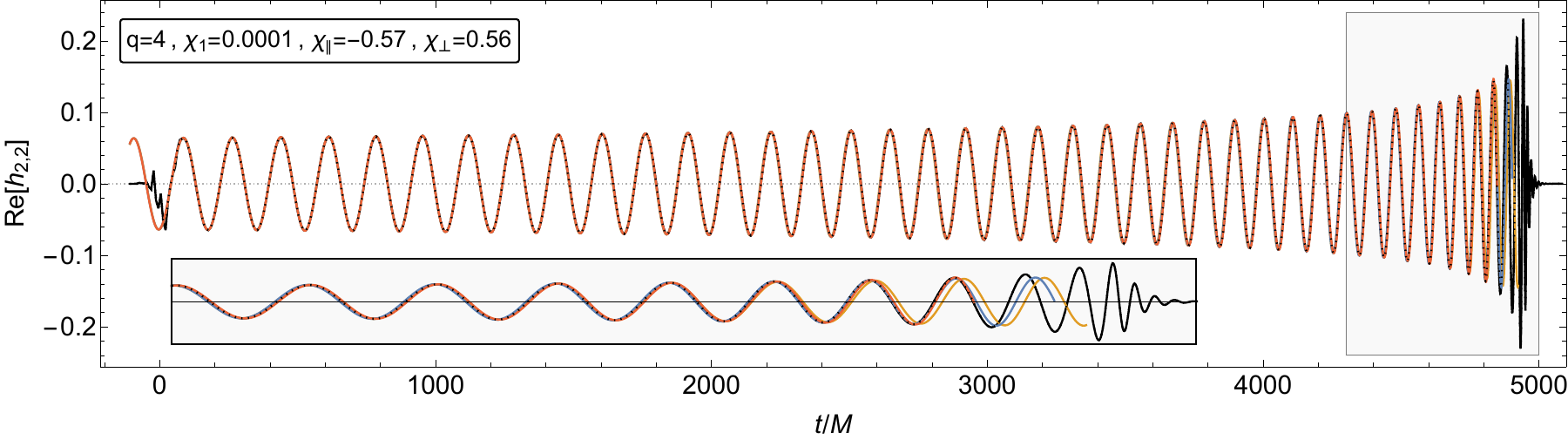}
 \caption{\label{fig:preccomparison1} Waveform comparison for a binary with large, precessing rapid spin on the secondary. In black is an NR simulation~\cite{sxs_collaboration_2019_3273407}. In blue is the corresponding 1PAT1e-$a$ waveform. In orange is the 1PAT1e-$\chi$ waveform. In dashed, red is the 1PAT1R waveform. Both 1PAT1e varieties show slightly more dephasing against the NR waveform earlier in the inspiral than the comparison in Fig.~\ref{fig:1PAT1vs1PAT1e}, but we highlight the initial mass ratio is now $\ee=1/q=1/4$. Note the precession modulations in the $(\ell,m)=(2,2)$ mode are too small to be visible in the plot, a fact that we foreshadowed in Fig.~\ref{fig:RelAmps}.} 
 \end{figure*}

\begin{figure*}[htb!]
\center
\includegraphics[width=.8625\textwidth]{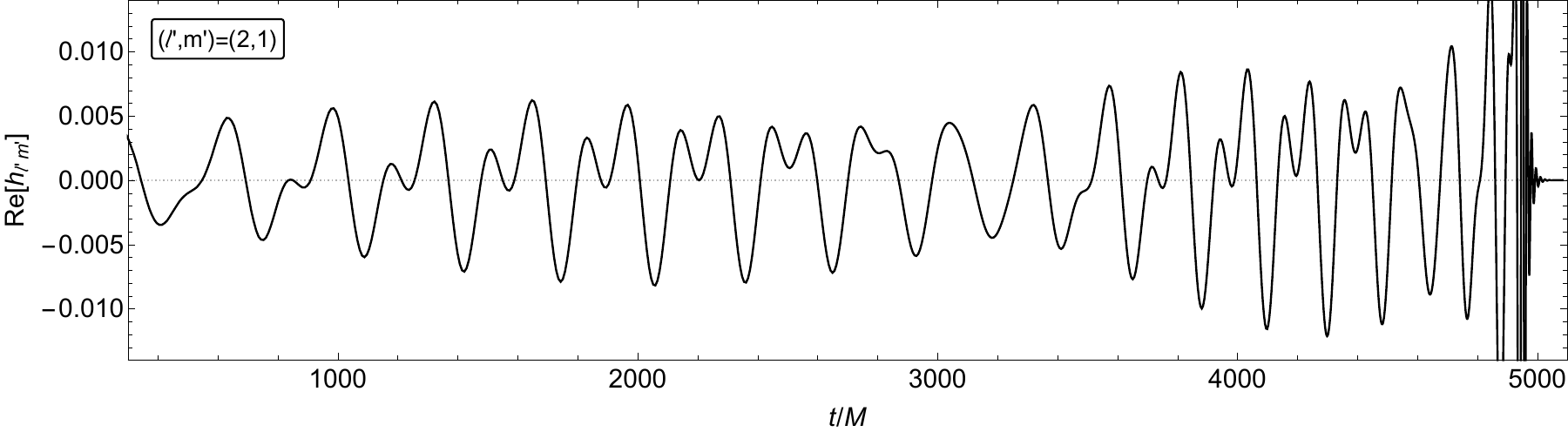}

\includegraphics[width=.8625\textwidth]{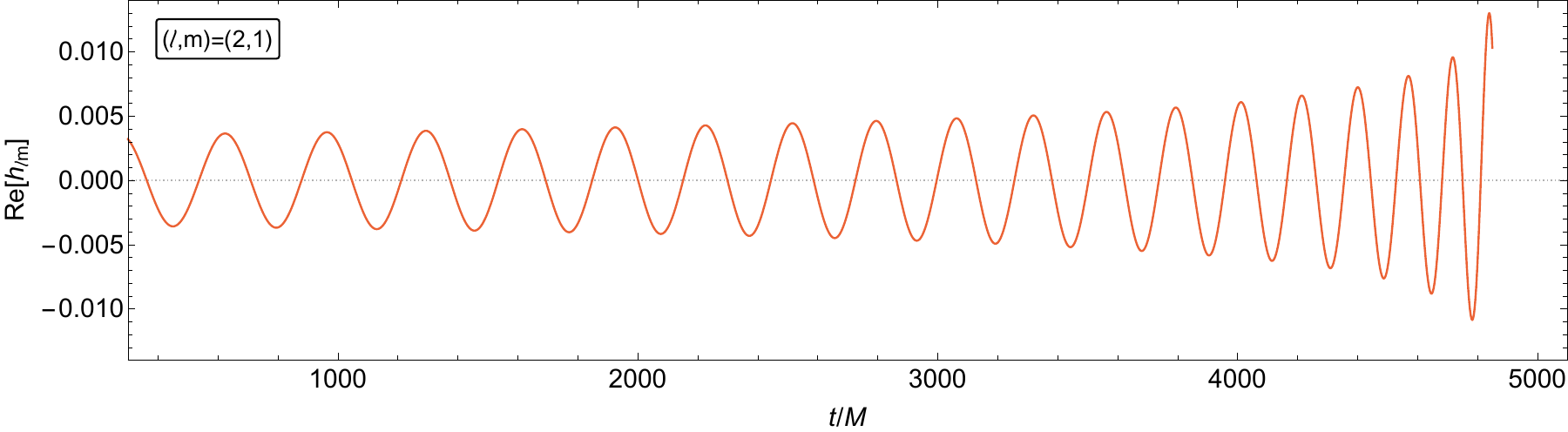}
\caption{\label{fig:precess21} 
An illustration of the frame difference between the self-force and NR precessing waveforms. \textbf{Top:} The $(\ell',m')=(2,1)$ mode of the same NR simulation as in Fig.~\ref{fig:preccomparison1}~\cite{sxs_collaboration_2019_3273407} with $q=4$ and a precessing secondary spin. \textbf{Bottom:} The $(\ell,m)=(2,1)$ mode of the 1PAT1R model (in red) corresponding to the same binary configuration.}
\end{figure*}

Our results for each model center around comparisons with NR waveforms from the SXS catalog \cite{Boyle:2019kee, Scheel:2025jct}. We examine binaries with mass ratios $q=1/\ee$ between 1 and 10, primary spins $|\chi_1|\leq0.12$, and secondary spins $|\chi_2|\leq0.8$. Of the five models we have defined, we find the general (somewhat expected) accuracy hierarchy for moderately asymmetric-mass binaries:
\begin{align*}
&\text{1PAT1-}\chi \sim \text{1PAT1e-}\chi\\
\lesssim &\text{1PAT1-}a \sim \text{1PAT1e-}a\\
\lesssim &\text{1PAT1R},
\end{align*}
with the re-summed model demonstrating the greatest accuracy.

In Fig.~\ref{fig:1PAT1vs1PAT1e}, we demonstrate that the original 1PAT1 model's approximation of neglecting the evolution of the primary's mass and spin was robust. In Fig.~\ref{fig:1PAT1vs1PAT1R}, we show that the 1PAT1R model offers a notable improvement on the original non-spinning 1PAT1 model for an equal-mass binary.

In Figs.~\ref{fig:spinaligned1},~\ref{fig:spinaligned2} and ~\ref{fig:spinaligned3}, we consider 3 different binaries with (anti-)aligned spins and consecutively increasing spin on the primary. Across all three simulations, the 1PAT1e-$a$ waveform is slightly more faithful to the NR waveform than the 1PATe-$\chi$ model. The 1PAT1R model performs best, with moderate gains when the spin of the primary is small and significant gains as the primary's spin increases. It is clear from Fig.~\ref{fig:spinaligned1} that all three models accurately capture the effects of the secondary's spin. 
However, as we increase the spin of the primary, our models struggle to stay in phase with the NR waveforms. Reference~\cite{Warburton:2021kwk} found good agreement between the self-force gravitational energy flux and the flux computed from the NR simulation used in Fig.~\ref{fig:spinaligned3}. 
Despite this, the 1PAT1e models dephase significantly with respect to the NR waveform. 
We speculate that the main source of the dephasing is from the approximation of $\frac{\partial E}{\partial\Omega}$ when including only linear order in the primary's spin. This could also explain why the 1PAT1R model offers a larger improvement for larger spins: the location of the ISCO varies significantly with the primary's spin.

In Figs.~\ref{fig:spinaligned4},~\ref{fig:spinaligned5}, we compare with the most mass-asymmetric simulations available in the SXS catalog ($q=15$) while limiting the primary's spin to be slow. 

In Fig.~\ref{fig:preccomparison1}, we find similar phase accuracy hierarchy among the 1PAT1 models in comparison with an NR simulation with a very slowly spinning primary and a generic (precessing) spinning secondary. There is a caveat to the comparison: the 1PAT1 models and NR simulation are in  different frames. The precessing waveforms in the SXS catalog fix the $z$-axis along the direction of the initial PN orbital angular momentum. In the self-force formalism, our frame is inherited from the symmetries of the Kerr metric (modulo perturbations) with the $z$-axis fixed along the primary's spin axis (at least at leading order). However, because the opening angle between the primary's spin and the orbital angular momentum is small and the $(2,2)$ mode is dominant in each of the respective frames, the $(\ell,m)=(2,2)$ mode in the self-force frame is approximately equal to the $(\ell',m')=(2,2)$ in the NR frame. In Fig.~\ref{fig:precess21}, we plot both the $(\ell',m')=(2,1)$ mode of the NR simulation and the corresponding $(\ell,m)=(2,1)$ waveform mode of the 1PATR model to highlight the frame difference. A small frame rotation away from the self-force frame mixes some of the numerically dominant $(\ell,m)=(2,2)$ mode into the $(\ell',m')=(2,1)$ mode which appears as the higher-frequency modulations visible in the NR simulation. We leave the comparisons of the other modes with the appropriate frame transformation to followup work.

\subsection{Summary: waveform models}

In Sec.~\ref{sec:SFwaveform}, we provided an overview of the native self-force waveform model that follows directly from the multiscale analysis in Sec.~\ref{sec:tt}. 

In Sec.~\ref{sec:Rexpand}, we described five waveform models as straight-forward modifications of the native self-force model. In particular, the 1PAT1e-$a$ (1PAT1e-$\chi$) are simple re-expansions of the Taylor series in the native model from expansions in $\epsilon$ at fixed $\tilde{a}_i$ ($\chi_i$) to expansions in $\nu$ at fixed $\tilde{a}_i$ ($\chi_i$). The 1PAT1-$a$ and 1PAT1-$\chi$ are similarly designed but neglect the evolution of the primary black hole's mass and spin. Finally, we introduced the re-summed 1PAT1R model, which derives from an effective re-summation of the 1PAT1e-$a$ model inspired by the balance law in Sec.~\ref{sec:energy balance law}. 

In Sec.~\ref{sec:results}, we performed a series of comparisons of the five waveform models against NR simulations which indicated that the 1PAT1R model was most faithful to NR. Our key findings are summarized in our concluding remarks in the following section.

\section{Conclusions}
\label{sec:conclusions}

In this work, we have extended the multiscale analysis of Ref.~\cite{Mathews:2021rod} to binaries with a spinning secondary body with a generic precessing spin in a quasi-circular inspiral about a slowly spinning primary black hole, where the primary's spin has at most a small spin misalignment from the orbital angular momentum. Or equivalently, we specialized the generic framework of Paper I to this binary configuration. We went on to describe how we obtain the 1PA inspiral evolution using the first law of binary black hole mechanics. There we also highlighted sources of error and recently discovered missing ingredients in our waveform, which will be the subject of future work. We provided an overview of the necessary offline computations for our 1PA waveform models, primarily consisting of  solving the perturbative Einstein equations in multiscale form. We then defined and implemented five waveform models, all equivalent to one another at 1PA order, and assessed their accuracy against NR simulations of comparable or moderately asymmetric-mass binaries.

Our comparisons with NR indicate that self-force theory accurately models the effects of the secondary's spin, even for rapid spins. We showed that using a simple re-summation in the 1PAT1R model significantly improved the accuracy of self-force models for lower values of the mass ratio $q$ and higher values of $\chi_1$. When completed with the plunge, merger and ringdown, it is quite possible that self-force waveforms can be used to directly model binaries with much lower values of $q$ than previously anticipated.

We found that the slow primary spin condition is unsurprisingly quite restrictive, underlining the need to develop second-order self-force calculations with a Kerr background metric. That being said, companion papers~\cite{Honet:2025gge,Honet:2025lmk} will show that there is significant promise in effectively replacing second-order dissipative effects with PN information; see also Refs.~\cite{Nagar:2022fep,Albertini:2022dmc,Albertini:2023aol}.

There are many possible avenues for extending this work. First and foremost, a followup paper in preparation will compare NR and self-force waveforms with precessing secondary spins in the same source frame, which allows a proper comparison of modes other than our approximate comparison of the $(\ell, m)=(2,2)$ mode here. Although the effect of precession in our model is very small (2PA, as opposed to the 1PA effects we focus on), comparison to NR simulations will be a valuable first step toward comparisons of more generic precessing systems, as well as a probe of how significant the secondary spin's prececession might be for comparable-mass systems. There have been an increasing number of NR simulations of spinning binaries (some including precession) performed for larger values of $q$ ~\cite{Boyle:2019kee, Healy:2022wdn, Ferguson:2023vta, Hamilton:2023qkv}, and the companion paper~\cite{Honet:2025lmk} includes further comparisons with these in the non-precessing case. Reference~\cite{MacUilliam:2024oif} highlighted that for many effective waveform models of precessing binaries, their faithfulness to NR decreases with increasing~$q$, motivating further development of precessing self-force waveforms to calibrate other models at higher values of~$q$.

Secondly, it is also desirable to perform a more quantitative exploration of the 1PAT1R model's accuracy at decreasing $q$ and increasing $|\chi_1|$. However, such analysis must be careful not to be muddied by the breakdown of the multiscale expansion at the ISCO and would be best performed with self-force models extended past the inspiral stage of the waveform. An obvious avenue to do so would be via  comparisons with state-of-the-art effective-one-body (EOB) models such as \texttt{SEOBNR} \cite{Pompili:2023tna,Ramos-Buades:2023ehm, Leather:2025nhu} on top of extensions of the existing comparisons with \texttt{TEOBResumS} ~\cite{Albertini:2022rfe,Albertini:2022dmc,Albertini:2023aol, Albertini:2024rrs, Gamba:2021ydi} and by new comparisons with the latest \texttt{Phenom} models \cite{Pratten:2020ceb, Pratten:2020fqn, Garcia-Quiros:2020qpx, Estelles:2021gvs, Hamilton:2021pkf, Thompson:2023ase, Yu:2023lml} and NR surrogates (see Ref.~\cite{Varma:2019csw} for example).

Thirdly, the 1PA models we have presented are technically valid for all sufficiently compact secondaries, since their quadrupolar structure first enters the waveform at 2PA order. To date, 1PA self-force waveforms have only been benchmarked for BH-BH binaries. It would be intriguing to investigate our consistency with BH-NS models such as those of Refs.~\cite{Thompson:2020nei,Matas:2020wab}.

In terms of model improvements, it is possible to relax the restriction of the small opening angle between the primary's spin and orbital angular momentum to extend the 1PA models to generic precession of both spins. All the required strong-field self-force data is either readily available or easy to compute with existing codes~\cite{BHPToolkit}. However, we would still need to restrict the primary's spin to be slowly spinning until 2SF fluxes are available with a Kerr background spacetime. 

The 1PAT1R and 1PAT1e-$a$ models are publicly available in the \texttt{WaSABI} package~\cite{BHPT_WaSABI} in the Black Hole Perturbation Toolkit, though limited at the time of writing to the non-precessing case until a future update.

\begin{acknowledgments}

JM thanks Jonathan Thompson for helpful discussion on source frame transformations of precessing waveforms and Loïc Honet for assistance in model consistency checks and \texttt{WaSABI} development. We also thank Soichiro Isoyama and Alvin Chua for feedback on earlier drafts of this work.

JM acknowledges support from the Irish Research Council under grant GOIPG/2018/448 and by the NUS Faculty of Science, under the research grant 22-5478-A0001. AP acknowledges the support of a Royal Society University Research Fellowship and the ERC Consolidator/UKRI Frontier Research Grant GWModels (selected by the ERC and funded by UKRI [grant number EP/Y008251/1]). 
NW acknowledges support from a Royal Society - Research Ireland University Research Fellowship. This publication has emanated from research conducted with the financial support of Research Ireland under Grant numbers 16/RS-URF/3428, 17/RS-URF-RG/3490 and 22/RS-URF-R/3825. 
This work is supported by ERC grant EMRIWaveforms (\href{https://doi.org/10.3030/101200625}{https://doi.org/10.3030/101200625}).
This work makes use of the Black Hole Perturbation Toolkit~\cite{BHPToolkit}.

\end{acknowledgments}

\appendix

\begin{widetext}
\section{Dipole stress-energy components}
\label{app:Kcomps}
The non-zero \emph{symmetric} components of the tensors appearing in Eq.~\eqref{eq:ttdipoleSE} are
\begingroup\allowdisplaybreaks%
\begin{align}%
\label{Kcomps}
&K_{1}^{t t}=\frac{- \chi_\parallel \left(m_1^{(0)}\right)^{3 / 2}}{r_0\left(r_0-2 m_1^{(0)}\right) \sqrt{r_0-3 m_1^{(0)}}}, \quad K_{1}^{t \phi}=-\chi_\parallel \frac{m_1^{(0)}}{r_0^{5 / 2} \sqrt{r_0-3 m_1^{(0)}}}, \nonumber \\
&K_{1}^{r r}=-\chi_\parallel \frac{\left(m_1^{(0)}\right)^{1 / 2}\left(r_0-2 m_1^{(0)}\right) \sqrt{r_0-3 m_1^{(0)}}}{r_0^{3}}, \quad K_{1}^{\phi \phi}=- \chi_\parallel \frac{\left(m_1^{(0)}\right)^{1 / 2}\left(r_0-2 m_1^{(0)}\right)}{r_0^{4} \sqrt{r_{0}-3 m_1^{(0)}}}, \nonumber \\
&K_{1}^{r \theta}=-\chi_\perp \sqrt{\frac{f_0}{r_0^2}}\frac{\Omega}{2 u^t_0} \cos\tilde\psi_s, \quad K_{1}^{\theta \phi }= \chi_\perp \frac{(r_0 -3m_1^{(0)})}{2 \sqrt{f_0} r_0^3} \Omega \sin \tilde\psi_s, \nonumber \\
&K_{2}^{t r}=\chi_\parallel\frac{\sqrt{r_0-3 m_1^{(0)}}}{2 r_0^{3 / 2}}, \quad K_{2}^{r \phi}=\chi_\parallel\frac{\left(m_1^{(0)}\right)^{1/2}\sqrt{r_0-3 m_1^{(0)}}}{2 r_0^{3}},  \quad K_{2}^{t \theta}=\chi_\perp\frac{\cos \tilde\psi_s}{2 \sqrt{f_0} r_0^{2} u^t_0},  \quad K_{2}^{\theta \phi}= \Omega  K_{2}^{t \theta},  \nonumber \\
&K_{3}^{t t}=-\chi_\parallel\frac{\left(m_1^{(0)}\right)^{1/2}}{\sqrt{r_0-3 m_1^{(0)}}}, \quad K_{3}^{t \phi}=-\chi_\parallel\frac{(r_0-m_1^{(0)})}{2 r_0^{3 / 2} \sqrt{r_0-3 m_1^{(0)}}}, \quad K_{3}^{\phi \phi}=-\chi_\parallel\frac{\left(m_1^{(0)}\right)^{1/2}\left(r_0-2 m_1^{(0)}\right)}{r_0^{3} \sqrt{r_0-3 m_1^{(0)}}}, \nonumber \\
&K_{3}^{t \theta}=\chi_\perp\frac{ \sqrt{f_0} \sin \tilde\psi_s}{2 r_0}, \quad K_{3}^{\theta \phi}= \Omega K_{3}^{t \theta}, \quad K_{4}^{t t}=-\chi_\perp\frac{\Omega r_0 \cos \tilde\psi_s}{\sqrt{(r_0-2m_1^{(0)})(r_0-3m_1^{(0)})}}, \nonumber \\
&K_{4}^{t r}=-\chi_{\perp}\frac{\sqrt{f_0}\sin \tilde\psi_s}{2 r_0}, \quad K_{4}^{t \phi}=-\chi_\perp\frac{ (r_0-m_1^{(0)}) \cos \tilde\psi_s}{2 r_0^2 \sqrt{(r_0-2m_1^{(0)})(r_0-3m_1^{(0)})}}, \quad K_{4}^{r \phi}= \Omega K_{4}^{t r}, \nonumber \\
&K_{4}^{\phi \phi}= -\chi_{\perp}\frac{\Omega}{r_0^2}\sqrt{\frac{r_0-2m_1^{(0)}}{r_0-3m_1^{(0)}}}\cos \tilde\psi_s.
\end{align}%
\endgroup%
\end{widetext}

\section{Computing $h_{\alpha\beta}^{2(\chi_\perp)}$}
\label{app:chiperp}
Since $h_{\alpha\beta}^{2(\chi_\perp)}$ is a linear perturbation to a Schwarzschild background metric, it can be readily computed using the Regge-Wheeler-Zerilli (RWZ) formalism~\cite{Martel:2005ir,Pound:2021qin}. The calculation involves only minor modifications to the frequency-domain approach in Sec.~V of Ref.~\cite{Mathews:2021rod}. The only modifications are the presence of additional terms in the RWZ source and the extension of the frequency spectrum with $k=\pm1$ modes (as well as  changes to the $\ell=0$ and $\ell=1$ modes of the metric). We summarise the calculation in this appendix.

Both the Zerilli-Moncrief (ZM) and the Cunningham-Price-Moncrief (CPM) master functions satisfy a wave equation (the RWZ equation), which in the frequency domain is of the form
\begin{equation}
\label{eq:RWZfreq}
\left[\dfrac{\partial^{2}}{\partial r_{*}^{2}} -V_{\ell}(r) + \omega^{2} \right]\psi_{\ell m \omega}(r)=Z_{\ell m \omega}(r),
\end{equation}
where $r_{*}$ is the usual Schwarzschild tortoise coordinate. The expressions for the potential $V_{\ell}(r)$ and the source term $Z_{\ell m \omega}$ are different in each parity sector. In any case, the source term for each master function derives from the stress-energy. In the pole-dipole approximation when the leading motion of the secondary is circular and approximately equatorial, the source is of the form
\begin{equation}
 \label{eq:explicitsource}
 Z_{\ell m\omega} = \left( \bar{G}_{\ell m \omega}\delta_{r_0} + \bar{F}_{\ell m \omega} \delta_{r_0}' + \bar{H}_{\ell m \omega} \delta_{r_0}'' \right),
\end{equation}
where we use the shorthand $\delta_{r_0}\equiv\delta(r-{r_0})$. $\bar{G}_{\ell m\omega}$, $\bar{F}_{\ell m\omega}$ and $\bar{H}_{\ell m\omega}$ depend only on ${r_0}$, $\chi_{\parallel}$, $\chi_{\perp}$, $m_2$ and $m_1^{(0)}$; we derive explicit expressions for these in the following subsections as well as defining the potentials, master functions and sources. 

Substituting the source~\eqref{eq:explicitsource}, we obtain the solutions to Eq.~\eqref{eq:RWZfreq} in terms of the basis of homogeneous solutions via variation of parameters and integrating by parts;
\begin{multline}
\label{eq:formR}
\psi_{\ell m\omega}(r) = C^{+}_{\ell m\omega}R_{\ell m\omega}^+(r)\Theta^+_{r_0} \\+ C^{-}_{\ell m\omega}R_{\ell m\omega}^-(r)\Theta^-_{r_0} + \frac{\bar{H}_{\ell m \omega}}{f_0^2}\delta_{r_0}, 
\end{multline} 
having introduced the shorthand for the Heaviside step functions $\Theta^{\pm}_{r_0}\equiv\Theta(\pm(r-r_0))$. In this work, we compute the homogeneous solutions $R_{\ell m\omega}^\pm(r)$ using the \texttt{ReggeWheeler} package of the Black Hole Perturbation Toolkit~\cite{BHPToolkit}.  Meanwhile, the matching coefficients are given by 
\begin{align}
C_{\ell m\omega}^{\pm}  = & \frac{1}{W_{\ell m\omega}} \frac{R_{\ell m\omega}^{\mp}(r_0) }{f_0}\bar{G}_{\ell m\omega}\nonumber \\ 
&-\frac{1}{W_{\ell m\omega}} \frac{d}{dr}\left. \left(\frac{R_{\ell m\omega}^{\mp}(r) }{f(r)}\right) \right\rvert_{r_0} \bar{F}_{\ell m\omega}\nonumber  \\
&+ \frac{1}{W_{\ell m\omega}} \frac{d^2}{dr^2}\left. \left(\frac{R_{\ell m\omega}^{\mp}(r) }{f(r)}\right) \right\rvert_{r_0} \bar{H}_{\ell m\omega} \label{eq:MatchingCalc},
\end{align}
and we have defined the Wronskian
\begin{equation}
W_{\ell m\omega}\equiv R_{\ell m\omega}^-\frac{dR_{\ell m\omega}^+}{dr_*}-R_{\ell m\omega}^+\frac{dR_{\ell m\omega}^-}{dr_*}.
\end{equation}

Once the master function is determined, the metric may be reconstructed in the frequency domain via differential operators acting upon $\psi_{\ell m \omega}$~\cite{Pound:2021qin,Hopper:2010uv}. In practice, we extract the GW strain amplitudes~\eqref{eq:waveformmodes} directly from the master variables~\cite{Pound:2021qin}:
\begin{equation}
R_{\ell m k} = 
\frac{1}{2}\sqrt{\frac{(\ell+2)!}{(\ell-2)!}} \left(C_{\ell m \omega}^{+ (\mathrm{ZM})}
-i C_{\ell m \omega}^{+(\mathrm{CPM})}\right),
\end{equation}
where 
$C_{\ell m \omega}^{+ (\mathrm{ZM})}$ is the matching coefficient in the ZM master fuction computed in the even parity sector and $C_{\ell m \omega}^{+ (\mathrm{CPM})}$ is the matching coefficient in the CPM master function computed in the odd parity sector.

\subsection{Even-parity perturbations}
In the even-parity sector, the spherical harmonic decomposition of the metric perturbation reads 
\begin{subequations}
\label{eq:evenmetric}
\begin{align}
h_{a b}&=\sum_{\ell, m} h_{a b}^{\ell m} Y^{\ell m}, \\
h_{a B}&=\sum_{\ell, m} j_{a}^{\ell m} Y_{B}^{\ell m}, \\
h_{A B}&=r^{2} \sum_{\ell, m}\left(K^{\ell m} \Omega_{A B} Y^{\ell m}+G^{\ell m} Y_{A B}^{\ell m}\right),
\end{align}
\end{subequations}
where $Y_{A B}^{\ell m}$, $Y_{B}^{\ell m}$ and $Y^{\ell m}$ are the even-parity tensor, vector and scalar spherical harmonics as defined in Ref.~\cite{Martel:2005ir}, and $\Omega_{AB}={\rm diag}(1,\sin^2\theta)$ is the metric on the unit two-sphere. 
Note that we use the same convention as Ref.~\cite{Martel:2005ir}, with uppercase Latin indices for coordinates on the two-sphere ($\theta$ and $\phi$ in standard Schwarzschild coordinates) and lowercase Latin indices (except $\ell$ and $m$) for coordinates that span the remaining submanifold of the Schwarzschild metric (with coordinates $t$ and $r$, for example). In Eq.~\eqref{eq:evenmetric}, $\ell$ is implied to be summed over all positive integers in the scalar sector, all integers with $\ell \geq 1$ in the vector sector and $\ell \geq 2$ in the tensor sector and $m$ is summed over all integers such that $-\ell\leq m \leq \ell $. 

From hereafter we drop the `$\ell m$' sub/superscripts for ease of notation. The even-parity projections of the stress-energy tensor onto the same spherical harmonic basis are
\begin{subequations}
\label{eq:qsources}
\begin{align}
Q^{a b}(t, r) & \equiv 8 \pi \int T^{a b} Y^{*} d \Omega,\\
Q^{a}(t, r) & \equiv \frac{16 \pi r^{2}}{\ell(\ell+1)} \int T^{a B} Y_{B}^{*} d \Omega, \\
Q^{\flat}(t, r) & \equiv 8 \pi r^{2} \int T^{A B} \Omega_{A B} Y^{*} d \Omega, \\
Q^{\sharp}(t, r) & \equiv 32 \pi r^{4} \frac{(\ell-2) !}{(\ell+2) !} \int T^{A B} Y_{A B}^{*} d \Omega,
\end{align}
\end{subequations}
where $d \Omega= \sin \theta d\theta d\phi$, and a star denotes complex conjugation.

The Zerilli gauge condition sets $j_{a}^{\ell m}=0=G^{\ell m}$. The Zerilli-Moncrief master function is then given in terms of the nonzero harmonic coefficients \cite{Hopper:2010uv}:
\begin{equation}
\psi_{\mathrm{even}} \equiv \frac{2 r}{\ell(\ell+1)}\left[K+\frac{1}{\Lambda}\left(f^2 h_{r r}-r f \partial_r K\right)\right].
\end{equation}
It satisfies the RWZ equation~\eqref{eq:RWZfreq} with the Zerilli potential,
\begin{equation}
V_{\text{even}}\equiv \frac{f}{r^2 \Lambda^{2}}\Biggl[2\lambda^{2} \left(\lambda+1+\frac{3}{\bar r} \right)
+ \frac{18}{\bar{r}^{2}} \left(\lambda+\frac{1}{\bar{r}} \right) \Biggr],
\end{equation}
where $\bar{r}\equiv r/m_1^{(0)}$, $\lambda \equiv (\ell+2)(\ell-1)/2$ and $\Lambda \equiv \lambda + 3/\bar{r}$. The even-parity source is written in terms of the harmonic coefficients of the stress-energy tensor:
\begin{align}
\label{eq:evensource}
S_{even} \equiv& \frac{1}{(\lambda+1)}\biggl\{\frac{r^{2} f}{\Lambda}\left(f^{2} \partial_{r} Q^{t t}-\partial_{r} Q^{r r}\right) 
+\frac{r}{\Lambda}(\Lambda-f) Q^{r r}\nonumber \\
&\quad +\frac{r f^{2}}{\Lambda} Q^{\flat} -\frac{f^{2}}{r \Lambda^{2}}\biggl[\lambda(\lambda-1) r^{2}+(4 \lambda-9) m_1^{(0)} r \nonumber \\
&\quad +15 \left(m_1^{(0)}\right)^{2}\biggr] Q^{t t}\biggr\}
+\frac{2 f}{\Lambda} Q^{r}-\frac{f}{r} Q^{\sharp}.
\end{align}

After substituting the stress-energy, neglecting non-linear perturbation terms and taking the Fourier transform, our frequency-domain source has the form given in Eq.~\eqref{eq:explicitsource}. We split the non-spinning, aligned-spin and precessing spin contributions as 
\begin{subequations}
\label{SourceAmpSplit}
\begin{align}
\bar{H}_{\ell m \omega}&=0,\\
\bar{F}_{\ell m \omega}&=
\begin{cases} 
\ee F^{1}_{\ell m \omega} + \ee^2 \chi_\parallel F^{2 (\chi_\parallel)}_{\ell m \omega} & \omega=m \Omega,\\
\ee^2\chi_\perp F^{2 (\chi_\perp)}_{\ell m \omega} & \omega=m \Omega +k  \Omega_s^{(0)},
\end{cases}\\
\bar{G}_{\ell m \omega}&=
\begin{cases} 
\ee G^{1}_{\ell m \omega} + \ee^2\chi_\parallel G^{2 (\chi_\parallel)}_{\ell m \omega} & \omega=m \Omega,\\
\ee^2\chi_\perp G^{2 (\chi_\perp)}_{\ell m \omega} & \omega=m \Omega +k\Omega_s^{(0)},
\end{cases}
\end{align}
\end{subequations}
for integer modes over $|k| \leq1$. Each term is given by
\begin{widetext}
\begin{subequations}
\begin{align}
F^{1}_{\ell m \omega} &= \frac{8 \pi m_1^{(0)} \bar{r} f_0^3 u^t_0 }{(\lambda \bar{r} +3) (\lambda+1) }Y_{\ell m}\left(\frac{\pi}{2},0\right), \\
F^{2 (\chi_\parallel)}_{\ell m \omega} &=-\frac{8 \pi m_1^{(0)}   f_0 (\bar{r}-3 ) u^t_0 }{(\lambda \bar{r} +3) (\lambda+1) \bar{r}^{5/2}}\left[ 3 \left(1-m^2\right) \frac{1}{\lambda}+ \bar{r} \left(\lambda -m^2\right)+\frac{\left(-7- \bar{r}+\bar{r}^2\right)}{(\bar{r}-3 ) } \right]Y_{\ell m}\left(\frac{\pi}{2},0\right), \\
F^{2 (\chi_\perp)}_{\ell m \omega} &=-\frac{8 \pi m_1^{(0)} k^2 f_0^{5/2} \sqrt{\bar{r}} }{(\lambda +1) u^t_0 (3+\lambda  \bar{r})}\sqrt{2 (\lambda +1)-m(m+1)}Y_{\ell (m+1)}\left(\frac{\pi }{2},0\right), \\
G^{1}_{\ell m \omega} &=\frac{8 \pi  f_0 u_0^t}{ (\lambda +1) \bar{r}^2
   (3 +\lambda  \bar{r})^2}  Y_{\ell m}\left(\frac{\pi}{2},0\right)\times\nonumber\\
  &\quad \left\{3 \left[5  +(3/\lambda)(m^2-1)\right]+2   \bar{r} \left(2
   \lambda +3 m^2-9\right) +\lambda \bar{r}^2 \left(\lambda +m^2-4\right)-\lambda (\lambda +1) \bar{r}^3\right\}, \\
G^{2 (\chi_\parallel)}_{\ell m \omega} &= -\frac{8 \pi  u_0^t}{\lambda  (\lambda +1) \bar{r}^{9/2} (3 +\lambda  \bar{r})^2} Y_{\ell m}\left(\frac{\pi}{2},0\right) \left\{(4 -\bar{r}) \left[9 \left(m^2-1\right) -3 \lambda   \left(1+ \bar{r}+\bar{r}^2\right) +\lambda^3 \bar{r}^2 (1-\bar{r})\right] \right. \nonumber\\
& \qquad\qquad \qquad \qquad\qquad \qquad \left. +\lambda ^2 \bar{r} \left[-m^2 \bar{r} \left(2 -4  \bar{r}+\bar{r}^2\right)+4 -10  \bar{r}-2  \bar{r}^2+\bar{r}^3\right]+3 \lambda  m^2  \bar{r} \left(2 +3 \bar{r}-\bar{r}^2\right)\right\},\\
G^{2 (\chi_\perp)}_{\ell m \omega} &=
\frac{4\pi k f_0^{3/2}  \left[3  (k m u^t_0+4)+\lambda   \bar{r} (k m u^t_0+5)+\lambda  (\lambda
   +1) \bar{r}^2\right]}{(\lambda +1) u^t_0 (3 +\lambda  \bar{r})^2 \bar{r}^{3/2}}\sqrt{2 (\lambda +1)-m (m+1)}Y_{\ell (m+1)}\left(\frac{\pi }{2},0\right).
\end{align}
\end{subequations}
\end{widetext}
It is worth noting that the modes of the master function in the non-precessing sector satisfy $\psi_{\ell m \omega}=(-1)^m \psi_{\ell (-m) \omega}^*$. They are also only non-zero when $\ell + m$ is even due to their proportionality to $Y_{\ell m}\left(\frac{\pi}{2},0\right)$, which indicates the sector is symmetric about the equatorial plane. In contrast, modes in the precessing sector are only non-zero when $\ell + m$ is odd due to their proportionality to $Y_{\ell (m+1)}\left(\frac{\pi}{2},0\right)$, corresponding to the precessing sector's anti-symmetry about the equatorial plane. If we label the two parts of the precessing spectrum as $ \omega_{+}= m \Omega +\Omega_s^{(0)}$ and $ \omega_{-}= m \Omega -\Omega_s^{(0)}$, then we will have a symmetry in the precessing part of the master function such that $\psi_{\ell m \omega_{+}}=(-1)^m \psi_{\ell (-m) \omega_{-}}^*$ and $\psi_{\ell (-m) \omega_{+}}=(-1)^m \psi_{\ell m \omega_{-}}^*$. So again,  we only need to compute modes with positive $m$ values.

\subsection{Odd-parity perturbations}

In the odd-parity sector, the spherical harmonic decomposition of the metric perturbation is
\begin{subequations}
\begin{align}
\label{eq:oddmetric}
h_{a b}&=0,\\
h_{a B}&=\sum_{\ell, m} h_{a}^{\ell m} X_{B}^{\ell m},\\
h_{A B}&=\sum_{\ell, m} h_{2}^{\ell m} X_{A B}^{\ell m},
\end{align}
\end{subequations}
where $X_{A B}^{\ell m}$ and $X_{B}^{\ell m}$ are the odd-parity tensor and vector harmonics as defined in Ref.~\cite{Martel:2005ir}. Again, $\ell$ is implied to be summed over all integers with $\ell \geq 1$ in the vector sector and $\ell \geq 2$ in the tensor sector, and $m$ is summed over all integers such that $-\ell\leq m \leq \ell $.  The odd-parity projections of the stress-energy tensor onto the same spherical harmonic basis are
\begin{equation}
\label{pequation}
P^{a}(t, r) \equiv \frac{16 \pi r^{2}}{\ell(\ell+1)} \int T^{a B} X_{B}^{*} d \Omega,
\end{equation}
and we again drop ``$\ell m$" sub/superscripts where convenient. 

The Regge-Wheeler gauge condition is  $h_2^{\ell m}=0$. Given this gauge choice, the Cunningham-Price-Moncrief master function is defined as
\begin{equation}
\psi_{\mathrm{odd}} \equiv \frac{r}{\lambda}\left(\partial_r h_t-\partial_t h_r-\frac{2}{r} h_t\right),
\end{equation}
and satisfies the RWZ equation~\eqref{eq:RWZfreq} with the Regge-Wheeler potential,
\begin{equation}
V_{\text{odd}}\equiv \frac{f}{r^2}\left[\ell(\ell+1) - \frac{6}{\bar{r}} \right].
\end{equation}

The odd-parity source term is given by
\begin{equation}
\label{defnoddsource}
S_{odd}(t, r) \equiv \frac{r f}{\lambda}\left(\frac{1}{f} \partial_{t} P^{r}+f \partial_{r} P^{t}+\frac{2 m_1^{(0)}}{r^{2}} P^{t}\right).
\end{equation}
The analysis of the frequency-domain source in the odd-parity sector is precisely analogous to the even-parity sector with different coefficients for the distributional terms:
\begin{widetext}
\begin{subequations}
\begin{align}
H^{1}_{\ell m \omega}&=0,\\
H^{2 (\chi_{\parallel})}_{\ell m \omega}&=  -\frac{4 \pi  f_0^2}{\lambda (\lambda+1) u_0^t }\left(m_1^{(0)}\right)^2 \sqrt{(\ell-m)(1+\ell+m)} Y_{\ell (m+1)}\left(\frac{\pi}{2},0\right), \\
H^{2 (\chi_{\perp})}_{\ell m \omega}&= 
\frac{-4 \pi   f_0^{5/2} m k }{\lambda  l (l+1)}\left(m_1^{(0)}\right)^2Y_{\ell m}\left(\frac{\pi}{2},0\right),\\
F^{1}_{\ell m \omega}&= \frac{8 \pi m_1^{(0)} f_0^2 u_0^t}{\lambda (\lambda+1) \sqrt{\bar{r}}}\sqrt{(\ell-m)(1+\ell+m)} Y_{\ell (m+1)}\left(\frac{\pi}{2},0\right), \\
F^{2 (\chi_\parallel)}_{\ell m \omega}&= -\frac{8 \pi   m_1^{(0)} f_0 u_0^t}{\lambda (\lambda+1)}\frac{\left( 1 +3 \bar{r} -\bar{r}^2\right)}{\bar{r}^3}\sqrt{(\ell-m)(1+\ell+m)} Y_{\ell (m+1)}\left(\frac{\pi}{2},0\right), \\
F^{2 (\chi_{\perp})}_{\ell m \omega}&=  
\frac{4 \pi m_1^{(0)} k^2  f_0^{3/2}  \left[\bar{r} \left(\ell^2+\ell-2 m^2\right)+6 m k  u^t_0\right]}{\lambda  \ell (\ell+1) \bar{r}^2 u^t_0}Y_{\ell m}\left(\frac{\pi}{2},0\right),\\
G^{1}_{\ell m \omega}&=- \frac{8 \pi f_0 u_0^t}{\lambda (\lambda+1) \bar{r}^{3/2}}\sqrt{(\ell-m)(1+\ell+m)} Y_{\ell (m+1)}\left(\frac{\pi}{2},0\right), \\
G^{2 (\chi_\parallel)}_{\ell m \omega}&= \frac{4 \pi u_0^t  \left[m^2 \bar{r} +(2-3m^2)\right]}{\lambda (\lambda+1) \bar{r}^4}\sqrt{(\ell-m)(1+\ell+m)} Y_{\ell (m+1)}\left(\frac{\pi}{2},0\right),\\
G^{2 (\chi_{\perp})}_{\ell m \omega}&=
-\frac{4 \pi k^2 \sqrt{f_0}  \left\{k m \left[\left(\ell^2+\ell+6\right)  u_0^t-m^2  u_0^t+ k m (-2 \bar{r})-2 \bar{r} u_0^t\right]+\ell (\ell+1) \bar{r} \right\}}{\lambda  \ell (\ell+1) \bar{r}^3 u_0^t}Y_{\ell m}\left(\frac{\pi}{2},0\right).
\end{align}
\end{subequations}
\end{widetext}

The master function in the non-precessing and precessing sectors has the same respective symmetries in $m$ as in the even-parity case. However, in the odd-parity, non-precessing sector the modes of the master function are only non-zero when $\ell + m$ is odd due to their proportionality to $Y_{\ell (m+1)}\left(\frac{\pi}{2},0\right)$ (because of the equatorial symmetry) and the precessing sector is only non-zero for modes in which $\ell + m$ is even due to their proportionality to $Y_{\ell m}\left(\frac{\pi}{2},0\right)$ (because of the equatorial anti-symmetry).

\bibliography{WaveformsSpin}

\end{document}